%% file: main.tex
\documentclass[12pt,a4paper]{article}

\input{preamble/format}
\input{preamble/preamble}

\input{preamble/math}

\usepackage{adjustbox}

\title{Bayesian Empirical Bayes: Simultaneous Inference from Probabilistic Symmetries}
\author{Bohan Wu\,$^{1}$ \and
Eli N. Weinstein$^{2}$ \and
David M. Blei$^{1,3}$}
\begin{document}
\maketitle
\footnotetext[1]{Department of Statistics, Columbia University, New York, NY}
\footnotetext[2]{Department of Chemistry, Technical University of Denmark, Denmark}
\footnotetext[3]{Department of Computer Science, Columbia University, New York, NY \\
\indent\indent Contact: \href{mailto:bw2766@columbia.edu}{bw2766@columbia.edu}, \href{mailto:enawe@dtu.dk}{enawe@dtu.dk}, \href{mailto:david.blei@columbia.edu}{david.blei@columbia.edu}}

\input{abstract.tex}
\input{introduction.tex}
\input{related_work.tex}
\input{method.tex}

\input{tools.tex}

\section{Simulation Studies} \label{sec-simulation}
We evaluate the new Bayesian empirical Bayes methods using synthetic data.
\begin{itemize}
\item In \Cref{sec-sim-EBMR}, we assess empirical Bayes matrix recovery (EBMR; \Cref{sec-EBMR}) across 60 settings by varying sample size, noise level, and the data-generating mechanism; we implement EBMR with a separately exchangeable prior (EBMR-Sep) and compare against NPMLE and empirical Bayes matrix factorization (EBMF; \cite{Wang2021EBMF}).
\item  In \Cref{sec-sim-CIEB}, we evaluate the covariate-informed empirical Bayes (CIEB; \Cref{sec-caeb}) method for denoising matrices with row- and column-specific covariates and compare it to EBMR-Sep, EBMF, and NPMLE.
\item  In \Cref{sec-sim-EBSR}, we apply empirical Bayes spatial regression (spatial EB; \Cref{sec-spatial}) to synthetic time-series data. We show that the MSE of EB spatial regression vanishes with increasing sample size, whereas the NPMLE plateaus at a constant risk.
\end{itemize}

In each study, we generate observations $\bx_{S_n}$ on a finite domain $S_n$ (e.g. $S_n = [n]$) from a known data-generating process with latent truth $\bz^\star_{S_n}$. We report the relative MSE to assess how well the BEB posterior mean $\hat \bz_{S_n}$ recovers the true latent values $\bz^\star_{S_n}$. The relative MSE measures the MSE of the estimator as a percentage of the MSE of the MLE (out of $100$):
\begin{equation*}
\textrm{R-MSE} = \frac{1}{\left|S_n\right|}\sum_{\omega \in S_n} \left(\hat z_\omega - z_\omega^\star \right)^2 \times 100 \times \tau, 
\end{equation*}
where $\tau$ is the precision of the noise distribution.

A better method achieves lower R-MSE. 
\subsection{Empirical Bayes Matrix Recovery} \label{sec-sim-EBMR}
In our first study, we simulate observations from the normal means model $x_{ij} \sim \cN \left(z_{ij}^\star, \tau^{-1}\right)$ with known precision $\tau > 0$. The true matrix $\bz_{n,p}^\star$ is a single draw from the Aldous--Hoover representation with a pre-specified function $T_0$, that is
\begin{equation*}
    u_i, v_j, u_{ij} \iid \unif[0,1], \qquad z_{ij}^\star = T_0\left(u_i, v_j, u_{ij}\right).
\end{equation*}

We consider five choices of $T_0$, and vary the precisions $\tau \in \{0.1, 0.25, 1, 4\}$ and the sample sizes $(n,p) \in \{(20,20),\ (20,50),\ (50,50)\}$. The different $T_0$ are as follows:
\begin{itemize}[leftmargin=1.5em]
  \item \textbf{Linear:} $T_0(U,V,W)=U+V+W$;
  \item \textbf{Sine-log:} $T_0(U,V,W)=\sin(\pi U V)+\log(1+W)$;
  \item \textbf{Sine-cos:} $T_0(U,V,W)=\dfrac{\sin(\pi U)\cos(\pi V)}{1+W^2}$;
  \item \textbf{Tanh:} $T_0(U,V,W)=\tanh(U+V+W)$;
  \item \textbf{Reciprocal:} $T_0(U,V,W)=\dfrac{1}{1+\lvert U+V+W\rvert}$.
\end{itemize}
 We examine all combinations of $(\tau,n,p,T_0)$ for a total of $60$ settings.  For each setting, we run the experiment 20 times to calculate the R-MSE. We compare EBMR with a separately exchangeable prior (EBMR-Sep, \Cref{alg:EBMR-sep}) against three benchmarks:
\begin{itemize}
\item \textbf{Benchmark 1: MLE.} The MLE is $\hat z_\omega = x_\omega$; by construction its R-MSE equals $100$.

\item \textbf{Benchmark 2: NPMLE.} NPMLE-based methods are state of the art for normal-means denoising problems \citep{Jiang2009,Koenker2014,Soloff2021}. In the context of matrix recovery, it is equivalent to EBMR with a fully exchangeable prior, $z_\omega \iid \rmg$. We estimate $\rmg$ via the nonparametric MLE using sequential quadratic programming \citep{Kim2020NPMLE}, implemented with \texttt{ebnm} in \textsf{R}.

\item \textbf{Benchmark 3: Empirical Bayes Matrix Factorization (EBMF).}
EBMF \citep{Wang2021EBMF} models the latent matrix as low rank with exchangeable loadings and factors,
\begin{equation*}
    \bz_{n,p} = \sum_{k=1}^C \ell_k f_k^\top,\qquad \ell_k \in \R^n,\ f_k \in \R^p,
\end{equation*}
with $\{\ell_k\}_{k=1}^C \iid$ from an unknown prior $\rmg_\ell$ and $\{f_k\}_{k=1}^C \iid$ from an unknown prior $\rmg_f$. The priors $(\rmg_\ell,\rmg_f)$ are fit by variational empirical Bayes under the normal-means likelihood for $\bx_{n,p}$, and we use the posterior mean of $\bz_{n,p}$ as the estimator. Unless noted otherwise, we set $C=10$ and implement the method using the \texttt{flashier} package in \textsf{R}.
\end{itemize}

To implement \Cref{alg:EBMR-sep}, we set $K=10$ and parameterize $\rmg_{\sep}:[0,1]^3 \mapsto \R$ as a fully connected feed-forward MLP with two hidden layers, ReLU activations, and 5 units per layer. We choose a simple network configuration because the sample sizes are small.

Table~\ref{tab:simul_EBMR} shows that EBMR-Sep attains the lowest R-MSE among the three empirical Bayes methods in $50$ out of the $60$ settings. When $T_0$ introduces nonlinear dependencies across rows and columns, for example the sine-log function, the error reduction of EBMR-Sep relative to NPMLE and EBMF is large. The exceptions occur at the smallest matrices $(n=20)$, where NPMLE is occasionally the best of the three, together with the largest sine-cos setting, where the near low-rank structure favours EBMF. Empirical Bayes matrix factorization is itself a special case of our framework that restricts the separately exchangeable prior to a low-rank parametric form, which lowers the effective number of parameters and makes it competitive when the latent matrix is close to low rank, as in the sine-cos settings.
\input{Tables/mse_table2}

\subsection{Covariate-Informed Empirical Bayes}\label{sec-sim-CIEB}
We evaluate the posterior-mean estimator of covariate-informed empirical Bayes (CIEB; \Cref{sec-caeb}) for matrix recovery with side information. Data are generated according to
\begin{equation*}
    z_{ij}^\star = T_0(y_i, a_j, u_i, v_j, w_{ij}), \qquad x_{ij} \sim \cN(z_{ij}^\star, 1),
\end{equation*}
where $y_i \in \R^{3}$ and $a_j \in \R^{4}$ are global row- and column-specific covariates, $u_i$ and $v_j$ are latent factors, and $w_{ij}$ is an additional stochastic input.

We simulate $y_i$ as iid draws from a standard $t$-distribution with $5$ degrees of freedom and $a_j$ from a beta distribution with $a = 2$ and $b = 5$. We consider three choices of $T_0$:
\begin{itemize}[leftmargin=1.5em]
  \item \textbf{Linear:} $T_0(y, a, u, v, w) = \sum_{k=1}^3 y[k] + \sum_{k=1}^4 a[k] + uv + w$
  \item \textbf{Nonlinear:} $T_0(y, a, u, v, w) = \sum_{k=1}^3 y[k] + \sum_{k=1}^4 a[k] + \sin(\pi u)\cos(\pi v) + 0.5 w^2$
  \item \textbf{Logistic:} $T_0(y, a, u, v, w) = \sum_{k=1}^3 y[k] + \sum_{k=1}^4 a[k] + \text{sigmoid}(u+v) + w$
\end{itemize}
where $y[k]$ and $a[k]$ denote the $k$th coordinates of $y \in \R^3$ and $a \in \R^4$.

\textbf{Implementation of covariate-informed EB.} We compare CIEB, EBMR-Sep, NPMLE, EBMF, and MLE. We also include a covariate ridge baseline (Ridge), a non-EB competitor that predicts each entry by a ridge-penalized linear regression of $x_{ij}$ on an intercept, the row covariates $y_i$, the column covariates $a_j$, and their pairwise interactions $y_i a_j^\top$, so that $\hat z_{ij}=\beta_0 + y_i^\top\beta_y + a_j^\top\beta_a + \mathrm{vec}(y_i a_j^\top)^\top\beta_{ya}$ with the coefficients fit by $\ell_2$-penalized least squares. To implement CIEB, we parameterize $\rmg_{\rel}:\R^7 \times [0,1]^3 \mapsto \R$ as a two-layer ReLU neural network with 10 hidden units per layer. We perform the stochastic gradient step using Adam and use NUTS with a target acceptance probability of $0.8$ to sample from the posterior.

Table~\ref{tab:simul_CIEB} reports the median R-MSE of the competing methods. CIEB consistently outperforms the NPMLE. In some cases, covariate-informed EB achieves $10\%$--$20\%$ of the R-MSE of the NPMLE method. Thus, for simulated matrix data with covariates, covariate-informed EB provides a markedly better denoising procedure than NPMLE.

\begin{figure}[t!]
\centering
\includegraphics[width=\textwidth]{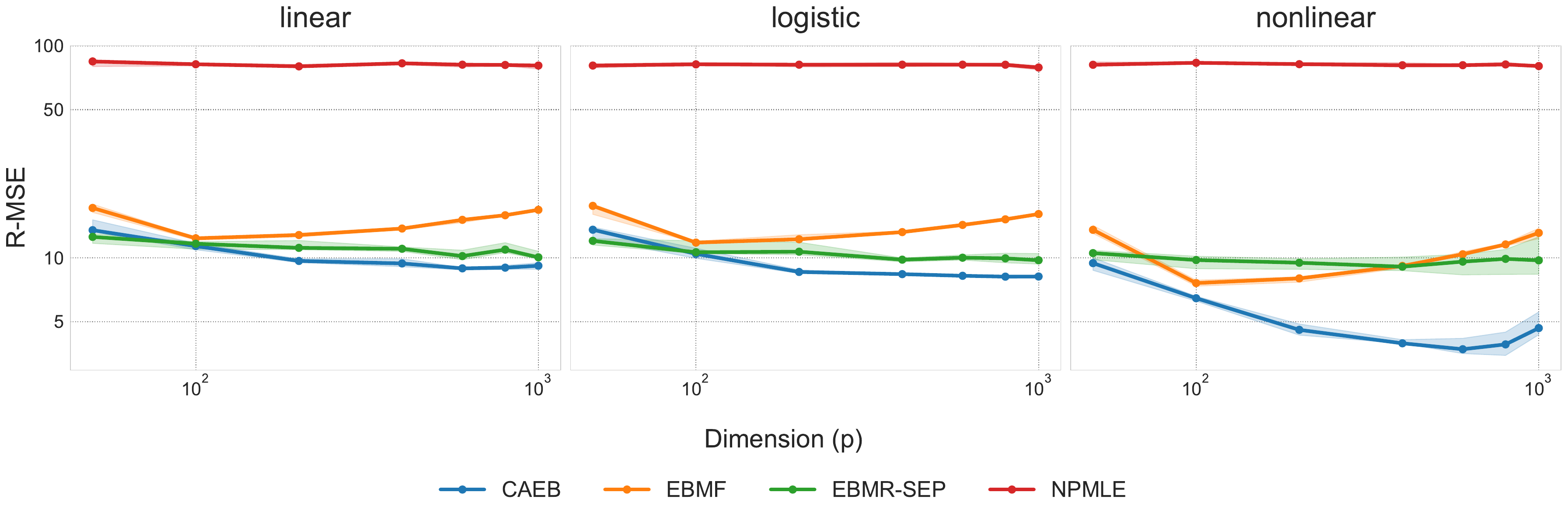}
\caption{\emph{Covariate-informed EB achieves the lowest risk in large-sample settings.} When $p$ is large, all methods achieve constant MSE that reflects the oracle risk as determined by the symmetry group. Covariate-informed EB achieves the lowest risk compared to other benchmarks.}
\label{fig:simul_caeb_mse_p}
\end{figure}

\Cref{tab:simul_CIEB} also shows that Bayesian EB methods achieve lower R-MSE than NPMLE. For smaller matrices $(n,p)\in\{(20,20),(20,50)\}$, EBMR-Sep beats CIEB. The reason is that estimating the 10-dimensional map $\rmg_{\rel}$ requires more data than the 3-dimensional $\rmg_{\sep}$. Once $n,p\ge 50$, CIEB is competitive with EBMR-Sep and overtakes it in the logistic and nonlinear settings. At $n=p=100$, the sample size suffices to fit the 10-dimensional network for $\rmg_{\rel}$, and the benefit of covariates becomes evident. \Cref{fig:simul_caeb_mse_p} further shows that with $n=100$ and $p$ increasing from $100$ to $1000$, CIEB dominates across all three generative settings.
\input{Tables/mse_table4}

We also assess the robustness of BEB methods to misspecification of the noise process. Specifically, we rerun the simulations with noise drawn from a $t$-distribution with degrees of freedom $5$ and scale parameter $1$ instead of the $N(0,1)$ distribution. While the absolute MSEs of the methods are slightly higher, the qualitative comparison among the methods is unchanged. See \Cref{fig:app:caeb_t_noise} in the Appendix for details.

\subsection{Empirical Bayes Spatial Regression}\label{sec-sim-EBSR}
Finally, we evaluate empirical Bayes spatial regression with simulated data. We generate spatial data on an evenly spaced grid $-10=\omega_1<\omega_2<\cdots<\omega_n=10$. The observed variables follow the model
\begin{equation*}
x_{\omega_i} = a_{\omega_i}^\top \beta^\star + z_{\omega_i}^\star + \varepsilon_{\omega_i},
\qquad i=1,\ldots,n,
\end{equation*}
with design vectors $a_{\omega_i}=(1,\xi_{\omega_i})\in\R^p$ where $\xi_{\omega_i}\sim\cN(0,I_{p-1})$, and noise
$\varepsilon_{\omega_i}\iid\cN(0,\tau^{-1})$.

We fix $p=3$ and set $\beta^\star=(0.5,-1.2,0.3)^\top$. The latent field $z^\star=(z_{\omega_1}^\star,\ldots,z_{\omega_n}^\star)$ is a stationary Gaussian process with
covariance kernel
\begin{equation*}
\Sigma^\star_{ij} = k_{w,\sigma,\mu}\left(\omega_i-\omega_j\right), \quad
w=(0.5,0.3,0.2),\quad \sigma=(1.0,0.2,0.35), \quad \mu=(0,0.15,-0.25),
\end{equation*}
where the formula for $k_{w,\sigma,\mu}$ is in \Cref{eq:GM-kernel}. Equivalently, it is a stationary kernel with a three-component Gaussian mixture
$0.5\cN(0,1^2)+0.3\cN(0.15,0.2^2)+0.2\cN(-0.25,0.35^2)$ as the spectral density.
We draw a single $\bz_n^\star\sim\cN(0,\Sigma^\star)$ as the ground truth.

To simulate $\bx_n$, we vary the precision $\tau\in\{0.25,1,4\}$ and the sample size $n$ from $100$ to $3000$. We fit spatial EB regression (\Cref{sec-spatial}) with $K\in\{3,4,5\}$, draw posterior samples of $\beta$ and $\bz_n$, and report the posterior mean of $\bz_n$ and its R-MSE.

As a baseline, we apply the NPMLE to residuals under the true coefficient $\beta^\star$:
\begin{equation*}
r_\omega = x_\omega - a_\omega^\top \beta^\star,\qquad
r_\omega \sim \cN\bigl(z_\omega^\star,\ \tau_\omega^{-1}\bigr),
\end{equation*}
and estimate $\bz_n^\star$ using the normal-means model with a fully exchangeable prior (\Cref{sec-NM}). The MLE for $z_\omega$ in this case is $r_\omega$.

\begin{figure}[t!]
\centering
\includegraphics[width=\textwidth]{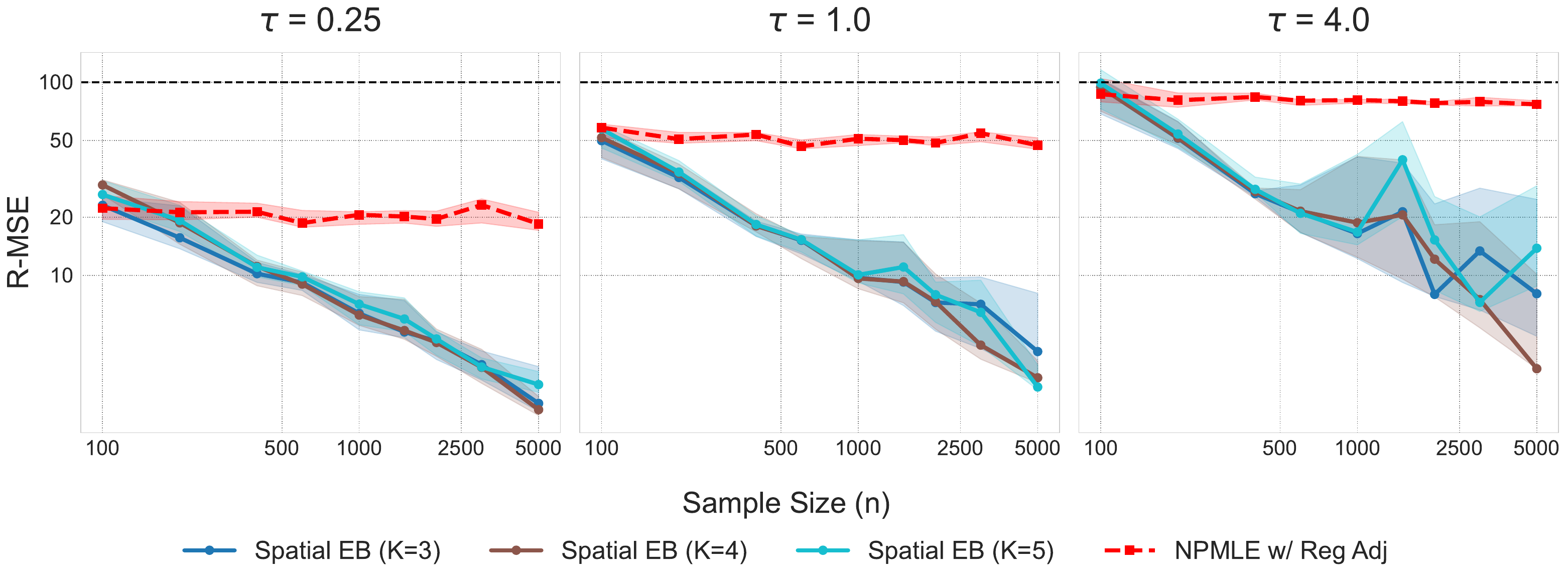}
\caption{\emph{EB spatial regression dominates NPMLE.} Relative MSE for recovering the latent series $\bz_n^\star$ across sample size $n$ and noise precision $\tau$. EB spatial regression is fit with $K\in\{3,4,5\}$ mixtures and is stable to modest over-specification of $K$. The NPMLE baseline uses residuals $r_\omega=x_\omega-a_\omega^\top\beta^\star$ with $r_\omega\sim\mathcal N(z_\omega,\tau_\omega^{-1})$ and a fully exchangeable prior. Points show medians over 10 replicates; bands show the interquartile range (25--75\%). EB error decreases with $n$, whereas NPMLE error remains roughly flat.}
\label{fig:simul_ebsr_mse_n}
\end{figure}

\Cref{fig:simul_ebsr_mse_n} shows the results. EB spatial regression uniformly dominates the regression-adjusted NPMLE, which indicates that stationarity is a more informative symmetry than full exchangeability for this example. At each noise level, the MSE of the EB spatial estimator of $\bz_n^\star$ decreases sharply with the sample size, while the MSE of the regression-adjusted NPMLE remains essentially flat as $n$ increases. The performance of the two methods is comparable only for small samples (e.g., $n = 100$); for larger $n$, the spatial EB estimator achieves substantially lower error. In \Cref{fig:simul_ebsr_mse_beta_n} of \Cref{sec:additional-simulation}, we show that the MSE of the posterior-mean estimator for $\beta$ also decreases with $n$ under the spatial EB model, which suggests that the posterior of $\beta$ satisfies posterior consistency.

In high-precision regimes, the NPMLE yields little improvement, with its R-MSE hovering around $80$. By contrast, the R-MSE of EB spatial regression drops below $20$ once $n \geq 2000$, albeit with higher cross-replicate variability than in lower-precision regimes.
\section{Real Data Examples} \label{sec-data}
For applications, we consider (i) two semisynthetic matrix-denoising studies, on a cancer gene-expression matrix and a brain connectome, where we add Gaussian noise at varying levels, and (ii) denoising and interpolating NYC air-quality data using EB spatial regression. In the first two examples, we evaluate our methods using R-MSE of the recovered matrix versus the ground truth. In the last example, we provide the full inference results to reason about the learned prior and posterior.

\subsection{Matrix Denoising: Gene Expression and Brain Connectome} \label{sec-cancer}
We apply empirical Bayes matrix recovery methods to a gene-expression dataset under varying levels of noise corruption. The dataset is a $251\times226$ gene-expression matrix analyzed by \citet{Carvalho2008} and \citet{Knowles2011}, containing measurements of 226 genes across 251 breast-cancer patients. Denoising and empirical-Bayes shrinkage of gene-expression data is an established problem \citep{Efron2001,Singh2002,Saha2020}. The statistical goal is to recover the latent gene-expression matrix from a noisy observation. It is reasonable to assume \emph{a priori} that genes are exchangeable and patients are exchangeable, so we apply the empirical Bayes matrix recovery method with a separately exchangeable prior to pool information across related gene-specific and patient-specific effects.

In the semisynthetic design, we add independent Gaussian noise with precision $\tau\in\{0.01,0.03,0.1,0.3,1,3,10\}$ to each entry and compare three empirical Bayes procedures: (i) EBMR with a separately exchangeable prior (EBMR-sep), (ii) nonparametric maximum likelihood estimation (NPMLE; \citealp{Jiang2009}), and (iii) empirical Bayes matrix factorization (EBMF; \citealp{Wang2021EBMF}), together with a non-empirical-Bayes matrix-denoising baseline, singular-value soft-thresholding (SVT; \citealp{Cai2010SVT}). We measure the performance by the relative mean-squared error (R-MSE) between the recovered and original matrices, averaged over 20 replicates and summarized by the median R-MSE.

For EBMF, we follow \citet{Wang2021EBMF} by setting $K=40$. For EBMR-sep, we discretize $[0,1]$ on a $K=10$ grid and parameterize $\rmg_{\mathrm{sep}}$ as a two-layer ReLU network with 10 hidden units per layer. We train $\rmg_{\sep}$ with Adam (learning rate $0.01$) on the variational objective in \Cref{alg:EBMR-sep} over 500 iterations with 50 gradient steps per iteration. The posterior sampling step uses NUTS with 300 warm-up and 300 retained samples (target acceptance probability $0.8$). Each EBMR-sep run requires approximately 5 minutes on a single NVIDIA A100 GPU. A full noise-level sweep takes two hours on two such GPUs.

\begin{figure}[t!]
\centering
\includegraphics[width=0.48\textwidth]{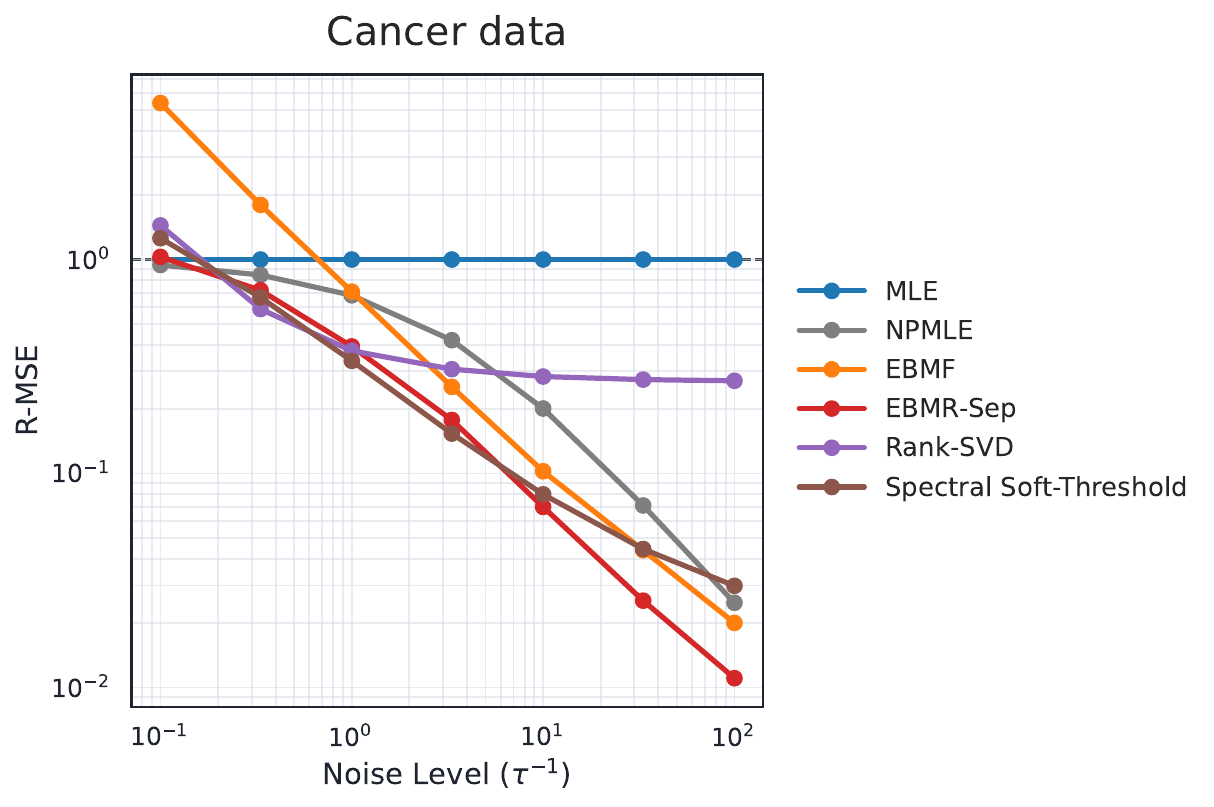}\hfill
\includegraphics[width=0.48\textwidth]{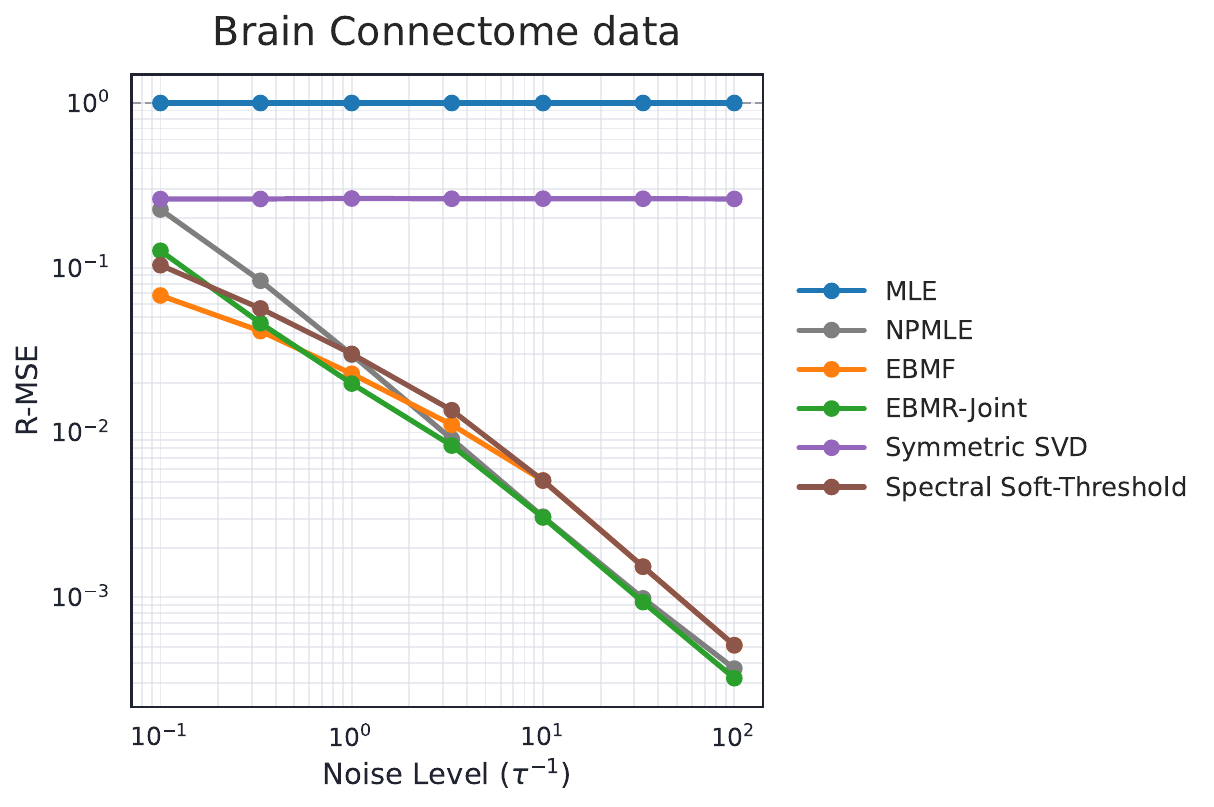}
\caption{\emph{Matrix denoising on two real matrices.} Left: breast-cancer gene-expression matrix ($251\times226$) under separate exchangeability, comparing MLE, NPMLE, EBMF, EBMR with a separately exchangeable prior, a low-rank SVD, and singular-value soft-thresholding. Right: Brainnetome connectome ($246\times246$) matrix denoising, with MLE, NPMLE, EBMF, EBMR with a jointly exchangeable prior, a symmetric low-rank SVD, and symmetric SVT. Relative MSE (as a fraction of the MLE risk) versus the noise level $\tau^{-1}$. EBMR with a separately exchangeable prior is the most accurate method on the cancer matrix at high noise, while EBMR with a jointly exchangeable prior attains the lowest relative MSE on the connectome at all but the lowest noise levels. There is a ``blessing of noise'': as the noise level increases, the empirical Bayes methods achieve lower relative MSE.}
\label{fig:cancer}
\end{figure}

\Cref{fig:cancer} shows the results. EBMR with a separately exchangeable prior attains lower R-MSE than EBMF at every noise level, and lower R-MSE than NPMLE at all but the lowest noise level, with the performance gap increasing under higher noise. The intuitive reason is that the separately exchangeable prior captures gene-level and patient-level sharing of information in a nonlinear way, whereas the other two methods either ignore the row/column-exchangeable structure (NPMLE) or assume a low-rank structure (EBMF). Among non-EB baselines, singular-value soft-thresholding (SVT) is the strongest competitor: it is marginally better than EBMR with a separately exchangeable prior at moderate noise levels, but the latter remains more accurate at both high and low noise. The symmetry-based prior helps most when the noise is large. Each entry then carries little information on its own, the MLE degrades quickly, and borrowing strength across the shared row and column structure dominates, so the relative MSE of EB falls as the noise grows. This is the ``blessing of noise'' visible in \Cref{fig:cancer}.

We repeat the exercise on a real brain connectome, a $246\times246$ structural connectivity matrix from the Brainnetome atlas \citep{Fan2016Brainnetome}. Here the rows and columns index the same set of brain regions and the matrix is symmetric, so joint exchangeability is the natural symmetry, and we fit EBMR with a jointly exchangeable prior (EBMR-joint). We add Gaussian noise at the same precision levels and compare against NPMLE, EBMF, and two non-EB baselines for symmetric matrices, a symmetric low-rank SVD and symmetric singular-value soft-thresholding \citep{Chatterjee2015USVT,Cai2010SVT}. The right panel of \Cref{fig:cancer} shows that EBMR with a jointly exchangeable prior achieves the lowest relative MSE at all but the lowest noise levels.

\subsection{Denoising and Interpolating Air Quality Data in NYC}
We apply empirical Bayes spatial regression (SpatialEB) to a dataset on fine particulate matter $\mathrm{PM}_{2.5}$ (in $\mu\mathrm{g}/\mathrm{m}^3$) in New York City. The data comprise hourly readings from 16 monitoring stations obtained from the NYC Department of Health and Mental Hygiene’s Real-Time Air Quality portal.\footnote{\url{https://a816-dohbesp.nyc.gov/IndicatorPublic/data-features/realtime-air-quality/}} We focus on four sites---Herald Square, Chinatown, the Lower East Side, and Queens North---using records from January 2019 to May 2025.\footnote{We select these four sites because they have the most complete records; for most other sites (including those in Brooklyn, Staten Island, and the Bronx), more than $60\%$ of weekly observations are missing.} To capture medium-term dynamics, we aggregate to weekly averages $x_{s,t}$ at site $s$ and week $t$ and assign sampling precision $\tau_{s,t}$ as the inverse within-week variances.

The weekly series contain gaps (outages/maintenance), occasional spikes, and heterogeneous noise. Our goals are to (i) denoise and interpolate the weekly average data for all four sites, (ii) estimate the latent weekly $\mathrm{PM}_{2.5}$ levels with uncertainty quantification, and (iii) characterize the spatio-temporal dependence structure via the fitted spectral density.

We model the data using the spatial regression model
\begin{equation*}
    x_{s,t}\mid z_{s,t}, \beta \sim \cN \bigl(z_{s,t} + \beta^\top a_s, \tau_{s,t}^{-1}\bigr),
\end{equation*}
where $z_{s,t}$ denotes the latent weekly mean and $a_s$ is a site-specific covariate vector including an intercept and site indicators. The regression component $\beta^\top a_s$ accounts for persistent location effects across monitoring sites. We fit the spatial regression model with $K=10$ Gaussian-mixture components for the spectral density, then apply the procedure in \Cref{sec-alg-spatial} to obtain posterior means $\hat z_{s,t}$, regression coefficients $\hat\beta$, $95\%$ credible bands for $\{z_{s,t}\}$, and kriged imputations for missing weeks. We also report site-adjusted means $\hat m_{s,t}=\hat z_{s,t}+\hat\beta^\top a_s$ with credible bands.

When we parameterize the spectral density, $\omega=(\omega_{\text{time}},\omega_{\text{space}})$ denotes temporal and spatial frequencies. The model learns a spectral density $\psi(\omega)$; by Bochner’s theorem this induces a stationary covariance $k(\Delta)$ in the $(t,\text{space})$ domain. Peaks in $\psi(\omega)$ correspond to dominant temporal cycles and spatial scales of variation in $z_{s,t}$.

\begin{figure}[t!]
\centering
\begin{minipage}[t]{0.65\textwidth}
  \vspace{0pt}
  \includegraphics[width=\textwidth]{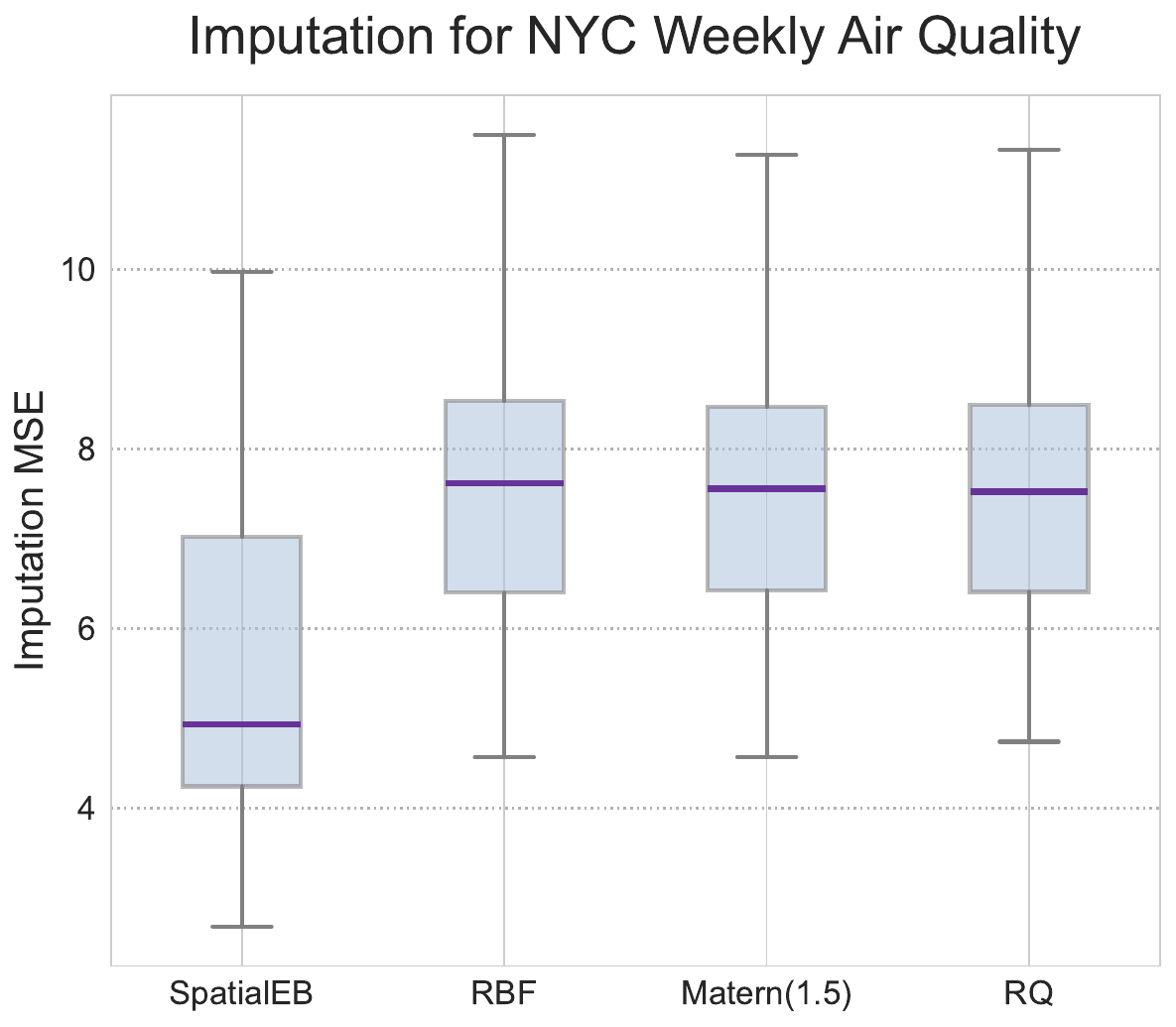}
\end{minipage}\hfill
\begin{minipage}[t]{0.33\textwidth}
  \vspace{0pt}
  \caption{\emph{Empirical Bayes spatial regression achieves substantially lower MSE than Gaussian-process baselines.}
  Boxplots show test MSE (median and quantiles) across 50 random masks that hide 20\% of observed entries in the NYC air-quality panel.
  Methods: EB spatial regression with $K=10$ Gaussian-mixture components for the spectral density, and GPR with RBF, Matérn ($\nu=3/2$), and Rational-Quadratic kernels.}
  \label{fig:pm25-masked-study}
\end{minipage}
\end{figure}

We assess the imputation accuracy of EB spatial regression against three Gaussian-process regressors on the same site-week features as our EB model, standardizing inputs and using per-observation noise variances $\alpha_i=\tau_i^{-1}$ from within-week precision. The models differ only by kernel---RBF (squared-exponential, smooth), Matérn $(\nu=3/2)$ (rougher trajectories), and Rational Quadratic (a scale-mixture kernel capturing multi-scale variability)---each with a small white-noise nugget ($10^{-3}$). Kernel hyperparameters are learned by maximizing the marginal likelihood. On held-out data, we report the MSE of the imputed series against the true weekly averages. \Cref{fig:pm25-masked-study} shows that EB spatial regression achieves substantially lower MSE than all GP baselines because the fitted spectral density adaptively captures the bimodal spatio-temporal variation. By comparing the four methods, we see that fitting a more flexible prior family that respects the shift invariance yields performance gain in imputation. Compared to the Gaussian-process baselines, EB spatial regression fits a flexible Gaussian-mixture spectral density by maximizing the marginal likelihood, recovering a multimodal kernel, whereas each baseline fixes a single parametric kernel. This allows the prior to capture the bimodal spatio-temporal structure.

\begin{figure}[t!]
\centering
\includegraphics[width=\textwidth]{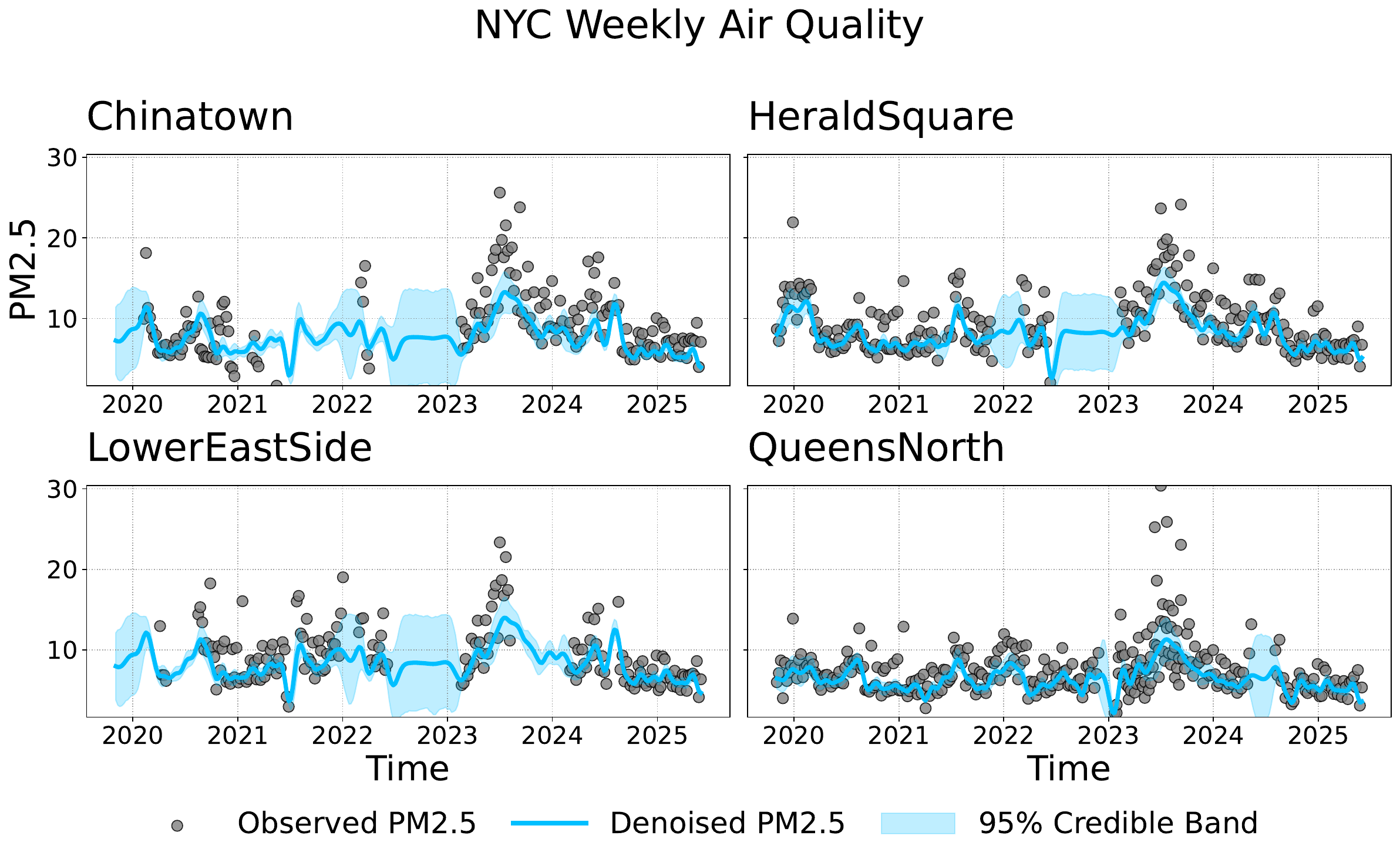}
\caption{\emph{Denoised and interpolated weekly $\mathrm{PM}_{2.5}$ in New York City using EB spatial regression.}
Posterior means with 95\% credible bands for three monitoring sites; gray points are observed weekly averages.
The model smooths noisy measurements and imputes missing weeks.}
\label{fig:pm25_denoised}
\end{figure}

\Cref{fig:pm25_denoised} shows empirical Bayes posterior means with $95\%$ credible bands. The posterior means shows the shrinkage and pooling effects of EB spatial regression: extreme weekly averages (e.g., mid-2023) are pulled toward the mean. Where observations are dense, the posterior uncertainty around the denoised weekly averages is narrow; where data are missing, it is wider. Within Manhattan (Herald Square, Chinatown, Lower East Side), the model borrows heavily across sites---for example, when Chinatown is entirely missing from mid-2021 to early-2022, the model interpolates from Herald Square and the Lower East Side to produce tight bands; similarly, during mid-2023 to early-2024, the model imputes the Lower East Side using the other two Manhattan sites, which recovers a decreasing, cyclical trend. By contrast, Queens North contributes less to interpolating Manhattan gaps (e.g., mid-2022 to early-2023). The reason for this different shrinking behavior is that the EB model finds two spatio-temporal regimes, one for Manhattan and one for northern Queens; see \Cref{fig:pm25_spectral} in \Cref{sec:additional-air}.

\section{Discussion} \label{sec-discussion}
We have extended \emph{Bayesian empirical Bayes}, the program of treating the unknown prior as an object of Bayesian inference \citep{Lindley1972,Deely1981}, from exchangeable sequences to general probabilistic symmetries. The resulting procedure gives accurate simultaneous posterior inference of latent variables when the prior respects such a symmetry, bringing the power of empirical Bayes to structured data beyond exchangeable sequences. Using Bayesian EB, we derived new EB models and inference algorithms for three settings: array data with separate and joint exchangeability, spatial data with shift invariance, and covariate-augmented data. We demonstrated the competitive performance of Bayesian EB over classical empirical Bayes. The methods are complementary. Empirical Bayes matrix recovery is preferable when the latent matrix is dense and admits a separately or jointly exchangeable structure, whereas empirical Bayes matrix factorization is preferable when the latent matrix is close to low rank, as in the settings of \citet{Wang2021EBMF}.

There are several avenues for further research.

First, we have focused here on the derivation and methods behind Bayesian EB. Its theoretical analysis remains an open area. Natural theoretical questions include the identifiability, consistency, and sample complexity of learning the invariant variable $\rmg$ via maximum marginal likelihood and more broadly, estimation under symmetry constraints.

Second, in this paper, we studied symmetry groups based on arrays and spatio-temporal constraints. In many scientific domains, however, the relevant symmetry groups are more complex, such as those arising from physical constraints (e.g., molecular or spacetime symmetries). Another interesting direction is to leverage geometric deep learning \citep{Bronstein2021} to design new BEB methods tailored to such applications.

Finally, while our initial evidence suggests that Bayesian EB remains competitive under heavy-tailed noise (see \Cref{fig:app:caeb_t_noise} in \Cref{sec:additional-covariates}), it would be valuable to develop sensitivity analysis for these new methods under other types of model misspecification. Sensitivity analysis for empirical Bayes has been studied in \citet{Berger1986,Buta2011,Doss2024}. 


\bibliographystyle{apalike}
\bibliography{bib}

\appendix
\section{Conditions on Groups} \label{sec-group}
We provide the regularity conditions on the group of transformations $\Phi$ acting on the space $\cZ^\Omega$. For $\phi_1, \phi_2 \in \Phi$, we write the composition $\phi_1 \phi_2(\cdot) \deq \phi_1\bigl(\phi_2(\cdot)\bigr)$. If $\phi \in \Phi$ and $A \subseteq \Phi$, we write $\phi A \deq \{ \phi \psi \mid \psi \in A\}$. If $K, A \subseteq \Phi$, we write $K A \deq \{\phi \psi \mid \phi \in K, \psi \in A\}$.

We equip $\Phi$ with the Borel $\sigma$-algebra $\mathcal{B}(\Phi)$. For every locally compact, second countable, and Hausdorff (lcscH) group, there exists a $\sigma$-finite measure $\lvert \cdot \rvert$ such that
\begin{equation*}
  \bigl\lvert \phi^{-1} A \bigr\rvert = \lvert A\rvert,
  \quad\text{for all }\phi \in \Phi,\ A \in \mathcal{B}(\Phi),
\end{equation*}
called a \emph{Haar measure} \citep{Folland2016}. The Haar measure is unique up to a positive scaling factor. If $A \subset \Phi$ is compact, then $\lvert A \rvert$ is finite.

A \emph{F{\o}lner sequence} in $\Phi$ is a sequence of compact sets $\bA_1, \bA_2,\dots$ such that, for every compact subset $K \subset \Phi$,
\begin{equation*}
  \frac{\lvert K \bA_n \cap \bA_n\rvert}{\lvert \bA_n\rvert}
  \xrightarrow[n\to\infty]{} 1.
\end{equation*}
If $\Phi$ is discrete, its compact sets are finite, and hence the condition specializes to the familiar finite F{\o}lner property. An lcscH group is called \emph{amenable} if it admits a F{\o}lner sequence.  

A F{\o}lner sequence $\{\bA_n\}$ is \emph{tempered} if there is some constant $c>0$ such that, for all $n\in\N$,
\begin{equation*}
  \bigl\lvert
    \bigcup_{k < n} \bA_k^{-1} \bA_n
  \bigr\rvert
  \le
  c \bigl\lvert \bA_n \bigr\rvert.
\end{equation*}
Not every F{\o}lner sequence is tempered, but every lcscH group that has a F{\o}lner sequence also has a tempered one \citep[Proposition~1.4]{Lindenstrauss2001}.

Finally, we give the definition of a "nice" group, which matches the standard definition in the literature of probabilistic symmetry \citep{Austern2022}. 
\begin{definition}[Nice group]\label{assum:G}
We call $\Phi$ a \emph{nice group} if
\begin{enumerate}[label=(\roman*)]
    \item $\Phi$ is a locally compact, second countable, Hausdorff (lcscH) topological group, and the group operations are continuous.
    \item $\Phi$ is amenable, i.e., it admits a F{\o}lner sequence $\left(\bA_n \right)_{n \in \N}$.
\end{enumerate}
\end{definition}
\section{Empirical Bayes Beyond Index Transformations}
In \Cref{sec-method}, we assumed $\Phi$ to be a group of transformations acting on the index space $\Omega$. There are many interesting groups beyond index transformations, for example the rotation group or the Heisenberg group. We can develop Bayesian empirical Bayes methods for these groups under an equivariant assumption on the likelihood.

Let $\bx \in \cX^{\Omega}$ be observed variables and $\bz \in \cX^\Omega$ local latent variables. A general EB setup is as follows:
\begin{equation*}
    \bz \sim \rmp(\bz), \quad \bx \sim \rmp(\bx \mid \bz).
\end{equation*}
The likelihood $\rmp(\bx \mid \cdot)$ is assumed to be $\Phi$-equivariant in the sense of \Cref{def-equivariance} (stated in \Cref{appendix-support-results}). Specifically,
\begin{equation*}
    \rmp(\bx \mid \phi(\bz)) = \rmp\left(\phi^{-1}(\bx) \mid \bz \right), \quad \forall \phi \in \Phi.
\end{equation*}
As a special case, the product likelihood $\rmp(\bx \mid \bz) = \prod_{\omega \in \Omega} \rmp(x_\omega \mid z_\omega)$ is equivariant with respect to any group of index transformations. The Gaussian likelihood is equivariant to rotations. This equivariance is a property of the conditional probability measure $\bz\mapsto\rmp(\cdot\mid\bz)$, a map into the distributions on $\cX^\Omega$, and should not be conflated with the function equivariance of \Cref{def-equivariance}.

By \Cref{assum:true-prior} and \Cref{assum:phi-prior}, there exists a true prior $\rmp^\star(\bz)$ that is $\Phi$-invariant. If $\Phi$ is a nice group in the sense of \Cref{assum:G}, the ergodic representation theorem holds for $\Phi$. Thus, there exists an invariant variable $\rmg$ such that the population distribution is
\begin{equation} \label{general-model}
    \rmg \sim \mu^\star, \quad \bz \sim \rmp(\bz \mid \rmg), \quad \bx \sim \rmp(\bx \mid \bz).
\end{equation}
The goal is to compute the posterior $\rmp^\star(\bz \mid \bx)$, given by
\begin{equation*}
    \rmp^\star(\bz \mid \bx) \propto \int_{\cG} \rmp(\bx \mid \bz)\, \rmp(\bz \mid \rmg)\, \dd \mu^\star(\rmg).
\end{equation*}

The next result shows that both the population posterior $\rmp^\star(\bz \mid \bx)$ and the population empirical Bayes posterior $\rmp(\bz \mid \bx, \hat\rmg)$ satisfy the equivariance property.
\begin{proposition}
The population posterior $\rmp^\star(\bz \mid \bx)$ is a $\Phi$-equivariant conditional distribution, in the sense that for any $\phi \in \Phi$,
\begin{equation*}
    \rmp^\star\!\left(\phi^{-1}(\bz) \mid \bx\right) = \rmp^\star\!\left(\bz \mid \phi(\bx)\right).
\end{equation*}
For any $\rmg \in \cG$, the conditional posterior $\rmp(\bz \mid \bx, \rmg)$ is also $\Phi$-equivariant, including the population EB posterior $\rmp(\bz \mid \bx, \hat\rmg)$.
\end{proposition}

This result shows that transforming the structured variable $\bz$ by $\phi$ under the posterior is equivalent to computing the posterior of $\bz$ after applying the transformation $\phi$ to the observed variables $\bx$.
\begin{proof}
We first consider the population posterior $\rmp^\star(\bz \mid \bx)$. For any $\phi \in \Phi$,
\begin{align*}
\rmp^\star\!\left(\phi(\bz) \mid \bx\right)
&= \int_{\cG} \rmp\!\left(\bx \mid \phi(\bz)\right)\, \rmp\!\left(\phi(\bz) \mid \rmg\right)\, \dd \mu^\star(\rmg) \\
&= \int_{\cG} \rmp\!\left(\bx \mid \phi(\bz)\right)\, \rmp\!\left(\bz \mid \rmg\right)\, \dd \mu^\star(\rmg) \\
&= \int_{\cG} \rmp\!\left(\phi^{-1}(\bx) \mid \bz\right)\, \rmp\!\left(\bz \mid \rmg\right)\, \dd \mu^\star(\rmg)
= \rmp^\star\!\left(\bz \mid \phi^{-1}(\bx)\right).
\end{align*}
The second equality follows from the $\Phi$-invariance of $\rmp(\bz \mid \rmg)$, and the third from the $\Phi$-equivariance of the likelihood $\rmp(\bx \mid \bz)$. An identical argument applies to $\rmp(\bz \mid \bx, \rmg)$ for any fixed $\rmg \in \cG$, and therefore also to the EB posterior $\rmp(\bz \mid \bx, \hat\rmg)$, where $\hat\rmg$ denotes the maximum marginal likelihood estimator.
\end{proof}

The result suggests that “transform then condition” and “condition then transform” commute for the posterior. The population posterior mean $\E{\bz \mid \bx} := \int_{\cX^{\Omega}} \bz\, \rmp^\star(\bz \mid \bx)\, \dd \bz$ therefore satisfies $\E{\phi(\bz) \mid \bx} = \E{\bz \mid \phi(\bx)}$ for any $\phi \in \Phi$ when $\Phi$ is a group of isometries. This suggests that, to learn the population Bayes estimator of transformed latent variables, it suffices to compute the posterior on transformed datasets. This observation connects empirical Bayes point estimation \citep{Jiang2009} to the common machine learning practice of enforcing symmetry via data augmentation \citep{gonzalez2018galaxy,lyle2019,chen2020contrastive}.

We also record the converse used at the end of \Cref{sec-prior}: an invariant population marginal forces the prior to be invariant.
\begin{proposition}[From an invariant marginal to an invariant prior]\label{prop:marginal-to-prior-invariance}
Suppose the likelihood is $\Phi$-equivariant, so that $\rmp(\phi(\bx)\mid\phi(\bz))=\rmp(\bx\mid\bz)$ for every $\phi\in\Phi$, and identifiable, so that the map from a prior to its induced marginal is injective. If the population marginal $\rmp^\star(\bx)$ is $\Phi$-invariant, then the true prior $\rmp^\star(\bz)$ is $\Phi$-invariant.
\end{proposition}
\begin{proof}
Fix $\phi\in\Phi$. Since $\phi$ permutes the index set and preserves the base measure, the change of variables $\bz=\phi(\bw)$ gives
\begin{equation*}
\rmp^\star(\phi(\bx))
= \int \rmp(\phi(\bx)\mid\bz)\,\rmp^\star(\bz)\,\dd\bz
= \int \rmp(\phi(\bx)\mid\phi(\bw))\,\rmp^\star(\phi(\bw))\,\dd\bw
= \int \rmp(\bx\mid\bw)\,\rmp^\star(\phi(\bw))\,\dd\bw,
\end{equation*}
where the last equality uses the equivariance of the likelihood. The right-hand side is the marginal induced by the pushed-forward prior $\bw\mapsto\rmp^\star(\phi(\bw))$. The $\Phi$-invariance of the population marginal gives $\rmp^\star(\phi(\bx))=\rmp^\star(\bx)=\int \rmp(\bx\mid\bw)\,\rmp^\star(\bw)\,\dd\bw$. The priors $\rmp^\star(\phi(\cdot))$ and $\rmp^\star(\cdot)$ therefore induce the same marginal, and identifiability yields $\rmp^\star(\phi(\bw))=\rmp^\star(\bw)$ for all $\bw$. Hence $\rmp^\star(\bz)$ is $\Phi$-invariant.
\end{proof}

\section{Support Results} \label{appendix-support-results}

\begin{definition} \label{def-exhaustible-by-compact-sets}
A space $\Omega$ is \emph{exhaustible by compact sets} if, for every open set $U \subseteq \Omega$, there exists an increasing sequence of compact subsets $S_1 \subseteq S_2 \subseteq \cdots$ such that
\[
U = \bigcup_{n=1}^\infty S_n,
\]
and, for every $n$, $S_n$ is contained in the interior of $S_{n+1}$.
\end{definition}

\begin{definition}[$\Phi$-equivariance]
\label{def-equivariance}
A function $t : \cX^\Omega \to \cX^\Omega$ is \emph{$\Phi$-equivariant} if
\begin{equation} \label{eqn-equivariance}
\phi(t(\bx)) = t(\phi(\bx)), \quad \text{for all } \phi \in \Phi.
\end{equation}
\end{definition}

Intuitively, a function $t$ is $\Phi$-equivariant if every action of $\Phi$ “commutes” with $t$. An equivariant function $t$ preserves $\Phi$-invariance, as $\phi(\bx) \overset{d}{=} \bx$ implies
\[
\phi(t(\bx)) = t(\phi(\bx)) \overset{d}{=} t(\bx).
\]
\section{Algorithms for Empirical Bayes Matrix Recovery}
\label{sec-algorithms-ebmr}
This appendix details the inference algorithms for the empirical Bayes matrix-recovery methods introduced in \Cref{sec-EB-models}. \Cref{sec-EBMR-separate} presents a \emph{variational} procedure for matrix recovery under a separately exchangeable prior (EBMR-sep; \Cref{alg:EBMR-sep}). \Cref{sec-EBMR-joint} develops the inference algorithm for EBMR with a jointly exchangeable prior (EBMR-joint; \Cref{alg:EBMR-jointly}), and \Cref{sec-CIEB-computation} provides the algorithm for covariate-informed EB (CIEB) which is analogous to the separately exchangeable case.

For all three algorithms, we use variational inference \citep{Blei2017} to approximate the nonparametric maximum marginal likelihood.

\textbf{Variational MMLE. } Let $\cQ \subseteq \cP(\cX^{S_n})$ be a family of distributions. By the Gibbs variational principle \citep[Lemma 4.10]{VanHandel2014}, one can lower bound the marginal likelihood by
\begin{equation} \label{variational-EB}
    \log \rmp\bigl(\bx_n \mid \rmg \bigr)
\geq - \ell(\bx_n; \rmg) \deq
\sup_{\rmq \in \cQ}
\EE{\rmq(\bz_n)}{
   \log \rmp(\bx_n \mid \bz_n) + \log \rmp(\bz_n \mid \rmg)
    -
    \log \rmq(\bz_n)
}, 
\end{equation}
where $-\ell(\bx_n; \rmg)$ is the maximum evidence lower bound (ELBO) over the family $\cQ$. 
\subsection{EBMR with a Separately Exchangeable Prior}\label{sec-EBMR-separate}

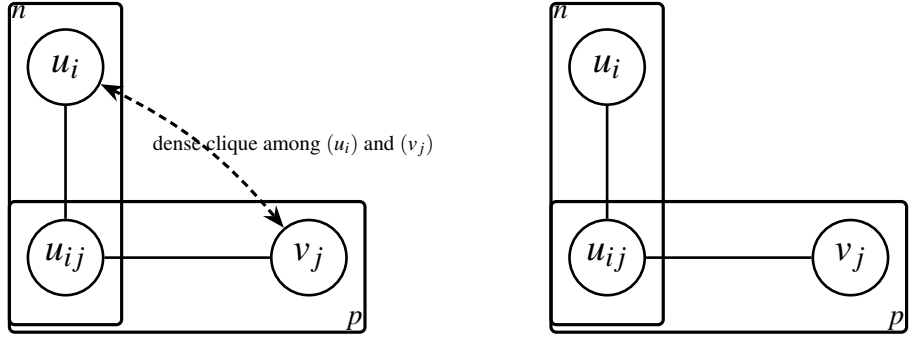
\begin{figure}[t]
\centering
\begin{tikzpicture}[>=Stealth, node distance=2cm]
  \tikzset{
    latent/.append style={minimum size=10mm, inner sep=2pt, font=\large},
    obs/.append style={minimum size=10mm, inner sep=2pt, font=\large},
    plate/.append style={line width=1.2pt, rounded corners=2pt},
    every path/.style={line width=1pt}
  }

\begin{scope}[xshift=0cm]
  \node[latent] (uiL) {$u_i$};
  \node[latent, below=1.5cm of uiL] (uijL) {$u_{ij}$};
  \node[latent, right=2.2cm of uijL] (vjL) {$v_j$};

  \path (uiL) edge[-] (uijL);
  \path (vjL) edge[-] (uijL);

  \path (uiL) edge[<->, densely dashed, very thick, bend left=12]
      node[midway, right=-0.8cm, font=\scriptsize, align=left]
      {dense clique among $(u_i)$ and $(v_j)$} (vjL);

  \plate[inner sep=6pt, yshift=0.1cm] {iplateL} {(uiL)(uijL)} {};
  \node[anchor=north west, inner sep=1pt] at (iplateL.north west) {\small $n$};

  \plate[inner sep=6pt] {jplateL} {(uijL)(vjL)} {};
  \node[anchor=south east, inner sep=1pt] at (jplateL.south east) {\small $p$};
\end{scope}

\begin{scope}[xshift=7.2cm]
  \node[latent] (uiR) {$u_i$};
  \node[latent, below=1.5cm of uiR] (uijR) {$u_{ij}$};
  \node[latent, right=2.2cm of uijR] (vjR) {$v_j$};

  \path (uiR) edge[-] (uijR);
  \path (vjR) edge[-] (uijR);

  \plate[inner sep=6pt, yshift=0.1cm] {iplate} {(uiR)(uijR)} {};
  \node[anchor=north west, inner sep=1pt] at (iplate.north west) {\small $n$};

  \plate[inner sep=6pt] {jplate} {(uijR)(vjR)} {};
  \node[anchor=south east, inner sep=1pt] at (jplate.south east) {\small $p$};
\end{scope}
\end{tikzpicture}
\caption{Variational approximation. \emph{Left}: exact posterior with a dense clique among $(u_i)$ and $(v_j)$ and local connections to $u_{ij}$. \emph{Right}: variational family $\rmq \in \cQ_{\sep}$ factorizes across $(u_i)$ and $(v_j)$ while preserving the local chain $u_i$---$u_{ij}$---$v_j$.}
\label{fig:vi-separately-exchangeable}
\end{figure}

To approximate the intractable marginal likelihood in \Cref{eq:ebmr-MMLE}, we posit the following structured variational family
\begin{equation} \label{variational-family}
    \cQ_{\sep} = \left\{\rmq \in \cP\left([0,1]^{ n + p + np}\right): \rmq\left(\bu_{n, p}\right) = \prod_{i = 1}^n \rmq_{\bu, i}(u_i)\,  \prod_{j = 1}^p \rmq_{\bv, j}(v_j)\, \prod_{i,j} \rmq_{i, j}(u_{ij} \mid u_i, v_j) \right\}.
\end{equation}
\Cref{fig:vi-separately-exchangeable} motivates $\cQ_{\sep}$. In the exact posterior $\rmp(\bu_{n,p}\mid \hat\rmg_{\sep}, \bx_{n,p})$, the variables $(u_i)_{i\in[n]}$ and $(v_j)_{j\in[p]}$ form a dense clique, while each $u_{ij}$ connects only to $(u_i, v_j)$. Our variational family replaces the dense clique with a product measure over $(u_i)$ and $(v_j)$ while preserving the local $u_i$---$u_{ij}$---$v_j$ chains.

Maximizing the ELBO over $\cQ_{\sep}$ yields a lower bound of the marginal likelihood, 
\begin{equation*}
\log \rmp\bigl(\bx_{n,p} \mid  \rmg_{\sep}\bigr)
\ge \sup_{\rmq \in \cQ_{\sep}} \ELBO{\rmq, \rmg_{\sep}} := \sup_{\rmq \in \cQ_{\sep}} \EE{\rmq(\bu_{n,p})}{
    \log \rmp\bigl(\bx_{n,p} \mid  \rmg_{\sep}, \bu_{n,p}\bigr) -
    \log \rmq(\bu_{n,p})}.
\end{equation*}

Let $\rmq_{\bu}=\bigotimes_{i=1}^n \rmq_{\bu,i}$ and $\rmq_{\bv}=\bigotimes_{j=1}^p \rmq_{\bv,j}$. We estimate $\rmg_{\sep}$ by minimizing the variational loss
\begin{equation} \label{EBMR-obj}
\hat \rmg_{\sep} \in \arg \min_{\rmg_{\sep}\in\cG} 
\ell(\bx_{n,p}; \rmg_{\sep})
\coloneqq
-\sup_{\rmq\in\cQ_{\sep}} \textsc{ELBO}(\rmq,\rmg_{\sep}).
\end{equation}
A direct calculation (see \Cref{sec-proof-inference}) shows that
\begin{equation} \label{ELBO-EBMR-separate}
\begin{aligned}
\ell \left(\bx_{n,p}; \rmg_{\sep} \right)
&=  \inf_{\rmq_{\bu} \in \cP([0,1])^{\otimes n}, \rmq_{\bv} \in \cP([0,1])^{\otimes p}}  
\Bigg\{-\sum_{i = 1}^n \sum_{j = 1}^p \EE{\rmq_{\bu, i}(u_i)\, \rmq_{\bv, j}(v_j)}{  \LSE\left(x_{ij}, \tau_{ij}, u_i, v_j, \rmg_{\sep} \right)} \\
&\hspace{20mm}
+ \sum_{i = 1}^n \EE{\rmq_{\bu, i}(u_i)}{\log \rmq_{\bu, i}(u_i)} 
+ \sum_{j = 1}^p \EE{\rmq_{\bv, j}(v_j)}{\log \rmq_{\bv, j}(v_j)} \Bigg\}, 
\end{aligned}
\end{equation}
where $\LSE\left(X, \tau, U, V, \rmg_{\sep}\right) \deq \log \int_0^1 \exp \left(-\frac{\tau}{2}\left(X  - \rmg_{\sep}(U, V, u) \right)^2  \right) \, du$.  

\paragraph{Alternating optimization.}
To fit the variational distributions, we discretize $\rmq_{\bu,i}$ and $\rmq_{\bv,j}$ on the $(K+1)$-point grid $u_k \deq k/K$ for $k=0,\ldots,K$ over $[0,1]$:
\begin{equation}
\rmq_{\bu,i} = \sum_{k=0}^K w_{\bu,ik}\,\delta_{u_k},\quad
\rmq_{\bv,j} = \sum_{k=0}^K w_{\bv,jk}\,\delta_{u_k},
\quad \text{where } u_k = k/K,\ \sum_{k=0}^K w_{\bullet,\cdot k}=1.
\end{equation}

Following \Cref{ELBO-EBMR-separate}, the empirical Bayes estimate minimizes the negative ELBO over the network and grid weights:
\begin{align}\label{discretized-EBMR-obj}
\begin{split}
\left( \hat \rmg_\sep, \hat \bw_{\bu}, \hat \bw_{\bv} \right)
\in \arg \min_{\rmg_{\sep}\in\cG_{\textrm{NN}},\ \bw_{\bu},\bw_{\bv}\in\Delta_{K+1}}
\Bigg\{ &-\sum_{i=1}^n\sum_{j=1}^p\sum_{k=0}^K\sum_{k'=0}^K
w_{\bu,ik}\,w_{\bv,jk'}\,
\widehat{\LSE}\bigl(x_{ij},\tau_{ij},u_k,u_{k'},\rmg_{\sep}\bigr) \\
&\quad + \sum_{i=1}^n\sum_{k=0}^K w_{\bu,ik}\log w_{\bu,ik}
+ \sum_{j=1}^p\sum_{k=0}^K w_{\bv,jk}\log w_{\bv,jk}\Bigg\},
\end{split}
\end{align}
where $\widehat{\LSE}\bigl(x,\tau,u,v,\rmg_{\sep}\bigr) \coloneqq 
\log \sum_{k = 0}^K \exp\!\left\{-\tfrac{\tau}{2}\bigl(x-\rmg_{\sep}(u,v,k/K)\bigr)^2\right\} - \log (K+1).$

We minimize \eqref{discretized-EBMR-obj} by applying a block-coordinate descent algorithm:
(i) update the row weights $\{\bw_{\bu,i}\}$,
(ii) update the column weights $\{\bw_{\bv,j}\}$,
(iii) update the network parameters of $\rmg_{\sep}$ by stochastic gradient descent (Adam; \citealp{Kingma2014Adam}).
\Cref{alg:EBMR-sep} provides the explicit updates. The weight updates can be parallelized over $i$ and $j$.
Each iteration costs $O(npK^2)$ time and $O((n+p)K)$ memory.

\begin{algorithm}[H]
\caption{EBMR with a Separately Exchangeable Prior (EBMR--sep)}\label{alg:EBMR-sep}
\SetAlgoLined
\KwIn{Data matrix $\bx_{n,p}$, precision matrix $\boldsymbol{\tau}_{n,p}$, neural network class $\cG_{\textrm{NN}}$, grid size $K$, tuning parameters (epochs, learning rate)}
\textbf{Initialize:} neural network parameters; probability weights $\bw_{\bu, 1},\ldots,\bw_{\bu, n},\ \bw_{\bv, 1},\ldots,\bw_{\bv, p} \in \Delta_{K+1}$.\\[2pt]
\Repeat{\textnormal{convergence}}{
\For{$i=1$ \KwTo $n$ \textnormal{(parallelizable)}}{
\[
\begin{aligned}
\rho_{\bu,ik} &\leftarrow
\sum_{j=1}^p \sum_{k_2=0}^K
w_{\bv,jk_2}\,
\widehat{\LSE}(x_{ij}, \tau_{ij}, u_k, u_{k_2}, \rmg_{\mathrm{sep}}),
&& k=0,\ldots,K,\\
\bw_{\bu,i} &\leftarrow \textsc{softmax}(\rho_{\bu,i0}, \ldots, \rho_{\bu,iK}). &&
\end{aligned}
\]
}
\For{$j=1$ \KwTo $p$ \textnormal{(parallelizable)}}{
\[
\begin{aligned}
\rho_{\bv,jk} &\leftarrow \sum_{i=1}^n \sum_{k_1=0}^K 
w_{\bu,ik_1}\,
\widehat{\LSE}(x_{ij}, \tau_{ij}, u_{k_1}, u_k, \rmg_{\mathrm{sep}}),
&& k=0,\ldots,K,\\
\bw_{\bv,j} &\leftarrow \textsc{softmax}(\rho_{\bv,j0}, \ldots, \rho_{\bv,jK}). &&
\end{aligned}
\]
}
Update the neural network parameters by stochastic gradient descent:
\begin{equation*}
\rmg_{\mathrm{sep}}\leftarrow
\arg\max_{\rmg_{\mathrm{sep}}\in \cG_{\textrm{NN}}}
\sum_{i=1}^n \sum_{j=1}^p \sum_{k_1=0}^K \sum_{k_2=0}^K
w_{\bu,ik_1}w_{\bv,jk_2}
\widehat{\LSE}\bigl(x_{ij},\tau_{ij},u_{k_1},u_{k_2},\rmg_{\mathrm{sep}}\bigr).
\end{equation*}
}
\KwOut{Fitted weights $\bw_{\bu,1},\ldots,\bw_{\bu,n},\ \bw_{\bv,1},\ldots,\bw_{\bv,p}$ and network $\rmg_{\mathrm{sep}}$.}
\end{algorithm}

\begin{proposition}\label{prop:alg-EBMR-sep}
\Cref{alg:EBMR-sep} performs block-coordinate ascent on the variational inference problem~\eqref{discretized-EBMR-obj}. 
\end{proposition}

\paragraph{Approximate posterior computation.}
The last step of \Cref{alg:BEB} is to sample from the empirical Bayes posterior
\begin{equation}
    \hat \rmp(\bz_n  \mid \bx_n) \propto \rmp(\bx_n \mid \bz_n) \rmp(\bz_n  \mid \hat \rmg). 
\end{equation}
We take the default sampler to be the No-U-Turn Sampler (NUTS) \citep{Hoffman2014}, an adaptive variant of Hamiltonian Monte Carlo \citep{Neal2010} implemented in the \texttt{Pyro} probabilistic programming package \citep{Bingham2019}. NUTS automatically tunes the step size and trajectory length parameters. We run NUTS for a fixed number of iterations that includes an initial warmup phase and thinning. 

We use the NUTS samples by default and trigger the variational fallback only when standard diagnostics (divergences or a Gelman--Rubin $\hat R$ far from one) flag non-convergence. When the full MCMC fails to converge, an alternative is to draw surrogate samples from variational approximation $\hat \rmq(\bz_n)$ obtained in \Cref{alg:EBMR-sep}. This requires using the fitted variational distributions from \Cref{alg:EBMR-sep} to approximately sample from the posterior $\rmp \left(\bz_{n,p} \mid \hat{\rmg}_{\sep}, \bx_{n, p}\right)$. Suppose that \Cref{alg:EBMR-sep} returns $\hat \rmq_{\bu, i} = \sum_{k = 0}^K \hat w_{\bu, ik}\, \delta_{u_k}$ and $\hat \rmq_{\bv, j} = \sum_{k = 0}^K \hat w_{\bv, jk}\, \delta_{u_k}$. To sample $z_{ij}$, we first draw $u_i\sim \hat \rmq_{\bu,i}$ and $v_j\sim \hat \rmq_{\bv,j}$. Conditional on $(u_i,v_j)$, we draw $u_{ij}\in\{0,1/K,\ldots,1\}$ with probability
\begin{equation*}
\hat \rmq_{ij}\big(u_{ij}= k/K \mid u_i,v_j,\bx_{n,p}\big)
=
\frac{\exp\!\left\{-\tfrac{\tau_{ij}}{2}\big(x_{ij}- \hat \rmg_{\sep}(u_i,v_j,k/K)\big)^2\right\}}
     {\sum_{\ell=0}^K \exp\!\left\{-\tfrac{\tau_{ij}}{2}\big(x_{ij}- \hat \rmg_{\sep}(u_i,v_j,\ell/K)\big)^2\right\}},
\quad k=0,\ldots,K.
\end{equation*}
This yields a sample $z_{ij}= \hat \rmg_{\sep}(u_i,v_j,u_{ij})$ from the variational posterior. 
\subsection{EBMR with a Jointly Exchangeable Prior} \label{sec-EBMR-joint}
We now develop empirical Bayes matrix recovery under a jointly exchangeable prior. Assume that \Cref{assum:true-prior} and \Cref{assum:phi-prior} hold with the true prior being jointly exchangeable. By the Aldous--Hoover theorem, the population distribution for $\bx$ has the form
\begin{align}
    &\rmg_{\mathrm{joint}} \sim \mu^\star, \label{eq:T-distribution} \\
    &u_i,\, u_{ij} \iid \unif[0,1], \label{eq:U-uniform} \\
    &z_{ij} = \rmg_{\mathrm{joint}}(u_i, u_j, u_{ij}), \label{eq:Z-definition} \\
    &x_{ij} \sim \cN\!\left(z_{ij}, \tau_{ij}^{-1}\right), \quad (i,j) \in \N^2. \label{eq:nm}
\end{align}
For finite population, we require the observed matrix to have the same number of rows and columns to apply joint exchangeability. For $n \ge 1$, let $\bx_n = (x_\omega)_{\omega \in [n]^2}$ denote an observed $n \times n$ matrix. By \Cref{prop:finite-EB}, $\bx_n$ follows the marginal distribution of \eqref{eq:U-uniform}--\eqref{eq:nm} on the index set $S_n = [n]^2$.

Let $\bz_n = (z_\omega)_{\omega \in S_n}$ denote the corresponding local latent variables. For the invariant variable $\rmg_{\mathrm{joint}}$, the conditional distribution
$\rmp\left(\bz_n \mid \rmg_{\mathrm{joint}} \right)$ is given by
\eqref{eq:U-uniform}--\eqref{eq:Z-definition}. By \Cref{prop:finite-EB}, the finite-sample generative process is described by the same equations \eqref{eq:T-distribution}--\eqref{eq:nm} with $(i,j) \in [n]^2$.
\subsubsection{Fitting procedure}
Let $\bu_n = (u_i, u_{ij})_{(i,j) \in [n]^2}$ denote the set of row-specific and entry-specific latent variables.  
\Cref{alg:BEB} estimates $\hat \rmg_{\mathrm{joint}}$ by maximizing the marginal likelihood
\begin{equation}
\log \rmp\!\left(\bx_n \mid \rmg_{\mathrm{joint}}\right) \deq 
\log \int_{[0,1]^{\,n + n^2}} \prod_{i = 1}^n \prod_{j = 1}^n  
\rmp \!\left(x_{ij} \mid  \rmg_{\mathrm{joint}}\!\left(u_i, u_j, u_{ij}\right) \right)\, \dd \bu_n. 
\end{equation}
Directly maximizing the marginal likelihood over all functions $\rmg_{\mathrm{joint}}: [0,1]^3 \mapsto \R$ is intractable, due to the $(n + n^2)$-dimensional integral. So we approximate the integral using variational inference. 

\subsubsection{Variational approximation}
Consider the posterior $\rmp\bigl(\bu_n \mid \bx_n, \rmg_{\mathrm{joint}}\bigr)$ under model~\eqref{eq:U-uniform}--\eqref{eq:nm}. 

In the exact posterior $\rmp\bigl(\bu_n \mid \bx_n, \rmg_{\mathrm{joint}}\bigr)$, the variables $(u_{ij})_{i,j\in[n]}$ are conditionally independent given $(u_i)_{i\in[n]}$. To capture the graphical structure locally, we posit the variational family
\begin{equation} \label{variational family joint}
    \cQ_{\mathrm{joint}} 
    = \Bigl\{\rmq:\ 
    \rmq(\bu_n) 
    = \prod_{i = 1}^{n} \rmq_{\bu,i}(u_i)\, \prod_{i,j} \rmq_{i,j}(u_{ij} \mid u_i, u_j) \Bigr\}.
\end{equation}
Let $\rmq_{\bu} = \bigotimes_{i=1}^n \rmq_{\bu,i}$. The variational method maximizes the evidence lower bound (ELBO) subject to $\rmq\in\cQ_{\mathrm{joint}}$. 

\begin{proposition} \label{prop-EBMR-joint}
Let $\cQ$ be the variational family in \eqref{variational family joint}. The variational loss for $\rmg_{\mathrm{joint}}$ is
\begin{equation} \label{ELBO-EBMR-joint}
\begin{aligned}
\ell \bigl(\bx_n; \rmg_{\mathrm{joint}} \bigr) 
\deq \inf_{\rmq_{\bu}\in \cP ([0, 1])^{\otimes n}} \Bigg\{ 
& - \sum_{i = 1}^n \sum_{j = 1}^n 
\EE{\rmq_{\bu,i}(u_i)\, \rmq_{\bu,j}(u_j)}{ 
\LSE\!\left(x_{ij}, \tau_{ij}, u_i, u_j, \rmg_{\mathrm{joint}} \right)}  \\
&\qquad\qquad
+ \sum_{i = 1}^{n} \EE{\rmq_{\bu,i}(u_i)}{\log \rmq_{\bu,i}(u_i)} 
\Bigg\},    
\end{aligned}
\end{equation}
where 
\[
\LSE\!\left(x_{ij}, \tau_{ij}, u_i, u_j, \rmg_{\mathrm{joint}} \right) 
\deq \log \int_0^1 \exp \!\left(-\frac{\tau_{ij}}{2}\left[x_{ij}  - \rmg_{\mathrm{joint}}(u_i, u_j, u) \right]^2 \right)\! \dd u. 
\]
\end{proposition}
We parametrize $\rmg_{\mathrm{joint}}$ with a neural network class $\cG_{\mathrm{NN}}$ and fit it by minimizing the variational loss:
\begin{equation} \label{EBMR-obj-joint}
\hat \rmg_{\mathrm{joint}} \in \arg \min_{\rmg_{\mathrm{joint}} \in \cG_{\mathrm{NN}}} \ell\bigl(\bx_n; \rmg_{\mathrm{joint}}\bigr).
\end{equation}
Since $\rmq_{\bu,i} \in \cP([0,1])$, we discretize it as 
\begin{equation*}
\rmq_{\bu,i} = \sum_{k = 0}^K w_{ik}\, \delta_{u_k}, 
\qquad u_k = \frac{k}{K}, \qquad \sum_{k = 0}^K w_{ik} = 1.
\end{equation*}
\Cref{alg:EBMR-jointly} implements EBMR with a jointly exchangeable prior. 

With the estimated function $\hat \rmg_{\mathrm{joint}}$ from \Cref{alg:EBMR-jointly}, we use the No-U-Turn sampler (NUTS) to sample from the posterior $\rmp\!\left(z_{ij} \mid \hat \rmg_{\mathrm{joint}}, \bx_n \right)$ of the EBMR model, given by
\begin{equation}
(u_i, u_j, u_{ij}) \sim \rmp\!\left(u_i, u_j, u_{ij} \mid \bx_n, \hat \rmg_{\mathrm{joint}} \right),  \quad z_{ij} = \hat \rmg_{\mathrm{joint}}(u_i, u_j, u_{ij}). 
\end{equation}

\begin{algorithm}[H]
\caption{EBMR with a Jointly Exchangeable Prior (EBMR-Joint)}\label{alg:EBMR-jointly}
\SetAlgoLined
\KwIn{Data matrix $\bx_n$, precision matrix $\tau_n=(\tau_{ij})_{i,j\in[n]}$, network class $\cG_{\mathrm{NN}}$, grid size $K$, optimization hyperparameters (epochs, learning rate)}
\textbf{Initialize:} network $\rmg_{\mathrm{joint}}\in \cG_{\mathrm{NN}}$; weights $\bw_1, \ldots, \bw_n \in \Delta_{K+1}$.\\[2pt]

\Repeat{\textnormal{convergence}}{

\For{$i=1$ \KwTo $n$ \textnormal{(parallelizable)}}{
\begin{align*}
\rho_{ik} &\leftarrow
\widehat{\LSE}\!\left(x_{ii},\tau_{ii},u_k,u_k,\rmg_{\mathrm{joint}}\right)
+\sum_{j\neq i}\sum_{k_2=0}^{K} w_{jk_2}\,
\widehat{\LSE}\!\left(x_{ij},\tau_{ij},u_k,u_{k_2},\rmg_{\mathrm{joint}}\right),
&& k = 0,\ldots,K,  \\
\bw_i &\leftarrow \textsc{softmax}\bigl(\rho_{i0},\ldots,\rho_{iK}\bigr).
\end{align*}
}

Update the neural network parameters with stochastic gradient ascent:
\begin{equation*}
\rmg_{\mathrm{joint}}\leftarrow
\arg\max_{\rmg\in \cG_{\mathrm{NN}}}
\sum_{i=1}^{n}\sum_{k=0}^{K} w_{ik}\!\left[
\widehat{\LSE}\!\left(x_{ii},\tau_{ii},u_k,u_k,\rmg\right)
+\sum_{j\neq i}\sum_{k_2=0}^{K} w_{jk_2}\,
\widehat{\LSE}\!\left(x_{ij},\tau_{ij},u_k,u_{k_2},\rmg\right)
\right].
\end{equation*}
}
\KwOut{Fitted weights $\bw_1,\ldots,\bw_n$ and network $\rmg_{\mathrm{joint}}$.}
\end{algorithm}

\begin{corollary} \label{prop:alg-EBMR-jointly}
\Cref{alg:EBMR-sep} performs block-coordinate ascent on the variational inference problem~\eqref{discretized-EBMR-obj}.
\end{corollary}
\subsection{Covariate-Informed Empirical Bayes}
\label{sec-caeb}

We extend empirical Bayes matrix recovery with a separately exchangeable prior to incorporate covariates on rows and columns. Let $\Omega=\N^2$ and $\cX=\R$. We observe a matrix $\bx_{n,p}=(x_{ij})_{(i,j)\in[n]\times[p]}$ with row covariates $\by_n=(y_i)_{i=1}^n$ with $y_i\in\cY$ and column covariates $\ba_p=(a_j)_{j=1}^p$ with $a_j\in\cA$, generated via
\begin{equation*}
    x_{ij}\sim \cN(z_{ij},\tau_{ij}^{-1}),\qquad \tau_{ij}>0. 
\end{equation*}
The statistical goal is to recover the latent array $\bz_{n,p}$. On the population level, we impose \emph{relative exchangeability} with respect to the observed covariates: the infinite array $\bz$ is said to be $\by$--$\ba$ exchangeable if its law is invariant under relabelings that preserve covariate values,
\begin{equation*}
    \Phi_{\by,\ba}:=
\bigl\{\sigma\in\SS: y_{\sigma(i)}=y_i\ \forall i\bigr\}
\times
\bigl\{\pi\in\SS: a_{\pi(j)}=a_j\ \forall j\bigr\}. 
\end{equation*}
Under $\by$--$\ba$ exchangeability, an extension of Aldous--Hoover theorem yields a measurable representation \citep{Crane2018}: there exists a random function $\rmg_{\rel}:  \cY \times \cA \times [0,1]^3 \to \R$ such that, for iid $u_i,v_j,u_{ij}\sim\unif[0,1]$,
\begin{equation*}
    u_i,v_j,u_{ij}\iid \unif[0,1],\qquad
z_{ij}=\rmg_{\rel}\bigl(y_i,a_j,u_i,v_j,u_{ij}\bigr).
\end{equation*}
 The covariate-informed array model is
\begin{equation}\label{eqn:CIEB}
    u_i,v_j,u_{ij}\iid \unif[0,1],\quad
z_{ij}=\rmg_{\textrm{rel}}\left(y_i,a_j,u_i,v_j,u_{ij}\right),\quad
x_{ij}\sim \cN\left(z_{ij},\tau_{ij}^{-1}\right). 
\end{equation}
We call the Bayesian empirical Bayes method for this model \emph{covariate-informed empirical Bayes} (CIEB). The estimation procedure mirrors EBMR with a separately exchangeable prior: we maximize a variational lower bound to the marginal likelihood over a structured family for $(u_i,v_j,u_{ij})$. We parameterize the invariant variable $\rmg_{\rel}$ as a neural network
that takes $(y_i,a_j,u_i,v_j,u_{ij})$ as inputs. With $\hat\rmg_{\rel}$ in hand, we draw from the plug-in posterior $\rmp(\bz_{n,p}\mid \hat\rmg_{\rel},\bx_{n,p})$. In the next section, we derive the variational lower bound to learn the invariant variable $\rmg_{\rel}$. 




Related covariate-dependent approaches include that of \citet{Ignatiadis2019},
which fits a normal-means sequence model with the prior’s mean and variance modeled
as functions of covariates, and \citet{denault2025covariate}, which develops
covariate-dependent empirical Bayes matrix factorization. In the latter model, the loading and factor components each follow i.i.d.\ priors parameterized by row/column covariates.


\subsubsection{Fitting procedure} \label{sec-CIEB-computation}
Let $\bu_{n,p}\coloneqq \bigl(u_i,v_j,u_{ij}\bigr)_{(i,j)\in[n]\times[p]}$. 
To follow \Cref{alg:BEB}, the inference steps of covariate-informed EB involve
\begin{enumerate}
\item Compute $\hat \rmg_{\rel}$ by maximizing the marginal likelihood
\begin{equation}\label{EBMR-rel}
\log \rmp\bigl(\bx_{n,p}\mid\rmg_{\rel}\bigr)
\coloneqq
\log \int_{[0,1]^{n+p+np}}
\exp\left\{-\sum_{i=1}^n\sum_{j=1}^p \frac{\tau_{ij}}{2}\bigl[x_{ij}-\rmg_{\rel}(y_i,a_j,u_i,v_j,u_{ij})\bigr]^2\right\}\, d\bu_{n,p}.
\end{equation}
\item Sample $z_{ij}$ from the posterior $\rmp\left(z_{ij}\mid \hat \rmg_{\rel}, \bx_{n,p}\right)$ for $i \in [n]$ and $j \in [p]$.
\end{enumerate}
As in EBMR-sep, the MMLE in \eqref{EBMR-rel} is intractable for large \(n,p\). We therefore adopt the same structured variational family \(\cQ_{\sep}\) from \eqref{variational-family} for the latent variables \(\bu_{n,p}\) and discretize each variational marginal on a \(K\)-point grid over \([0,1]\). We parameterize \(\rmg_{\rel}\) with a neural network class \(\cG_{\mathrm{NN}}\) and minimize the variational loss
\begin{equation*}
    \hat \rmg_{\rel} \in \arg\min_{\rmg_{\rel}\in \cG_{\mathrm{NN}}}\ell(\bx_{n,p}; \rmg_{\rel}), 
\end{equation*}
where, with discretized weights,
\begin{equation*}
\begin{aligned}
\ell(\bx_{n,p}; \rmg_{\rel})
=\;
\inf_{\bw_{\bu}, \bw_{\bv}\in \Delta_{K+1}}
\Bigg\{&
-\sum_{i=1}^n\sum_{j=1}^p\sum_{k=0}^K\sum_{k'=0}^K
\bw_{\bu,ik}\,\bw_{\bv,jk'}\,
\LSE\left(x_{ij},\tau_{ij},y_i,a_j,\tfrac{k}{K},\tfrac{k'}{K},\rmg_{\rel}\right)
\\
&\quad + \sum_{i=1}^n\sum_{k=0}^K \bw_{\bu,ik}\log \bw_{\bu,ik}
+ \sum_{j=1}^p\sum_{k=0}^K \bw_{\bv,jk}\log \bw_{\bv,jk}
\Bigg\},
\end{aligned}
\end{equation*}
with
\begin{equation*}
    \LSE(X,\tau,Y,A,U,V,\rmg_{\rel})
\coloneqq
\log \int_0^1 \exp\left\{-\tfrac{\tau}{2}\bigl(X-\rmg_{\rel}(Y,A,U,V,u)\bigr)^2\right\}\,du.
\end{equation*}
Weight updates are similar to those in EBMR-sep:
\begin{align}
\bw_{\bu,ik} &\propto
\exp\left(
\sum_{j=1}^p \sum_{k'=0}^K \bw_{\bv,jk'}\,
\log \sum_{k''=0}^K \exp\left\{-\tfrac{\tau_{ij}}{2}\Big[x_{ij}-\rmg_{\rel}\bigl(y_i,a_j,\tfrac{k}{K},\tfrac{k'}{K},\tfrac{k''}{K}\bigr)\Big]^2\right\}
\right), \label{CIEB-wu-update}
\\[4pt]
\bw_{\bv,jk} &\propto
\exp\left(
\sum_{i=1}^n \sum_{k'=0}^K \bw_{\bu,ik'}\,
\log \sum_{k''=0}^K \exp\left\{-\tfrac{\tau_{ij}}{2}\Big[x_{ij}-\rmg_{\rel}\bigl(y_i,a_j,\tfrac{k'}{K},\tfrac{k}{K},\tfrac{k''}{K}\bigr)\Big]^2\right\}
\right), \label{CIEB-wv-update}
\end{align}
normalized to lie in $\Delta_{K+1}$. With fixed weights, update $\rmg_{\rel}$ by
\begin{equation}\label{CIEB-g-update}
\hat\rmg_{\rel} \in \arg\max_{\rmg_{\rel}\in \cG_{\textrm{NN}}}
\sum_{i=1}^n \sum_{j=1}^p \sum_{k=0}^K \sum_{k'=0}^K
\bw_{\bu,ik}\,\bw_{\bv,jk'}\,
\log \sum_{k''=0}^K \exp\left\{-\tfrac{\tau_{ij}}{2}\Big[x_{ij}-\rmg_{\rel}\bigl(y_i,a_j,\tfrac{k}{K},\tfrac{k'}{K},\tfrac{k''}{K}\bigr)\Big]^2\right\}.
\end{equation}
The algorithm fits $\hat \rmg_{\rel}$ by iterating steps~\eqref{CIEB-wu-update}--\eqref{CIEB-g-update} until convergence.  With the estimate $\hat \rmg_{\rel}$, we sample from the posterior $\rmp\!\left(z_{ij} \mid \hat \rmg_{\rel}, \bx_{n, p}\right)$ of model~\eqref{eqn:CIEB}, given by
\begin{equation}
(u_i, u_j, u_{ij}) \sim \rmp\!\left(u_i, u_j, u_{ij} \mid \hat \rmg_{\rel}, \bx_{n, p} \right),  \quad z_{ij} = \hat \rmg_{\rel}(u_i, u_j, u_{ij}). 
\end{equation}

\section{Algorithms for Empirical Bayes Spatial Regression}
\label{sec-alg-spatial}

Recall that the invariant variable is the spectral density $\psi\in\cP_{ac}(\R^d)$. Following \citet{Wilson2013}, we further parametrize $\psi$ as a Gaussian mixture,
\begin{equation}
  \psi_\theta(s)=\sum_{k=1}^K w_k\,\phi\!\left(s;\mu_k,\Sigma_k\right), 
  \qquad w_k\ge0,\ \sum_{k=1}^K w_k=1,
  \label{eq:GM-spectral}
\end{equation}
where each $\Sigma_k=\mathrm{diag}(\sigma_{k,1}^2,\ldots,\sigma_{k,d}^2)$ is diagonal. By Bochner’s theorem, \Cref{eq:GM-spectral} induces the kernel
\begin{equation}
  k_\theta(\Delta)
  =\sum_{k=1}^K w_k
  \prod_{j=1}^d
  \cos\!\bigl(2\pi\,\mu_{k,j}\,\Delta_j\bigr)\,
  \exp\!\bigl(-2\pi^2\,\sigma_{k,j}^2\,\Delta_j^2\bigr), 
  \label{eq:GM-kernel}
\end{equation}
for lags $\Delta\in\R^d$. The invariant variable is thus
$\theta=(w_{1:K},\mu_{1:K},\Sigma_{1:K})$, which governs the stationary prior
$\GP\!\left(0,k_\theta\right)$ for the latent process.

With this parametrization, the empirical Bayes spatial regression model is
\begin{equation}
\bz \sim \GP(0,k_\theta), 
\qquad \beta \sim \pi(\beta)\propto1, 
\qquad x_\omega \sim \cN\!\bigl(\beta^\top a_\omega+z_\omega,\ \tau_\omega^{-1}\bigr).
\label{eq:EB-spatial-reg}
\end{equation}

\paragraph{Posterior Inference.}
We estimate the kernel parameters $\theta$ by maximizing the marginal likelihood,
\begin{equation}
\hat\theta
\in
\argmax_{\theta\in\Theta}
\log
\int
\Bigl[\prod_{\omega\in S_n}\rmp(x_\omega\mid z_\omega,\beta)\Bigr]\,
\rmp(\bz_n\mid\theta)\,\pi(\beta)\;d\beta\,d\bz_n.
\end{equation}
This fits the spectral density (equivalently, the kernel) that best explains the marginal distribution of the observed spatial data. Given $\hat\theta$, the posterior $\rmp(\beta,\bz_n\mid\bx_n,\hat\theta)$ is Gaussian with known mean and covariance. 

Let $S_n=(\omega_1,\ldots,\omega_n)$, $\bA_n=[a_{\omega_1},\ldots,a_{\omega_n}]^\top\in\R^{n\times p}$, and
$D_\tau=\mathrm{diag}(\tau_{\omega_1},\ldots,\tau_{\omega_n})$. With
$\Sigma_\theta=[k_\theta(\omega_i-\omega_j)]_{(i,j) \in [n]^2}$, define $S_\theta=\Sigma_\theta+D_\tau^{-1}$ and
$W_\theta=S_\theta^{-1}$. Assuming $\bA_n$ has full column rank and $\pi(\beta)\equiv1$, the MMLE objective is
\begin{equation}\label{spatial-objective}
\ell(\bx_n;\theta)=\log\det S_\theta+\log\!\bigl|\bA_n^\top W_\theta\bA_n\bigr|
+\bx_n^\top W_\theta\bx_n-\bx_n^\top W_\theta\bA_n\hat\beta_\theta,
\end{equation}
where $\hat\beta_\theta = (\bA_n^\top W_\theta\bA_n)^{-1}\bA_n^\top W_\theta\bx_n$ is the weighted least-squares estimator.

We minimize \eqref{spatial-objective} using stochastic gradient descent \citep{Robbins1951}. The computational bottleneck is to solve linear systems in $S_\theta$; we avoid forming $W_\theta$ explicitly by using a Cholesky factorization $S_\theta=LL^\top$. The resulting costs are $O(n^3)$ time and $O(n^2)$ space, which are tractable if the number of sites $n$ is in the thousands. Scalable alternatives are available, including Nyström approximations and variational low-rank Gaussian processes \citep{hensman2013GP}.

With $S_{\hat\theta}=\Sigma_{\hat\theta}+D_\tau^{-1}$, $W_{\hat\theta}=S_{\hat\theta}^{-1}$, and
$\hat\beta_{\hat\theta}=(\bA_n^\top W_{\hat\theta}\bA_n)^{-1}\bA_n^\top W_{\hat\theta}\bx_n$,
\begin{equation}\label{spatial-posterior-beta}
\beta\mid \bx_n \sim \cN\!\left(\hat\beta_{\hat\theta},\,(\bA_n^\top W_{\hat\theta}\bA_n)^{-1}\right),
\end{equation}
and, conditional on $\beta$,
\begin{equation}\label{spatial-posterior-z}
\bz_n\mid \beta,\bx_n \sim \cN\!\left((\Sigma_{\hat\theta}^{-1}+D_\tau)^{-1} D_\tau(\bx_n-\bA_n\beta),\,
(\Sigma_{\hat\theta}^{-1}+D_\tau)^{-1}\right).
\end{equation}
Posterior samples are obtained by Monte Carlo: sample $\beta^{(s)}$ from \eqref{spatial-posterior-beta}, then
$\bz_n^{(s)}$ from \eqref{spatial-posterior-z} with $\beta=\beta^{(s)}$.

\paragraph{Kriging.}
To predict at new sites $S_n'\supset S_n$, define $Q_\theta\in\R^{|S_n'|\times|S_n'|}$ by
$Q_{\theta,ii}=[\Sigma_\theta^{-1}]_{ii}+\tau_i$ and $Q_{\theta,ij}=[\Sigma_\theta^{-1}]_{ij}$ for $i\neq j$,
where $\tau_i$ denotes the noise precision at site $\omega_i$. Let $L\in\R^{|S_n'|}$ have entries
$L_{S_n}=D_\tau\bigl(\bx_{S_n}-\bA_{S_n}\beta\bigr)$ and zeros elsewhere. Then
\begin{equation}\label{kriging-posterior}
\bz_{S_n'}\mid \beta,\bx_n \sim \cN\!\left(Q_{\hat\theta}^{-1}L,\;Q_{\hat\theta}^{-1}\right).
\end{equation}

\section{Proofs}

\subsection{Proofs of \Cref{sec-method}}
\begin{proof}[Proof of \Cref{prop:finite-EB} and \Cref{prop:finite-invariant}]
Let $S_n^c \coloneqq \Omega \setminus S_n$.  
The true distribution $\rmp^\star(\bx_n)$ is the $S_n$-marginal of $\rmp^\star(\bx)$:
\begin{equation*}
    \rmp^\star(\bx_n)
    = \int \rmp^\star(\bx)\, \dd \bx_{S_n^c}.  
\end{equation*}
By \Cref{assum:true-prior},
\begin{equation*}
    \rmp^\star(\bx)
    = \int_{\cZ^\Omega} \rmp^\star(\bz)\,
      \prod_{\omega \in \Omega} \rmp(x_\omega \mid z_\omega)\, \dd\bz.
\end{equation*}
Applying Fubini’s theorem to interchange the integrals over $\bz$ and $\bx_{S_n^c}$ gives
\begin{align*}
\rmp^\star(\bx_n)
&= \int_{\cX^{S_n^c}} \int_{\cZ^\Omega}
      \rmp^\star(\bz)\,
      \prod_{\omega \in \Omega} \rmp(x_\omega \mid z_\omega)\,
      \dd\bz\, \dd \bx_{S_n^c} \\
&= \int_{\cZ^\Omega} \rmp^\star(\bz)\,
     \prod_{\omega \in S_n} \rmp(x_\omega \mid z_\omega)
     \left(\int_{\cX^{S_n^c}}
       \prod_{\omega \in S_n^c} \rmp(x_\omega \mid z_\omega)\,
       \dd \bx_{S_n^c} \right)\dd\bz \\
&= \int_{\cZ^\Omega} \rmp^\star(\bz)\,
     \prod_{\omega \in S_n} \rmp(x_\omega \mid z_\omega)\, \dd\bz, 
\end{align*}
which establishes \Cref{prop:finite-EB}. 

For \Cref{prop:finite-invariant}, we integrate the representation \eqref{invariance} over $\bx_{S_n}$:
\begin{align*}
\rmp^\star(\bx_n)
&= \int_{\cZ^{S_n}} \int_{\cG}
   \!\left(\int_{\cZ^{S_n^c}} \rmp(\bz \mid \rmg)\, \dd \bz_{S_n^c} \right)
   \dd \mu^\star(\rmg)\,
   \prod_{\omega \in S_n} \rmp(x_\omega \mid z_\omega)\, \dd \bz_{S_n} \\
&= \int_{\cZ^{S_n}}
   \left(\int_{\cG} \rmp(\bz_n \mid \rmg)\, \dd \mu^\star(\rmg)\right)
   \prod_{\omega \in S_n} \rmp(x_\omega \mid z_\omega)\, \dd \bz_{S_n}.
\end{align*}
\end{proof}

\subsection{Proofs of the Algorithms}\label{sec-proof-inference}

\begin{proof}[Proof of \Cref{ELBO-EBMR-separate}]
For $\rmq \in \cQ_{\sep}$ we can decompose the ELBO as
\begin{align*}
\begin{split}
\ELBO{\rmq, \rmg_{\sep}}
=\, & \sum_{i=1}^n \sum_{j=1}^p
      \EE{\rmq_{\bu,i}(u_i)\, \rmq_{\bv,j}(v_j)}{
        \EE{\rmq_{ij}(u_{ij} \mid u_i, v_j)}{
          \log \rmp\!\left(x_{ij} \mid \rmg_{\sep}(u_i,v_j,u_{ij})\right)}} \\
&\quad
    - \sum_{i=1}^n \EE{\rmq_{\bu,i}(u_i)}{\log \rmq_{\bu,i}(u_i)}
    - \sum_{j=1}^p \EE{\rmq_{\bv,j}(v_j)}{\log \rmq_{\bv,j}(v_j)} \\
&\quad
    - \sum_{i=1}^n \sum_{j=1}^p
      \EE{\rmq_{\bu,i}(u_i)\, \rmq_{\bv,j}(v_j)}{
        \EE{\rmq_{ij}(u_{ij}\mid u_i,v_j)}{
          \log \rmq_{ij}(u_{ij}\mid u_i,v_j)}}.
\end{split}
\end{align*}
Introduce the reference distribution
\[
\rmq_{\mathrm{ref}}(u_{ij}\mid u_i,v_j)
\propto
\exp\!\left\{-\frac{\tau_{ij}}{2}\bigl(x_{ij}-\rmg_{\sep}(u_i,v_j,u_{ij})\bigr)^2\right\}.
\]
Then
\begin{equation}
\begin{aligned}
\ELBO{\rmq, \rmg_{\sep}}
 &= \sum_{i=1}^n \sum_{j=1}^p
      \EE{\rmq_{\bu,i}(u_i)\, \rmq_{\bv,j}(v_j)}{
        \LSE(x_{ij},\tau_{ij},u_i,v_j,\rmg_{\sep})} \\
&\quad - \sum_{i=1}^n \EE{\rmq_{\bu,i}(u_i)}{\log \rmq_{\bu,i}(u_i)}
      - \sum_{j=1}^p \EE{\rmq_{\bv,j}(v_j)}{\log \rmq_{\bv,j}(v_j)} \\
&\quad
    - \sum_{i=1}^n \sum_{j=1}^p
      \EE{\rmq_{\bu,i}(u_i)\, \rmq_{\bv,j}(v_j)}{
        \KL{\rmq_{ij}(\cdot\mid u_i,v_j)}{\rmq_{\mathrm{ref}}(\cdot\mid u_i,v_j)}},
\end{aligned}
\end{equation}
where
\[
\LSE(X,\tau,U,V,\rmg_{\sep})
\coloneqq
\log\!\int_0^1
\exp\!\left\{-\frac{\tau}{2}\bigl[X-\rmg_{\sep}(U,V,u)\bigr]^2\right\}\dd u.
\]
Maximizing over $\rmq_{ij}$ for fixed $\rmq_{\bu},\rmq_{\bv}$ gives
\[
\rmq_{ij}^\star(u_{ij}\mid u_i,v_j)
= \rmq_{\mathrm{ref}}(u_{ij}\mid u_i,v_j),
\]
and evaluating at $\rmq^\star$ yields
\begin{align*}
\ELBO{\rmq^\star, \rmg_{\sep}}
&= \sum_{i=1}^n \sum_{j=1}^p
      \EE{\rmq^\star_{\bu,i}(u_i)\, \rmq^\star_{\bv,j}(v_j)}{
        \LSE(x_{ij},\tau_{ij},u_i,v_j,\rmg_{\sep})} \\
&\quad - \sum_{i=1}^n \EE{\rmq^\star_{\bu,i}(u_i)}{\log \rmq^\star_{\bu,i}(u_i)}
      - \sum_{j=1}^p \EE{\rmq^\star_{\bv,j}(v_j)}{\log \rmq^\star_{\bv,j}(v_j)}.
\end{align*}
\end{proof}

\begin{proof}[Proof of \Cref{prop:alg-EBMR-sep}]
Recall that $u_k=k/K$ for $k=0,\ldots,K$.  
The discretized form of~\eqref{ELBO-EBMR-separate} is
\begin{equation}\label{discrete-ELBO-EBMR-joint}
\begin{aligned}
\hat{\rmg}_{\mathrm{sep}}
\in \arg\max_{\rmg_{\mathrm{sep}}\in\cG_{\mathrm{NN}}}
\ \sup_{\bw\in\Delta_{K+1}^{\,n+p}}
\Bigg\{&
\sum_{i=1}^n \sum_{j=1}^p \sum_{k_1=0}^K \sum_{k_2=0}^K
w_{\bu,ik_1}\, w_{\bv,jk_2}\,
\widehat{\LSE}\!\left(x_{ij},\tau_{ij},u_{k_1},u_{k_2},\rmg_{\mathrm{sep}}\right) \\
&\quad
- \sum_{i=1}^n \sum_{k=0}^K w_{\bu,ik}\log w_{\bu,ik}
- \sum_{j=1}^p \sum_{k=0}^K w_{\bv,jk}\log w_{\bv,jk}
\Bigg\},
\end{aligned}
\end{equation}
where $\bw$ collects all weight vectors
$\bw_{\bu,1},\ldots,\bw_{\bu,n},\bw_{\bv,1},\ldots,\bw_{\bv,p}$ and
$\widehat{\LSE}$ is the discretized form of $\LSE$.

We fit $\bw_{\bu},\bw_{\bv}$ and $\rmg_{\mathrm{sep}}$ by alternating updates.  
For $j\in[p]$ and $k=0,\ldots,K$,
\begin{equation}
w_{\bv,jk}
=
\frac{\exp\!\left(\sum_{i=1}^n \sum_{k_1=0}^K
w_{\bu,ik_1}\,
\widehat{\LSE}\!\left(x_{ij},\tau_{ij},u_{k_1},u_k,\rmg_{\mathrm{sep}}\right)\right)}
{\sum_{k'=0}^K \exp\!\left(\sum_{i=1}^n \sum_{k_1=0}^K
w_{\bu,ik_1}\,
\widehat{\LSE}\!\left(x_{ij},\tau_{ij},u_{k_1},u_{k'},\rmg_{\mathrm{sep}}\right)\right)}.
\end{equation}
Similarly, for $i\in[n]$ and $k=0,\ldots,K$,
\begin{equation}
w_{\bu,ik}
=
\frac{\exp\!\left(\sum_{j=1}^p \sum_{k_2=0}^K
w_{\bv,jk_2}\,
\widehat{\LSE}\!\left(x_{ij},\tau_{ij},u_k,u_{k_2},\rmg_{\mathrm{sep}}\right)\right)}
{\sum_{k'=0}^K \exp\!\left(\sum_{j=1}^p \sum_{k_2=0}^K
w_{\bv,jk_2}\,
\widehat{\LSE}\!\left(x_{ij},\tau_{ij},u_{k'},u_{k_2},\rmg_{\mathrm{sep}}\right)\right)}.
\end{equation}

Finally, we train the neural network $\rmg_{\mathrm{sep}}$ by maximizing the weighted objective
\begin{equation}
\rmg_{\mathrm{sep}} \in
\arg\max_{\rmg_{\mathrm{sep}}\in\cG_{\mathrm{NN}}}
\sum_{i=1}^n \sum_{j=1}^p \sum_{k_1=0}^K \sum_{k_2=0}^K
w_{\bu,ik_1}\, w_{\bv,jk_2}\,
\widehat{\LSE}\!\left(x_{ij},\tau_{ij},u_{k_1},u_{k_2},\rmg_{\mathrm{sep}}\right).
\end{equation}
\end{proof}

\begin{proof}[Proof of \Cref{prop-EBMR-joint}]
By the Gibbs variational principle,
\begin{equation} \label{Gibbs-VP-joint}
\begin{aligned}
\log \rmp\!\left(\bx_n \mid \rmg_{\mathrm{joint}} \right) 
= \sup_{ \rmq \in \cP ([0,1]^{\,n + n^2})} 
\Bigg\{&
\sum_{i = 1}^n \sum_{j = 1}^n 
\EE{\rmq(u_i, u_j, u_{ij}) }{ 
\log \rmp \!\left(x_{ij} \mid \rmg_{\mathrm{joint}} (u_i, u_j, u_{ij}) \right)}  - \EE{\rmq(\bu_n)}{\log \rmq(\bu_n)} \Bigg\}.
\end{aligned}
\end{equation}
Leveraging the structure of $\rmq \in \cQ$ and rearranging yields
\begin{equation*}
\textsc{ELBO} = \sup_{\rmq \in \cQ} \Bigg\{
- \sum_{i = 1}^n \sum_{j = 1}^n 
\EE{\rmq_{\bu,i}(u_i)\, \rmq_{\bu,j}(u_j)}{
\KL{\rmq_{ij}(\cdot \mid u_i, u_j)}{\rmq_{\mathrm{ref}}(\cdot \mid u_i, u_j)}}
- \sum_{i = 1}^n \EE{\rmq_{\bu,i}(u_i)}{\log \rmq_{\bu,i}(u_i)} 
\Bigg\},
\end{equation*}
where $\rmq_{\mathrm{ref}}(u_{ij}\mid u_i,u_j)
\propto \exp \!\left(-\frac{\tau_{ij}}{2} \bigl[x_{ij} - \rmg_{\mathrm{joint}}(u_i, u_j, u_{ij}) \bigr]^2 \right)$.  Then, maximizing the ELBO over $\rmq_{ij}$ for fixed $\rmq_{\bu}$ yields the stated objective.
\end{proof}

\begin{proof}[Proof of \Cref{spatial-objective}]
Integrating over $\bz_n \sim \cN(0,\Sigma_\theta)$ gives
\[
\bx_n \mid \beta,\theta \sim \cN\bigl(\bA_n\beta, S_\theta\bigr).
\]
With $\pi(\beta)\equiv1$, integrating over $\beta$ yields
\[
\log \rmp(\bx_n\mid\theta)
= -\tfrac12\Bigl\{
  \log\det S_\theta
  + \log\bigl|\bA_n^\top W_\theta\bA_n\bigr|
  + (\bx_n-\bA_n\hat\beta_\theta)^\top
    W_\theta(\bx_n-\bA_n\hat\beta_\theta)
  \Bigr\} + C,
\]
where $C$ is constant in $\theta$.  
The objective $\ell(\bx_n;\theta)$ equals $-\log\rmp(\bx_n\mid\theta)$ up to an additive constant.
\end{proof}

\begin{proof}[Proof of \Cref{prop:alg-EBMR-jointly}]
The discretized version of \eqref{ELBO-EBMR-joint} is
\begin{align*}
\sup_{\bw \in \Delta_{K+1}^{\,n}}
\ \sup_{\rmg_{\mathrm{joint}} \in \cG_{\mathrm{NN}}}\ 
\sum_{i = 1}^{n}  \sum_{k = 0}^K  w_{ik} \Bigg[
& \widehat{\LSE}\!\left(x_{ii}, \tau_{ii}, u_k, u_k, \rmg_{\mathrm{joint}} \right)
+ \sum_{j \neq i} \sum_{k_2 = 0}^K w_{j k_2}\,
   \widehat{\LSE}\!\left(x_{ij}, \tau_{ij}, u_{k}, u_{k_2}, \rmg_{\mathrm{joint}}\right) 
- \log w_{ik}\Bigg], 
\end{align*}
where
\[
\widehat{\LSE}\!\left(x_{ij}, \tau_{ij}, u_i, u_j, \rmg_{\mathrm{joint}}\right) 
\deq \log \sum_{k = 0}^K \exp \!\left(-\frac{\tau_{ij}}{2}\left[x_{ij}  - \rmg_{\mathrm{joint}}(u_i, u_j, u_k) \right]^2  \right) - \log (K+1).
\]
Alternating maximization over $\bw$ (softmax updates) and over $\rmg_{\mathrm{joint}}$ (stochastic gradient step) yields the stated algorithm.
\end{proof}

\begin{proof}[Proof of \Cref{kriging-posterior}]
Conditional on $\beta$, the joint model for $\bz_{S_n'}$ and $\bx_{S_n}$ is
\[
\bz_{S_n'} \sim \cN\bigl(0,\Sigma_{\hat\theta}\bigr),
\qquad
\bx_{S_n} \sim \cN\bigl(\bA_{S_n}\beta+\bz_{S_n},D_\tau^{-1}\bigr).
\]
By Bayes’ rule,
\begin{align*}
\rmp\bigl(\bz_{S_n'} \mid \bx_{S_n},\beta\bigr)
&\propto
\exp\!\left(
   -\tfrac12 \bz_{S_n'}^\top \Sigma_{\hat\theta}^{-1}\bz_{S_n'}
   -\tfrac12 \bigl(\bx_{S_n}-\bA_{S_n}\beta-\bz_{S_n}\bigr)^\top
              D_\tau \bigl(\bx_{S_n}-\bA_{S_n}\beta-\bz_{S_n}\bigr)
\right) \\
&\propto
\exp\!\left(
   -\tfrac12 \bz_{S_n'}^\top Q_{\hat\theta}\bz_{S_n'}
   + \bz_{S_n'}^\top L
\right),
\end{align*}
where $Q_{\hat\theta}$ and $L$ are defined in the main text.
Hence the posterior is Gaussian with mean $Q_{\hat\theta}^{-1}L$ and covariance $Q_{\hat\theta}^{-1}$.
\end{proof}
\section{Additional Empirical Results} \label{sec:additional-simulation}
\subsection{Matrix Recovery Results}
Table~\ref{tab:simul_EBMR_full} reports the empirical Bayes matrix recovery results across all simulation settings described in \Cref{sec-sim-EBMR}.
\input{Tables/mse_table_full_2lines}

\subsection{Covariate-Informed Empirical Bayes with Model Misspecification} \label{sec:additional-covariates}

To assess the robustness of Bayesian EB methods, we compare covariate-informed empirical Bayes to competing methods under a heavy-tailed noise. For each $(n,p,T_0)$ design, we generate independent noise $\varepsilon_{ij}$ from a scaled Student--$t$ distribution with $\nu=5$ degrees of freedom:
\[
\varepsilon_{ij}\sim c_\nu\, t_\nu,\qquad  c_\nu=\frac{\sigma}{\sqrt{\nu/(\nu-2)}},
\]
where $c_\nu$ is chosen to ensure $\operatorname{Var}(\varepsilon_{ij})=\sigma^2$. Thus, the noise variance matches the target variance but the tail probabilities are heavier than Gaussian. 

\begin{figure}[t!]
\centering
\includegraphics[width=\textwidth]{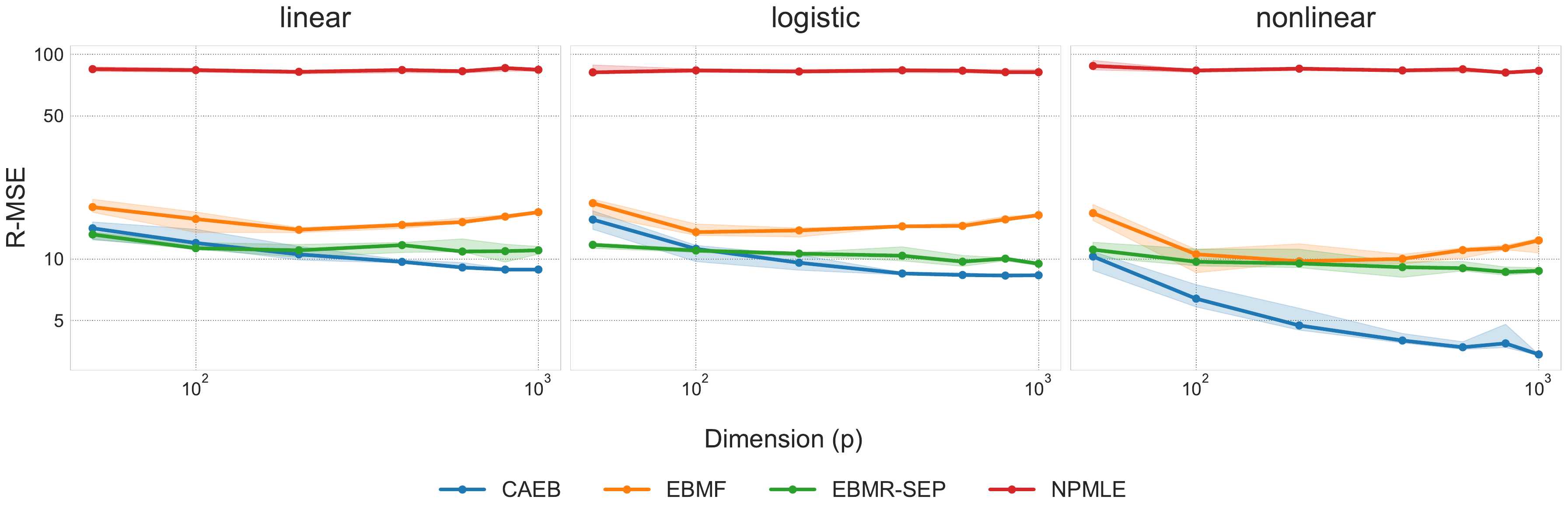}
\caption{\emph{CIEB under heavy-tailed noise.}
R-MSE of latent recovery versus dimension $p$ for three choices of $\rmg$: linear, logistic, and nonlinear.
Curves compare CIEB, EBMF, EBMR-sep, and NPMLE; points are medians and bands are IQR (25--75\%).
With noise following a $t_5$ distribution, CIEB attains the lowest error and continues to improve as the number of columns $p$ increases.}
\label{fig:app:caeb_t_noise}
\end{figure}

\Cref{fig:app:caeb_t_noise} plots RMSE of $\hat{\bz}_{n,p}$ versus $p$; points show medians across replicates and bands indicate interquartile ranges for each method.

\subsection{Additional Results for NYC Air Quality}
\label{sec:additional-air}

\begin{figure}[t!]
\centering
\begin{minipage}[b]{0.5\textwidth}
  \includegraphics[width=\linewidth]{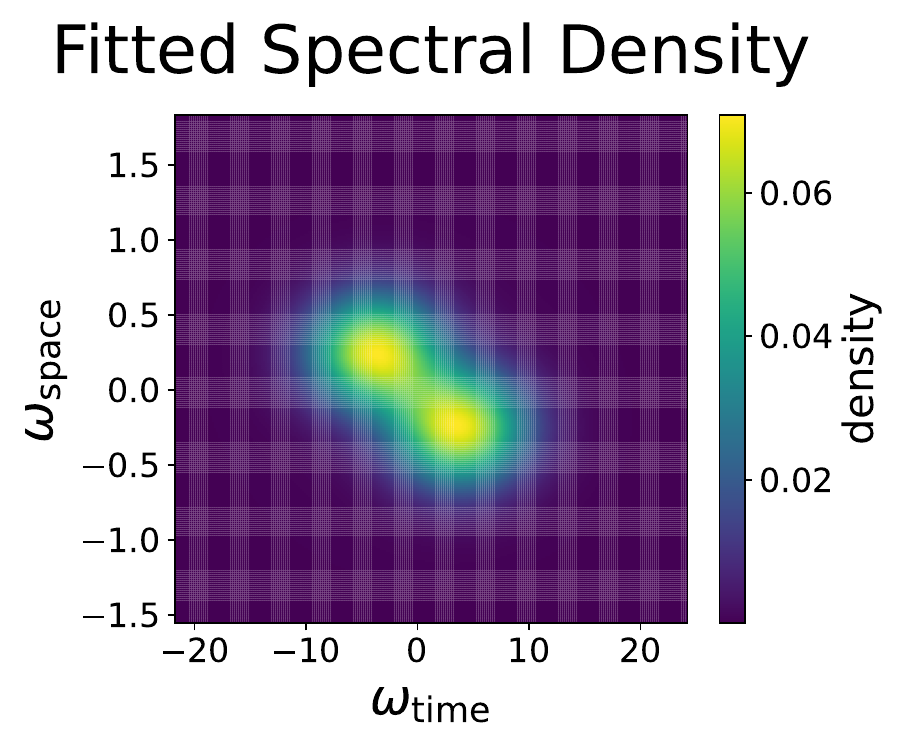}
\end{minipage}\hfill
\begin{minipage}[b]{0.4\textwidth}
  \captionof{figure}{\emph{Fitted spectral density for weekly $\mathrm{PM}_{2.5}$ in New York City, with $K = 10$.} The spectrum shows two dominant modes, one on the upper left corner and one on the lower right corner, which indicates two distinct spatio-temporal regimes. Based on the denoised series and site geography, one mode might correspond to the Manhattan sites (Herald Square, Chinatown, Lower East Side) and the other might correspond to northern Queens.}
  \label{fig:pm25_spectral}
\end{minipage}
\end{figure}

\begin{figure}[t!]
\centering
\includegraphics[width=1\textwidth]{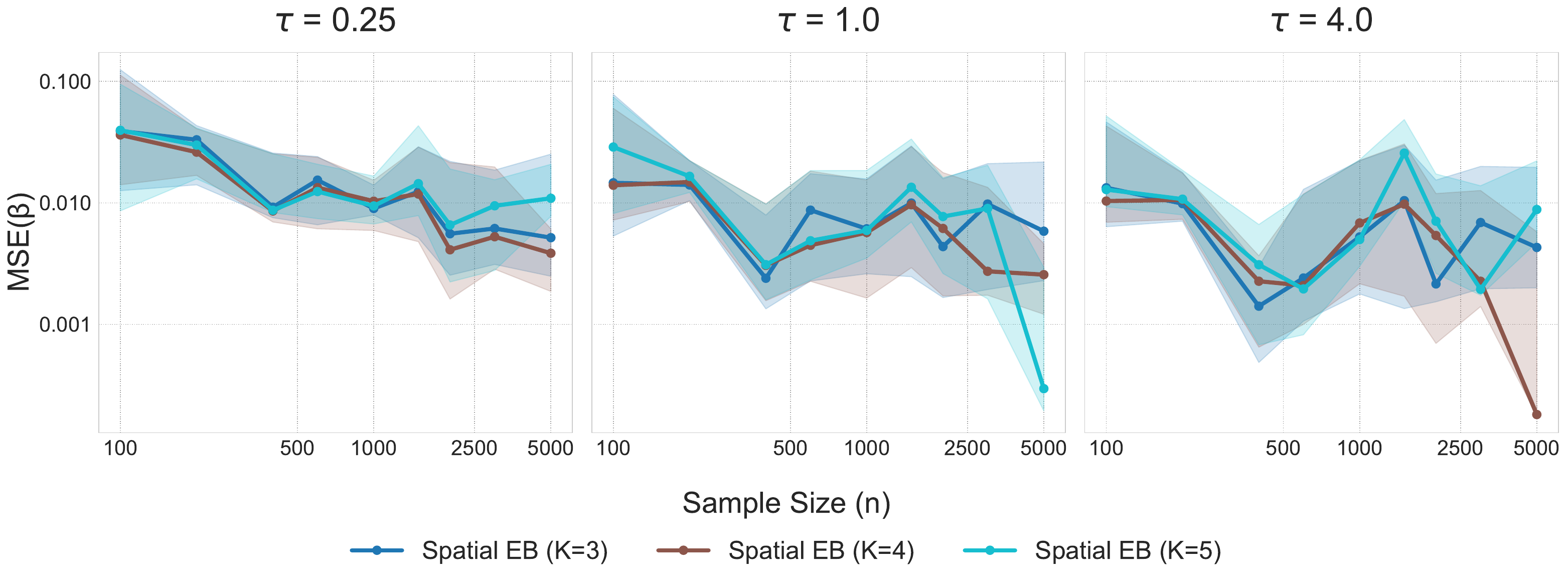}
\caption{\emph{Posterior mean estimation of $\beta^\star$ improves with increasing $n$.}
Mean squared error (MSE) for estimating $\beta^\star$ using the posterior mean under the EB spatial regression, across varying sample size $n$ and precision $\tau$.
We fit the spectral density using $K\in\{3,4,5\}$ Gaussian components; results are robust to moderate over-specification of $K$.
Both axes use logarithmic scales. The points denote medians over 10 replicates, while bands show interquartile ranges (25--75\%).}
\label{fig:simul_ebsr_mse_beta_n}
\end{figure}

\begin{figure}[t!]
\centering
\includegraphics[width=1\textwidth]{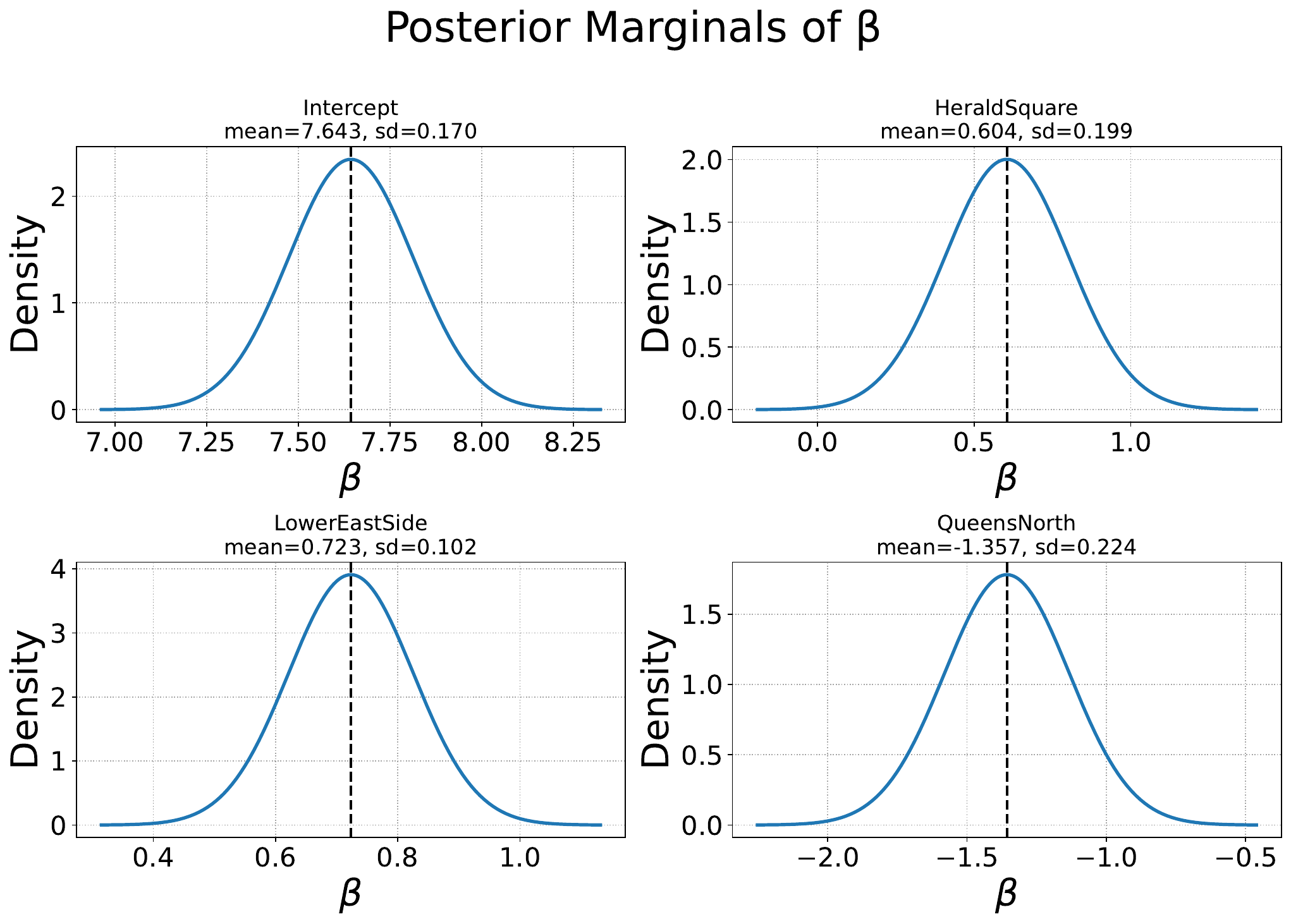}
\caption{\emph{Posterior distributions of location effects in EB spatial regression for NYC $\mathrm{PM}_{2.5}$.}
Baseline is Herald Square (the mean is $7.654 \,\mu\mathrm{g}/\mathrm{m}^3$).
The effect of measuring at Lower East Side is statistically indistinguishable from the baseline.
Queens North is lower by $1.357\,\mu\mathrm{g}/\mathrm{m}^3$ (posterior s.d.\ $0.224$), which provides strong evidence of better air quality there relative to Herald Square.}
\label{fig:airquality_beta}
\end{figure}




\end{document}

%% file: preamble/format.tex
\newif\ifmtpro
\newif\ifsimple
\newif\ifhipster

\mtprotrue

\ifmtpro
\usepackage[T1]{fontenc}
\usepackage[full]{textcomp}
\usepackage{newtxtext}
\usepackage{cabin} 
\usepackage[varqu,varl]{inconsolata} 
\usepackage[final,expansion=alltext]{microtype}
\usepackage[english]{babel}

\usepackage{amsmath}
\usepackage{amssymb}
\usepackage{mathptmx}
\usepackage{mathalpha}
\usepackage{bm}
\usepackage{fullpage}
\fi

\ifsimple
\usepackage[T1]{fontenc}
\usepackage{newtxtext}
\usepackage{bm}
\usepackage{amsthm} 
\usepackage{newtxmath}
\usepackage[bb=mt,cal=mt]{mathalpha}
\fi

\ifhipster
\usepackage{ccfonts}
\usepackage[T1]{fontenc}
\usepackage{textcomp}
\usepackage[final]{microtype}
\usepackage[english]{babel}

\usepackage[bb=mt,cal=mt]{mathalpha}
\usepackage{upgreek}

\usepackage{bm}

\fi

\usepackage{amsmath,amssymb,amsfonts,amsthm}
\usepackage{mathtools}
\usepackage{thmtools}
\usepackage{bm}

\usepackage[parfill]{parskip}
\usepackage{afterpage}
\usepackage{setspace}
\usepackage[utf8]{inputenc}

\apptocmd\normalsize{%
  \setlength\abovedisplayskip{2.5pt}
  \setlength\belowdisplayskip{2.5pt}
  \setlength\abovedisplayshortskip{2.5pt}
  \setlength\belowdisplayshortskip{2.5pt}
}{}{}
\usepackage{times}
\usepackage{paralist}
\usepackage[compact,tiny]{titlesec}
\titlespacing*{\section}{0pt}{0.5ex}{0.5ex}
\titlespacing*{\subsection}{0pt}{0.25ex}{0.25ex}
\titlespacing*{\subsubsection}{0pt}{0.25ex}{0.25ex}
\usepackage{fullpage}


%% file: preamble/preamble.tex
\usepackage[usenames,dvipsnames]{xcolor}

\usepackage{lineno}
\usepackage[colorinlistoftodos,
prependcaption,
textsize=footnotesize,
backgroundcolor=blue!10,
linecolor=black!10,
bordercolor=yellow!20]{todonotes}

\usepackage{graphicx}
\usepackage[labelfont=bf]{caption}
\usepackage[format=hang]{subcaption}

\usepackage[algoruled,ruled,vlined]{algorithm2e}

\usepackage{listings}
\usepackage{fancyvrb}
\fvset{fontsize=\normalsize}

\lstdefinestyle{mystyle}{
    commentstyle=\color{OliveGreen},
    keywordstyle=\color{BurntOrange},
    numberstyle=\tiny\color{black!60},
    stringstyle=\color{darkblue},
    basicstyle=\ttfamily,
    breakatwhitespace=false,
    breaklines=true,
    captionpos=b,
    keepspaces=true,
    numbers=left,
    numbersep=5pt,
    showspaces=false,
    showstringspaces=false,
    showtabs=false,
    tabsize=2
}
\usepackage{enumitem}

\usepackage{natbib}
\usepackage[colorlinks,linktoc=all]{hyperref}
\usepackage[all]{hypcap}
\hypersetup{citecolor=MidnightBlue}
\hypersetup{urlcolor=MidnightBlue}
\hypersetup{linkcolor=black}

\usepackage[nameinlink]{cleveref}
\creflabelformat{equation}{#1#2#3}
\crefname{equation}{eq.}{eqs.}
\Crefname{equation}{Eq.}{Eqs.}
\Crefname{section}{\S}{\S}
\Crefname{proposition}{Proposition}{Propositions}

\usepackage
[acronym,nowarn,section,nogroupskip,nonumberlist]{glossaries}
\glsdisablehyper{}
\setglossarystyle{long3col}
\makeglossaries

\usepackage{blindtext}
\usepackage{framed}

\usepackage{booktabs}
\usepackage{multirow}
\usepackage{longtable}
\usepackage{etoolbox,siunitx}
\robustify\bfseries
\colorlet{OLIVE}{olive}

\DeclareRobustCommand{\parhead}[1]{\textbf{#1}~}

\usepackage{tabularx}
\usepackage{booktabs}
\usepackage{multirow}
\usepackage{caption}
\usepackage{array}
\usepackage[most]{tcolorbox}
\usepackage{rotating}

%% file: preamble/math.tex
\newcommand{\dd}{\mathrm{d}}

\DeclareMathOperator*{\argmax}{arg\,max}

\newcommand{\g}{\,\vert\,}

\newcommand{\E}[1]{\mathbb{E}\left[#1\right]}
\newcommand{\EE}[2]{\mathbb{E}_{#1}\left[#2\right]}

\newcommand{\KL}[2]{D_{\textsc{KL}}\left( #1 \parallel #2\right)}

\newcommand{\rmp}{\mathrm{p}}

\newcommand{\LSE}{\operatorname{LSE}}

\newcommand{\unif}{\textrm{unif}}

\newcommand{\GP}{\mathrm{GP}}

\newcommand{\cA}{\mathcal{A}}

\newcommand{\cE}{\mathcal{E}}

\newcommand{\cG}{\mathcal{G}}

\newcommand{\cN}{\mathcal{N}}

\newcommand{\cP}{\mathcal{P}}
\newcommand{\cQ}{\mathcal{Q}}

\newcommand{\cX}{\mathcal{X}}
\newcommand{\cY}{\mathcal{Y}}
\newcommand{\cZ}{\mathcal{Z}}

\newcommand{\mbx}{\mathbf{x}}

\newcommand{\mbz}{\mathbf{z}}

\newcommand{\ba}{\bm{a}}

\newcommand{\bu}{\bm{u}}
\newcommand{\bv}{\bm{v}}
\newcommand{\bw}{\bm{w}}
\newcommand{\bx}{\bm{x}}
\newcommand{\by}{\bm{y}}
\newcommand{\bz}{\bm{z}}

\newcommand{\bA}{\bm{A}}

\newcommand{\ELBO}[1]{\operatorname{ELBO}\left(#1\right)}

\makeatother

\theoremstyle{plain}  \newtheorem{theorem}{\textbf{Theorem}}
\newtheorem{corollary}{\textbf{Corollary}} \newtheorem{assumption}{\textbf{Assumption}}

 \newtheorem{definition}{\textbf{Definition}}

\newtheorem{proposition}{\textbf{Proposition}}

\theoremstyle{definition}
\newtheorem{remark}{\textbf{Remark}}

\newcommand{\R}{\mathbb{R}}
\renewcommand{\SS}{\mathbb{S}}

\newcommand{\Z}{\mathbb{Z}}
\newcommand{\N}{\mathbb{N}}

\newcommand{\iid}{\stackrel{iid}{\sim}}

\newcommand{\deq}{:=}

\newcommand{\rmg}{\textrm{g}}
\newcommand{\rmq}{\textrm{q}}

\usepackage{cleveref}
\crefname{proposition}{Proposition}{Proposition}
\crefname{assumption}{Assumption}{Assumptions}
\crefname{equation}{eq.}{eqs.}
\Crefname{equation}{Eq.}{Eqs.}
\crefname{observation}{Observation}{Observation}
\Crefname{algocf}{Algorithm}{Algorithms}
\crefname{algocf}{algorithm}{algorithms}

\newcounter{problem}

\crefname{problem}{problem}{problems}
\Crefname{problem}{Problem}{Problems}

\newcommand{\sep}{\textrm{sep}}
\newcommand{\rel}{\textrm{rel}}

\newcommand{\MSE}{\textrm{MSE}}

\usepackage{tikz}
\usetikzlibrary{bayesnet,positioning,fit,arrows.meta,calc}


%% file: abstract.tex
\begin{abstract}
  Empirical Bayes (EB) improves the accuracy of simultaneous inference “by learning from the experience of others” \citep{Efron2012}. Classical EB theory focuses on latent variables that are iid draws from a fitted prior \citep{Efron2019OracleBayes}. Modern applications, however, feature complex structure, like arrays, spatial processes, or covariates. We propose a generalized approach to empirical Bayes based on \emph{probabilistic symmetry}. Our method pairs a simultaneous inference problem with an unknown prior to a symmetry assumption on the joint distribution of the latent variables. Each symmetry implies an ergodic decomposition, which we use to derive a corresponding empirical Bayes method. We call this method \emph{Bayesian empirical Bayes} (BEB). BEB recovers classical empirical Bayes methods, which implicitly assume exchangeability. We extend EB to other probabilistic symmetries: (i) EB matrix recovery for arrays and graphs; (ii) covariate-informed EB for conditional data; and (iii) EB spatial regression under shift invariance. We develop scalable algorithms based on variational inference and neural networks. In simulations, BEB outperforms existing denoising methods. On real data, we demonstrate BEB on cancer gene-expression and brain-connectivity matrices and NYC air-quality data.
\end{abstract}


%% file: introduction.tex
\section{Introduction}
\label{sec-intro}

Empirical Bayes (EB) is about \textit{simultaneous posterior inference}. We observe related data $x_1,\ldots,x_n$, where each data point is associated with a latent variable $z_i$, and assumed drawn from $x_i \sim \rmp(x \g z_i)$. The form of the conditional likelihood is known. The goal of EB is to use Bayesian inference to estimate all the $z_i$ together \citep{Robbins1956,Efron2015FreqBayes,Efron2019OracleBayes}.

In a traditional Bayesian analysis, we specify a prior $z_i \sim \rmg(z)$ and compute the posteriors $\rmp(z_i \g x_i)$. But in EB, we treat the prior $\rmg(z)$ as \textit{unknown} and we estimate it from the data. EB thus blends Bayesian and frequentist thinking: we reason in a frequentist way about the prior in order to form a collection of Bayesian estimates.

The classical approach to EB assumes $z_1,\ldots,z_n \iid \rmg$ and estimates $\hat{\rmg}$ so that the marginal distribution of $x$ maximizes the likelihood of the data. In the EB nomenclature, this is known as ``$\rmg$-modeling'' \citep{Efron2014}. Once the prior is fit, we calculate the posteriors $\rmp(z_i \g x_i, \hat{\rmg})$ to estimate the latent variables. Through the fitted prior, EB connects all $n$ inferences: information from each observation improves all the others.

The assumption $z_1,\ldots,z_n \iid g$ is central to EB thinking.  It says that the latent variables share a common prior, which justifies the EB procedure.  But many modern datasets violate this iid assumption.  Latent variables may come in rows and columns, they may lie on a spatial or temporal grid, or they may depend on known covariates. Such structure is the rule in modern applications. One may wish to denoise a gene-expression matrix indexed by genes and patients, smooth a spatial field of air-quality measurements across a city, or share information across related hypotheses that each carry covariates. In each case the goal is an accurate simultaneous estimate of the latent values, and the shared structure is what lets each unit borrow strength from the others. While these settings clearly violate the iid assumption, they still exhibit symmetries that tie the latent variables together.  In this paper, we develop a general way to use such probabilistic symmetries to extend EB to non-iid settings.

Stepping back, the assumption that $z_1,\ldots,z_n$ are iid from $\rmg$ is a special case of a broader principle: \textit{exchangeability}.  If the joint distribution of the latent variables is invariant to permutations, then---under mild conditions---it can be viewed as part of an infinitely exchangeable sequence.  The iid structure then arises through de~Finetti’s famous representation theorem \citep{definetti1929}. It states that any exchangeable sequence can be written as a mixture of iid laws,
\begin{equation}
      \rmp(z_1,\ldots,z_n)
  =\int \Bigl[\prod_{i=1}^n \rmp(z_i \mid \rmg)\Bigr]\, \dd \mu(\rmg),
  \label{eq:definetti}
\end{equation}
where $\rmg$ is a probability distribution on $\cZ$ and $\mu$ is a
distribution over such distributions.  De~Finetti's result is
important because it connects a symmetry assumption---namely, exchangeability---to classical Bayesian reasoning \citep{Savage1954,Robert2007Bayesianchoice}. If the $z_i$'s are exchangeable, then the prior $g$ appears as the latent variable that renders the $z_i$ conditionally independent.

Our insight is that empirical Bayes uses the same hierarchical structure as in \Cref{eq:definetti}, where $z_i$ are conditionally iid given $g$.  However, in place of integration with respect to $\mu$, EB substitutes a fitted value $\hat \rmg$ obtained from data.  Seen through de~Finetti’s theorem, we can view EB as a Bayesian inference under exchangeability, but with the mixing distribution collapsed to a point estimate.

Using this perspective, we extend classical EB by relaxing its exchangeability assumption to other probabilistic symmetries.  Latent variables often have structure, such as being organized in an array or indexed by space or time.  Such settings are not fully exchangeable, but they may satisfy weaker invariances.  An array may be invariant to permutations of its rows and columns; a spatial process may be invariant to shifts in location \citep{Kallenberg2005book}.  Each symmetry has an analogue of de~Finetti’s representation theorem.  We will use these representation theorems to extend EB to non-iid settings.

We call our approach \textit{Bayesian empirical Bayes} (BEB). The name reflects that BEB views empirical Bayes through a hierarchical Bayesian lens: the unknown prior is encoded by the invariant variable $\rmg$ in the ergodic decomposition implied by the assumed symmetry, so that estimating the prior amounts to inference about this global latent variable. The probabilistic symmetry is the device that makes $\rmg$ well defined. It is not itself a Bayesian assumption. We begin with a dataset $\mbx$ and an assumed probabilistic symmetry.  We posit that each observation arises from a known likelihood conditioned on a local latent variable $\rmp(x_i \g z_i)$, and our goal is to infer all $z_i$ simultaneously.  The key idea is that our assumed symmetry implies an \textit{ergodic decomposition} of the prior, i.e., a generalization of de~Finetti’s theorem \citep{Orbanz2015}.  Specifically, if the latent variables are invariant under a group $\Phi$ acting on its index set, then their joint distribution admits the form
\begin{equation}
      \rmp(\bz)
  = \int \rmp(\bz \mid \rmg)\, \dd \mu(\rmg),
\end{equation}
where $\rmg$ indexes the $\Phi$-ergodic family.  The chosen symmetry determines the form of $\rmg$ and the conditional distribution $\rmp(\bz \g \rmg)$.

Following the EB recipe, we fit $g$ by maximum marginal likelihood under the expanded model $\rmp(\bz,\bx)$,
\begin{equation}
      \hat{\rmg} \in \argmax_{\rmg} \ \log \int \rmp(\bx \g \bz)\, \rmp(\bz \g \rmg)\, \dd \bz.
\end{equation}
Finally, we use $\hat g$ to compute the empirical Bayes posteriors $\rmp(z_i \g x_i, \hat{\rmg})$. In practice, we can use approximate posterior inference both to fit $\hat{\rmg}$ and estimate $\rmp(z_i \g x_i, \hat \rmg)$.

This construction continues the program of \citet{Lindley1972} and \citet{Deely1981}. Treating the prior as the target of inference goes back to \citet{Lindley1972}, who analyzed the linear model by placing priors on the parameters of the prior and estimating them from data, and \citet{Deely1981} introduced the name ``Bayes empirical Bayes'' for placing a prior on the unknown prior and targeting its posterior. Bayesian empirical Bayes expands this idea from a parametric prior to the nonparametric invariant variable $\rmg$, and from exchangeable sequences to general probabilistic symmetries. The motivating thread is de~Finetti's motivation for Bayesian inference: exchangeability induces the prior through de~Finetti's theorem, and a general symmetry induces it through the ergodic decomposition, so $\rmg$ plays the role of the prior that we wish to infer. 

We illustrate BEB in several data analysis settings.  For array data, the relevant symmetry is separate or joint exchangeability, and the ergodic decomposition is given by the Aldous-Hoover theorem~\citep{Aldous1985}.  The invariant variable $\rmg$ is a measurable function on $[0,1]^3$, which we parameterize with a neural network. For spatial data, the relevant symmetry is shift invariance, and we can flexibly model $\rmg$ with stationary ergodic Gaussian processes. In both settings, we can additionally condition on covariates and derive covariate-dependent EB procedures.

Thus, Bayesian empirical Bayes produces a generalization of EB to any probabilistic symmetry. The invariant variable $\rmg$ replaces the classical prior, and the ergodic decomposition supplies the structure needed for inference. We treat the assumed symmetry as a modeling choice rather than a required property of the data-generating process. Even when the prior is not exactly invariant, positing a symmetry restricts inference to decision rules that are equivariant under it, which can be sensible in its own right \citep{Weinstein2021permutation}. We make this view precise in \Cref{sec-method}.

The rest of the paper is organized as follows. \Cref{sec-method} introduces the modeling assumptions and the general methodology of Bayesian empirical Bayes. \Cref{sec-EB-models} develops new inference methods from Bayesian EB principles in three settings: the classical normal sequence model, empirical Bayes matrix recovery, and empirical Bayes spatial regression. For normal means, we show that Bayesian EB with an exchangeable sequence model recovers the nonparametric maximum-likelihood method of \citet{Efron2015deconvolution}. For the latter two, we derive new EB procedures based on alternative invariances, namely joint, separate, and relative exchangeability for matrices and stationarity for spatio-temporal data. \Cref{sec-simulation} evaluates these methods in simulation studies. \Cref{sec-data} applies them to real datasets. \Cref{sec-discussion} concludes with discussion and future directions. Detailed algorithms for our inference procedures and all technical proofs appear in the Appendix.


%% file: related_work.tex
\subsection{Related Work}
\label{sec-related-works}
Empirical Bayes (EB) dates back to the seminal works of \citet{Robbins1951compound,Robbins1956} and has been developed in both parametric and nonparametric forms \citep{Efron1971,Efron1976,Efron2001,Brown2009,Jiang2009,Efron2011,Efron2012,Efron2014,Efron2015deconvolution,Efron2021,Soloff2021,Ignatiadis2022,Yang2024}. Much of the nonparametric theory centers on normal and Poisson sequence models. For normal means, \citet{Efron2015FreqBayes} surveys frequentist assessments of EB estimators and \citet{Efron2019OracleBayes} surveys oracle-Bayes viewpoints. Beyond $\rmg$-modeling (modeling the prior), \citet{Efron2014} described an alternative EB approach via $f$-modeling, which models the marginal data distribution; a key tool is Tweedie’s formula, linking the posterior mean to the derivative of the log marginal density. For the $f$-modeling methods, \cite{Saremi2019} obtains nonparametric EB estimators using flexible score functions, by minimizing the score-matching loss. Although \citet{ghosh2025stein} also minimize a score-matching objective, their criterion targets the prior rather than the marginal density, and is therefore more aligned with $\rmg$-modeling. EB also underpins large-scale multiple testing \citep{Efron2012,Stephens2017,Ignatiadis2021}. 
While it is impossible to exhaust the works on empirical Bayes for the normal sequence model, we refer to \citet{Ignatiadis2025EB} for a recent textbook overview. In this work we focus on $\rmg$-modeling and extend it beyond the iid \ prior setting to accommodate general invariant latent structures.

Nonparametric EB methods for non-sequence models have only appeared more recently. In high-dimensional linear models and variable selection, variational EB methods---such as SuSiE and related procedures---learn a nonparametric prior over coefficients \citep{Carbonetto2012,Wang2020susie,Kim2022}. The prior is exchangeable with respect to covariates \citep{trippe2021high}. The theoretical properties of these methods are beginning to be understood \citep{Mukherjee2023,Fan2023}.  For low-rank matrices, empirical Bayes matrix factorization and EB-PCA fit learned shrinkage priors for loadings and scores via approximate inference \citep{Wang2021EBMF,Zhong2022,denault2025covariate}. These new developments share the theme of fitting shared priors for latent variables to structure the borrowing of information.  Inspired by this line of work, we provide a unified framework for inferring many latent variables by learning a shared latent structure defined by probabilistic symmetries. This invariance-based perspective enables systematic extensions of EB methods. These structured settings also connect to broad literatures on modeling array and network data and on spatial and temporal processes, which provide natural comparison methods for the symmetry-based EB. We draw on these in our empirical studies. Empirical Bayes has likewise been extended beyond exchangeable sequences in other directions, for example through compound-decision approaches to covariance matrix estimation \citep{XinZhao2023}.

From a Bayesian perspective, EB concerns approximating a full Bayesian posterior by a point estimate of the hyperparameter (Ch.~5.6.4 of \citep{Bernardo1994}) and is closely connected to hierarchical Bayes \citep{Lindley1972,good1992introduction}. \citet{Deely1981} compare the statistical performance of parametric empirical Bayes and hierarchical Bayes on simultaneous inference problems. One can place additional layers of priors on EB hyperparameters; for example, \citet{Kass1989EB} introduce a hyperprior to capture uncertainty in EB estimation of prior hyperparameters in general hierarchical models. \cite{Rousseau2017} provide sufficient conditions for general consistency and posterior contraction rates of the EB posterior under parametric maximum marginal likelihood; in their setting, the results show that hierarchical and empirical Bayes posteriors obtain comparable posterior contraction rates. \citet{Rizzelli2024} analyze the asymptotic behavior of posteriors obtained after fitting such hyperparameters. \cite{Buta2011} and \cite{Doss2024} discussed the connection between empirical Bayes and Bayesian sensitivity analysis. While the Bayesian literature on EB has focused almost exclusively on parametric models, Bayesian empirical Bayes provides a principled approach to nonparametric empirical Bayes, motivated from Bayesian reasoning.

Many objects in classical statistics obey probabilistic symmetries (see \citet{Diaconis1988Book}; see also Chapter~3 of \citealp{Lehmann2006}). The notion of \emph{exchangeability}, for example, serves as the foundation for Bayesian theory \citep{definetti1929,Savage1954,Bernardo1994} and underlies much of practical Bayesian modeling \citep{Gelman1995}. Other applications of group symmetries include invariant testing---e.g., permutation-based testing methods \citep{eden1933,fisher1935,pitman1937,pesarin2001,ernst2004,pesarin2010,pesarin2012,good2006,anderson2001,kennedy1995,hemerik2018,ramdas2023}. Early advances for estimation and testing under general groups include \citet{lehmannstein1949} and \citet{hoeffding1952}; the celebrated Hunt-Stein theorem states conditions under which a minimax invariant test exists in hypothesis testing \citep{Lehmann1959}. Broader treatments appear in \citet{eaton1989,wijsman1990,giri1996}. In modern probabilistic modeling,  the ergodic decomposition theorem provides a foundation for building invariant representations and modeling \citep{Orbanz2015}. For example, \citet{Lloyd2012} study exchangeable arrays by placing a Gaussian process prior on the underlying random function. We build on the general theory of probabilistic symmetries as developed in \cite{Kallenberg2005book} and \cite{Orbanz2015}. The theory characterizes general invariant distributions under group actions using ideas from ergodic theory. It allows us to extend empirical Bayes to many different kinds of structured data settings, creating a foundation for new applications of EB.

%% file: method.tex
\section{Bayesian Empirical Bayes}
\label{sec-method}

Consider a \textit{structured dataset} $\bx = (x_\omega)_{\omega \in \Omega}$, where each $x_{\omega}$ is indexed by an element in an index set $\Omega$.  For example, if $\Omega = \N$, then $\bx$ is an infinite sequence; if $\Omega = \N^2$, then $\bx$ is an infinite array. (After describing a model of infinite data, we will reduce it to one of finite data.)



We posit that the data $\bx$ are noisy measurements of a matching set of \textit{structured latent variables} $\bz = (z_\omega)_{\omega \in \Omega}$.  In the likelihood, each $x_\omega$ is conditionally independent given its local latent $z_\omega$,
\begin{equation}
      \label{eq:likelihood}
  \rmp(\bx  \g  \bz) = \prod_{\omega \in \Omega} \rmp(x_\omega  \g  z_\omega).
\end{equation}
We assume the form of the noise distribution $\rmp(x_\omega  \g  z_\omega)$ is known. We write the joint likelihood as a product of likelihoods for concreteness. Each likelihood may depend on the index $\omega$, for instance through known, index-specific noise levels or covariances, and the general construction below applies to any likelihood $\rmp(\bx \g \bz)$. Only specific downstream results, such as the marginal likelihood in \Cref{prop:finite-EB} and the observation model in the hierarchical model~\eqref{finite-data-model}, rely on the product form in \Cref{eq:likelihood}. 

We place a prior distribution on the latent variables $\bz \sim \rmp(\bz)$. The joint distribution of $\bz$ and $\bx$ is now
\begin{equation}
      \label{generic model}
  \rmp(\bz, \bx) =
  \rmp(\bz) \prod_{\omega \in \Omega}
  \rmp(x_\omega  \g  z_\omega).
\end{equation}
Note, we do not assume independence in the prior. Throughout, we write $\rmp$ for a distribution and, when it exists, its density. The intended reference is clear from context.

Our goal is \emph{simultaneous probabilistic inference} of all local latent variables across $\Omega$. That is, we want to calculate the posterior $\rmp(\bz \g \bx)$. From this posterior, we can calculate summary statistics, such as the posterior means $\E{\bz  \g  \bx}$, or other quantities of interest, such as credible intervals, quantiles, posterior predictive distributions, or event probabilities. In short, we can denoise the data.

We now make the \textit{empirical Bayes assumption}: there exists a prior $\rmp^\star(\bz)$ such that the true data-generating distribution of $\bx$ is expressed by marginalizing over $\bz$.
\begin{assumption}[Empirical Bayes]
  \label{assum:true-prior}
  There exists a prior $\,\rmp^\star(\bz)$ such that the true data-generating distribution can be written as
\begin{equation*}
        \rmp^\star(\bx)
    = \int_{\cX^\Omega}
    \rmp^\star(\bz)
    \prod_{\omega \in \Omega}
    \rmp(x_\omega  \g  z_\omega)
    \, \dd \bz.
\end{equation*}
\end{assumption}
Of course, we do not know this true prior ahead of time - we just assume it exists.
We call this the empirical Bayes assumption because it links the Bayesian model to the true data-generating process.
  
\subsection{Priors with Probabilistic Symmetries}
\label{sec-prior}
How do we set the prior $\rmp(\bz)$? We will use the idea of a \textit{probabilistic symmetry}.  A symmetry describes how we can relabel or transform the indices of structured random variables without changing their joint distribution.

Formally, let $\upPhi$ denote a group of transformations acting bijectively on the index set $\Omega$, where each element $\phi \in \upPhi$ reindexes the coordinates of $\Omega$.   The action of $\phi$ on $\bz$ is the \emph{index transformation}
\begin{equation}
      (\phi(\bz))_\omega = z_{\phi^{-1}(\omega)}, \quad \omega \in \Omega.
\end{equation}
As examples, if $\Omega = \N$, then $\phi(\bz)$ reorders the elements of a sequence.  If $\Omega = \N^2$, then $\phi(\bz)$ could permute rows and columns of a matrix.  If $\Omega = \R^2$, then $\phi(\bz)$ could implement a spatial shift.

A distribution $\rmp(\bz)$ is \emph{$\upPhi$-invariant} if it does not change under these transformations. For example, an exchangeable sequence is invariant under permutations; a stationary process is invariant under spatial shifts.
\begin{definition}[$\upPhi$-invariance] \label{def-structured-var} A joint distribution $\rmp(\bz)$ of structured variables $\bz \in \cZ^\Omega$ is \emph{$\upPhi$-invariant} if
\begin{equation}
        \phi(\bz) \overset{d}{=} \bz \quad \text{for all } \phi \in \upPhi,
\end{equation}
  where equality is in distribution for all finite-dimensional marginals.
\end{definition}
A formal consequence of $\upPhi$-invariance is the \emph{ergodic decomposition theorem}, which states that any $\upPhi$-invariant distribution can be expressed as a mixture of \textit{ergodic distributions}. 
\begin{theorem}[Ergodic decomposition; \cite{Varadarajan1963}]
\label{ergodic-rep-thm}
Let $\upPhi$ be a nice group acting on $\cZ^\Omega$.  
There exists a family of distributions
\begin{equation}
      \cE \deq \left\{ \rmp(\bz  \g  \rmg) : \rmg \in \cG \right\} \subseteq \cP(\cZ^\Omega)
\end{equation}
such that any $\upPhi$-invariant distribution $\rmp(\bz)$ admits the unique decomposition
\begin{equation}
      \rmp(\bz) = \int_{\cG} \rmp(\bz  \g  \rmg)\, \dd \mu(\rmg),
\end{equation}
for some distribution $\mu$ on $\cG$.  The set $\cE$ is called the \emph{$\upPhi$-ergodic family} and its index $\rmg$ is called the \textit{invariant variable}.\footnote{It is also known as the directing random measure \citep{Kallenberg2005book}.}
\end{theorem}
A ``nice'' group is a technical condition; see \Cref{sec-group} of the Appendix for a formal definition. The $\upPhi$-ergodic family is determined by the group $\upPhi$ and forms the set of extreme points of the convex set of $\upPhi$-invariant distributions on $\cZ^\Omega$ \citep[Theorem~25.24]{Kallenberg1997foundations}. Once $\upPhi$ is fixed, so too is the form of $\rmp(\bz  \g  \rmg)$.

The ergodic decomposition theorem generalizes de Finetti’s representation \citep{definetti1929} from exchangeable sequences to arbitrary group symmetries. Under exchangeability, the ergodic measures are precisely the i.i.d.\ product distributions.




The ergodic decomposition describes the structure of any $\upPhi$-invariant distribution. We now apply this idea to the true prior. 
\begin{assumption}[True $\upPhi$-invariant prior]
  \label{assum:phi-prior}
  The true prior $\rmp^\star(\bz)$ is $\upPhi$-invariant.
\end{assumption}
By \Cref{ergodic-rep-thm}, there exists an invariant variable $\rmg \in \cG$ with distribution $\rmg \sim \mu^\star$ such that $\rmp^\star(\bz)$ is a $\mu^\star$-mixture of the $\upPhi$-ergodic family,
\begin{equation}
    \label{invariance}
\rmp^\star (\bz)
=
\int_{\cG}
\rmp (\bz \g \rmg)
\, \dd \mu^\star(\rmg).
\end{equation}
Together, \Cref{assum:true-prior,assum:phi-prior} state that among all models of the form in \Cref{generic model}, there exists a correctly specified version whose prior is $\upPhi$-invariant. In the Bayesian empirical Bayes model, the latent variables $\bz$ capture the structured, invariant part of the world, and the observations $\bx$ are noisy realizations through the shared likelihood $\rmp(x_\omega \g  z_\omega)$. 

An important consequence of Assumption~\ref{assum:phi-prior} is that the data-generating distribution $\rmp^\star(\bx)$ inherits the same symmetry as the prior. To see this, we first note that applying any transformation $\phi \in \upPhi$ to both $\bx$ and $\bz$ leaves the joint distribution unchanged,
\begin{equation}
\begin{aligned}
  \rmp^\star(\phi(\bx), \phi(\bz))
  &= \rmp^\star(\bz)
     \prod_{\omega\in\Omega}
     \rmp(x_{\phi^{-1}(\omega)}  \g  z_{\phi^{-1}(\omega)}) \\
  &= \rmp^\star(\bz)
     \prod_{\omega\in\Omega}
     \rmp(x_\omega  \g  z_\omega)
   = \rmp^\star(\bx,\bz).
\end{aligned}
\end{equation}
Then we marginalize out $\bz$ to give $\rmp^\star(\phi(\bx))=\rmp^\star(\bx)$.  

Alternatively, instead of assuming the true prior is invariant, we could assume that the marginal distribution of the data is invariant. Then, under mild regularity assumptions on the likelihood, the $\Phi$-invariance of the population distribution $\rmp^\star(\bx)$ also implies the $\Phi$-invariance of $\rmp^\star(\bz)$. We give a formal statement and proof in \Cref{prop:marginal-to-prior-invariance}.

\begin{remark}[Symmetry as a modeling choice]
Assumptions~\ref{assum:true-prior} and~\ref{assum:phi-prior} posit a true, $\upPhi$-invariant prior. An equivalent reading is decision-theoretic: positing a symmetry $\upPhi$ restricts decision rules to those that are equivariant under $\upPhi$, meaning that transforming the data by any group element $\phi\in\upPhi$ transforms the estimate by the same $\phi$, and BEB is a statistical method to construct such rules \citep{Weinstein2021permutation}. When the prior is $\upPhi$-invariant, the resulting rule is Bayes optimal. Otherwise the symmetry acts as a modeling constraint that trades flexibility for statistical efficiency, with weaker symmetries enlarging the admissible class of priors and rules. Several of our examples, including the MNIST digit example, can be understood as cases where the symmetry is imposed to obtain an equivariant estimator.
\end{remark}

\subsection{Examples of $\upPhi$-invariant priors}
\label{sec-invariant-priors}

We present three examples of $\upPhi$-invariant priors, which we will use throughout this paper. In each, we specify the index set $\Omega$, the symmetry group $\upPhi$, the form of the ergodic decomposition, and the corresponding invariant variable $\rmg$.


\parhead{Exchangeable sequences.} Let $\Omega = \N$, so that $\bz=(z_1,z_2,\ldots)$ is a sequence of random variables. The symmetry group is $\upPhi=\SS$, the group of permutations of $\N$. The $\upPhi$-invariant distributions on $\cZ^\N$ are the \textit{exchangeable sequences}.

In this case, the ergodic decomposition recovers de~Finetti’s representation theorem \citep{definetti1929},
\begin{equation}\label{de-Finetti}
  \rmp(\mbz)=
\int_{\cG_{{\rm perm}}}
\prod_{\omega \in \Omega}
\rmg_{\rm perm}(z_\omega) \,
\dd \mu(\rmg_{\rm perm}).
\end{equation}
The invariant variable $\rmg_{\mathrm{perm}}$ is a random probability measure on $\cZ$, and $\mu$ is a distribution over such random measures.
  
\parhead{Exchangeable arrays.} Let $\Omega=\N^2$, so that $\bz=(z_{ij})_{i,j\in\N}$ is an array of random variables. The array is \emph{separately exchangeable} if its distribution is invariant under the independent permutations of rows and columns. That is, the symmetry group is $\upPhi=\SS^2$, and
\begin{equation}
        (z_{ij}) \overset{d}{=} (z_{\pi(i),\pi'(j)}) \quad
    \text{for all } (\pi,\pi') \in \upPhi.
\end{equation}
  Under separate exchangeability, the Aldous--Hoover theorem provides the ergodic decomposition \citep{Aldous1981jointy,Aldous1985}.
  \begin{theorem}[Aldous--Hoover representation for separate exchangeability]\label{thm:aldous-hoover}
  A random array $\bz=(z_{ij})_{i,j\in\N}$ is separately exchangeable if and only if its distribution can be written as
\begin{equation}
        \rmp(\bz) =
    \int_{\cG_{\textrm{sep}}}
    \rmp(\bz \g \rmg_{\textrm{sep}}) \,
    \dd \mu(\rmg_{\textrm{sep}}).
\end{equation}
  The invariant variable is a random function
\begin{equation} \label{eq:array_g}
 \rmg_{\mathrm{sep}}:[0,1]^3\to\cZ
\end{equation}
  and the ergodic distribution $\rmp(\bz \g \rmg)$ is defined through uniform random variables
\begin{equation*}
  u_i, v_j, u_{ij}
\iid \unif(0,1)
\quad
z_{ij}
=
\rmg_{\mathrm{sep}}(u_i, v_j, u_{ij}) \quad \text{for all } i,j\in\N.
\end{equation*}
\end{theorem}
Through the shared row and column variables $(u_i)_{i\in\N}$ and $(v_j)_{j\in\N}$, the entries $z_{ij}$ exhibit dependence within rows and within columns.

Variants of this representation capture other symmetry assumptions. If the array is \textit{jointly exchangeable}---invariant when rows and columns are permuted together---then the same decomposition holds with
\begin{equation*}
      u_i, u_{ij} \iid \mathrm{Unif}(0,1) \quad
  z_{ij} = \rmg_{\mathrm{joint}}(u_i, u_j, u_{ij}) \quad
  \textit{for all } i, j \in \N.
\end{equation*}
In a binary array with joint exchangeability, i.e., a random graph, the Aldous--Hoover representation yields
\begin{equation*}
      z_{ij}=\mathbf{1}\{u_{ij}\le \rmg_{\mathrm{graph}}(u_i,u_j)\},
\end{equation*}
for a random symmetric graphon $\rmg_{\mathrm{graph}}:[0,1]^2\to[0,1]$. In \Cref{sec-caeb} of the Appendix, we discuss array exchangeability conditional on row- and column-specific covariates.  \citet{Orbanz2015} extends all these representations to $k$-arrays, where $\Omega=\N^k$.

\parhead{Stationary processes.} We now consider variables in time or space.  Let $\Omega=\R^d$ (continuous space) or $\Z^d$ (lattice), and $\cZ=\R$.  
A distribution $\rmp(\bz)$ on $\cZ^\Omega$ is \emph{stationary} if it is invariant under spatial shifts
$\upPhi_{\mathrm{shift}}=\{\phi_h:\omega\mapsto\omega+h\}$ with $h\in \Omega$:
\begin{equation*}
      \rmp(\phi_h^{-1}(A))=\rmp(A)
  \quad\text{for all measurable }A\subseteq\cZ^\Omega,\ h\in\Omega.
\end{equation*}
Each such $\rmp(\bz)$ is a \emph{stationary probability measure}. Here $\rmp(\bz)$ is the distribution of the random field on the path space $\cZ^\Omega$,where the shift group acts by reindexing the coordinates. Such shift-invariant laws are exactly the stationary processes (for example, i.i.d.\ noise or a stationary Gaussian process). This should not be confused with a shift-invariant measure on the state space $\cZ$ alone, which in general is not a probability measure. Among these, a measure $\rmg_{\mathrm{erg}}$ is called \emph{ergodic} if it cannot be written as a nontrivial mixture of other stationary measures. Equivalently, it is ergodic if $\rmg_{\mathrm{erg}}(A)\in\{0,1\}$ for every shift-invariant event $A$.

By the ergodic decomposition theorem \citep[Theorem~I.4.10]{Shields1996}, any stationary measure can be written uniquely as a mixture of ergodic stationary measures,
\begin{equation}
      \label{eq:spatial_decomp}
  \rmp(\bz)
  = \int_{\cG} \rmg_{\mathrm{erg}}(\bz)\, \dd \mu(\rmg_{\mathrm{erg}}),
\end{equation}
where $\cG$ is the space of all ergodic stationary measures on $\cZ^\Omega$.  

Compared with exchangeable models, stationary processes impose a weaker symmetry: the discrete shift group grows only polynomially with system size, while the permutation group grows super-exponentially \citep{Lubotzky2003}.  As a result, representation theorems for stationarity are less restrictive, but also less tractable.

\subsection{The finite population model}

We have defined our model in terms of the entire set $\upOmega$. But in practice we only observe finite data. We now specialize our population model to a finite domain.

Let $S_n \subseteq \Omega$ be a finite subset, with $S_1 \subseteq S_2 \subseteq \cdots$ forming a nested sequence that exhausts $\Omega$.  We observe finite data $\bx_n = (x_\omega)_{\omega \in S_n}$, each associated with a local latent variable $z_\omega$. We write $\bz_n = (z_\omega)_{\omega \in S_n}$. Intuitively, this means we observe a finite subset of the full population $\bx$ at selected index locations, i.e. certain entries of an array or certain locations in time or space).

We first state the finite-data analog of the empirical Bayes assumption, that there exists a true prior that captures the population distribution.

\begin{proposition}[Finite empirical Bayes model]
\label{prop:finite-EB}
Under \Cref{assum:true-prior}, the marginal distribution of the finite data $\bx_n$ can be written as
\begin{equation*}
      \rmp^\star(\bx_n)
  = \int_{\cZ^{S_n}}
  \rmp^\star(\bz_n)
  \prod_{\omega\in S_n}\rmp(x_\omega \g  z_\omega)\, d\bz_n,
\end{equation*}
for some prior $\rmp^\star(\bz_n)$ on the latent variables indexed by $S_n$.
\end{proposition}
We next connect this finite prior to the group symmetry and its ergodic decomposition.

\begin{proposition}[Finite $\upPhi$-invariant prior]
\label{prop:finite-invariant}
Suppose the population prior $\rmp^\star(\bz)$ is $\upPhi$-invariant as in \Cref{assum:phi-prior}. Then its marginal $\rmp^\star(\bz_n)$ also admits an ergodic representation,
\begin{equation*}
      \rmp^\star(\bz_n)
  = \int_{\cG} \rmp(\bz_n \g \rmg)\, d\mu^\star(\rmg),
\end{equation*}
where $\rmp(\bz_n \g \rmg)
= \int_{\cZ^{\Omega\setminus S_n}} \rmp(\bz \g \rmg)\, d\bz_{\Omega\setminus S_n}$ is the marginal of ergodic distribution $\rmp(\bz \g \rmg)$ on $S_n$.
\end{proposition}
Consider a dataset $\bx_n$ and assume it was drawn from $\bz_n$ with a $\upPhi$-symmetric prior. As a consequence of \Cref{prop:finite-EB,prop:finite-invariant}, its generative process is a hierarchical model:
\begin{equation}  \label{finite-data-model}
\begin{aligned}
  \rmg &\sim \mu^\star \\
  \bz_n &\sim \rmp(\bz_n  \g  \rmg) \\
  x_\omega &\sim \rmp(x \g z_\omega) \quad \omega \in S_n.    
\end{aligned}  
\end{equation}
The latent variables are $\bz_n$ and $\rmg$, and the observations are $\bx_n$. The only unknown factor is the prior $\mu^\star$, because the likelihood $\rmp(x \g z)$ is fixed and the form of $\rmp(\bz_n \g \rmg)$ is known and determined by the probabilistic symmetry.

\subsection{Bayesian empirical Bayes}

We are given a dataset $\bx_n$ and assume a probabilistic symmetry on its prior $\rmp(\bz_n)$. Consequently, our data comes from the model in \Cref{finite-data-model}, with the form of the invariant variable $\rmg$ and ergodic distribution $\rmp(\bz_n \g \rmg)$ determined by the assumed symmetry. Our goal is to compute the posterior $\rmp^\star(\bz_n \g \bx_n)$.

We first write the posterior as an integral
\begin{equation}
      \label{ideal posterior}
  \rmp^\star(\bz_n  \g  \bx_n)
  = \int_{\cG} \rmp(\bz_n \g \rmg, \bx_n)\, \rmp^\star(\rmg \g \bx_n)\, \dd \rmg. 
\end{equation}
Note the conditional distribution $\rmp(\bz_n  \g  \rmg, \bx_n)$ is proportional to the product of the likelihood $\rmp(\bx_n  \g  \bz_n)$ and the ergodic distribution $\rmp(\bz_n  \g  \rmg)$:
\begin{equation} \label{conditional posterior}
    \rmp(\bz_n  \g  \rmg, \bx_n)
    \propto \rmp(\bx_n  \g  \bz_n)\, \rmp(\bz_n  \g  \rmg),
\end{equation}
where both these terms are known. The only unknown component in
\Cref{ideal posterior} is $\rmp^\star(\rmg \g \bx_n)$, which depends
on the unknown $\mu^\star$.




We next approximate $\rmp^\star(\rmg  \g  \bx_n)$ by a point estimate $\delta_{\hat \rmg}$ at its value that maximizes the marginal likelihood of the data,
\begin{equation}
      \label{eqn:MMLE}
  \hat \rmg
  \in
  \argmax_{\rmg \in \cG}
  \log \int_{\cZ^{S_n}}
  \rmp(\bz_n  \g  \rmg) \,
  \prod_{\omega \in S_n} \rmp(x_\omega  \g  z_\omega)\,
  \dd \bz_n.
\end{equation}
The form of $\rmg \in \cG$ depends on the chosen symmetry. For example, in exchangeable data, $\rmg$ is a prior on $z$ and $\cG$ is the space of such priors. In separately-exchangeable array data, $\rmg$ are the functions described in \Cref{eq:array_g}; we will parameterize them with neural networks to capture a large space $\cG$.

This approximation is motivated by both asymptotic and computational considerations. The observed dataset $\bx_n$ is generated from a single draw $\rmg^\star$.  So asymptotically, we expect the ideal posterior $\rmp^\star(\rmg \g \bx_n)$ to concentrate at $\rmg^\star$ as $n \to \infty$. In particular, by \Cref{ergodic-rep-thm}, the law $\rmp(\bz \g \rmg)$ is $\Phi$-ergodic and hence identifiable (see, e.g., \cite[Lemma 27.1]{Kallenberg1997foundations} and \cite[Theorem 1.2]{Lindenstrauss2001}). We emphasize that the identified object is the ergodic law $\rmp(\bz \g \rmg)$, equivalently $\rmg$ up to measure-preserving transformations of its arguments. In the Aldous--Hoover case, for instance, the function $\rmg_{\mathrm{sep}}$ is identifiable only up to such transformations. Nonetheless, the induced prior of $\bz$ and the corresponding marginal of $\bx$ are fully identified.  Under a characteristic likelihood, such as the normal model with known variance, distinct priors $\rmp(\bz \g \rmg)$ induce distinct marginals $\rmp(\bx \g \rmg)$, so $\rmg$ is identifiable from $\rmp(\bx \g \rmg)$. Posterior concentration then follows from Doob’s theorem \cite[Ch.~6]{Ghosal2017}. Computationally, although the marginal likelihood in \Cref{eqn:MMLE} may be intractable, we can use variational inference to approximate it \citep{Blei2017}.

Finally, we substitute the point-estimate approximation of the prior into the posterior of \Cref{ideal posterior}. This gives the approximation
\begin{equation}
      \label{EB-posterior}
  \hat \rmp(\bz_n  \g  \bx_n) \deq \rmp(\bz_n  \g  \hat \rmg, \bx_n). 
\end{equation}
We estimate $\rmp(\bz_n \g \bx_n, \hat{g})$, given in \Cref{conditional posterior}, by exact or approximate methods. For example, we can use variational inference or MCMC.

Our procedure is called \emph{Bayesian empirical Bayes (BEB)}. We are given data $\bx_n$ and assume a symmetry group $\upPhi$ for the latent variables $\bz_n$. We use the ergodic representation theorem to form the expanded hierarchical model (\Cref{finite-data-model}), which includes the invariant variable \(\rmg\) and the corresponding ergodic family \(\rmp(\bz \g \rmg)\). We parameterize $\rmg$ and estimate it from data via maximum marginal likelihood (\Cref{eqn:MMLE}) or a suitable surrogate. Finally, we plug the estimate $\hat{\rmg}$ into the conditional posterior (\Cref{EB-posterior}).  \Cref{alg:BEB} presents this recipe.

\begin{algorithm}[t]
  \caption{Bayesian Empirical Bayes}
  \label{alg:BEB}
  \KwIn{Data $\bx_n$, likelihood model $\rmp(\bx_n  \g  \bz_n)$, group $\upPhi$.}
  \vspace{0.5em}
  \textbf{Steps:}
  \begin{enumerate}
  \item Derive the ergodic decomposition $\rmp(\bz_n  \g  \rmg)$ from $\upPhi$. 
  \item 
    Estimate $\hat \rmg$ by maximizing the marginal likelihood or a surrogate:
    \begin{equation*}
      \hat \rmg \in \arg\max_{\rmg \in \cG}  \log \rmp(\bx_n; \rmg). 
    \end{equation*}
  \item Approximate the posterior of the latent $\bz_n$:
\begin{equation*}
          \hat{\rmp}(\bz_n   \g  \bx_n) \propto \rmp(\bx_n  \g  \bz_n) \rmp(\bz_n   \g  \hat \rmg). 
\end{equation*}
  \end{enumerate}
  \vspace{0.5em}
  \KwOut{Approximate posterior $\hat{\rmp}(\bz_n \g \bx_n)$}
\end{algorithm}

%% file: tools.tex
\newcommand{\joint}{\textrm{joint}}
\newcommand{\perm}{\textrm{perm}}

\section{New Models for Empirical Bayes} \label{sec-EB-models}

We now use BEB to design new methods for empirical Bayes inference. In our first example, we use BEB to recover the classical empirical Bayes $\rmg$-modeling for the normal means model (\Cref{sec-NM}). This recovers classical EB methods~\citep{Efron2015deconvolution,Soloff2021}. We then extend EB to additional probabilistic symmetries: separate/joint exchangeability in arrays (\Cref{sec-EBMR}), stationarity in spatial data (\Cref{sec-spatial}), and relative exchangeability conditional on covariates (\Cref{sec-caeb} of the Appendix). In each case, we follow the recipe of \Cref{alg:BEB}. We start with a dataset, a symmetry group, and a likelihood. We then design an empirical Bayes model--tailored to the symmetry--and approximate the posterior of the latent structured variables.

\subsection{Classical Example Revisited: Normal Means Model}
\label{sec-NM}

Let $\Omega=\N$ and $\cX=\R^d$. The observed variables $\bx\in\cX^\Omega$ are generated from the heteroskedastic normal means model
\begin{equation} \label{NM}
\bz \sim \rmp^\star(\bz), \quad x_i \sim \cN\left(z_i, \Sigma_i \right),
\end{equation}
where $\Sigma_i\in\R^{d\times d}$ is known.


We assume the true prior $\rmp^\star(\bz)$ is exchangeable.  By de Finetti's theorem, we can write \Cref{NM} as the following generative process:
\begin{equation}
      \label{NM-exchangeable}
  \rmg_{\rm perm} \sim \mu^\star, \quad z_i \iid \rmg_{\rm perm}, \quad x_i \sim \cN \left(z_i, \Sigma_i \right). 
\end{equation}
Here $\rmg_{\rm perm}$ is a random prior over $z$. In short, de Finetti's representation specializes \Cref{ergodic-rep-thm} to sequences, with $\prod_{i = 1}^\infty \rmg_{\rm \perm}(z_i)$ the ergodic distribution. 

Under the expanded model, the posterior $\rmp^\star(\bz_n \mid \bx_n)$ has the form
\begin{equation}
      \label{eq:exch-z-post}
  \rmp^\star(\bz_n \mid \bx_n)
  = \int_{\cG}
  \prod_{i = 1}^n \rmp(z_i \mid \rmg_{\rm perm}, x_i) \, \rmp^\star(\rmg_{\rm perm} \mid \bx_n) \, \dd \rmg_{\rm perm}. 
\end{equation}
In Bayesian EB, we approximate $\rmp^\star(\rmg_{\rm perm} \mid
\bx_n)$ with its maximum marginal likelihood estimator (MMLE),
\begin{equation}
      \label{1d-NPMLE}
  \hat \rmg_{\rm perm} \in \argmax_{\rmg_{\rm perm} \in \cP(\R^d)}
  \log \int_{\cX^n} \prod_{i=1}^n \left(\rmp(x_i \mid z_i)  \rmg_{\rm
  perm}(z_i) \right) \dd \bz_n.
\end{equation}
If we restrict $\rmg_{\rm perm}$ to a parametric family, then this
recovers parametric empirical Bayes estimation, or type-II maximum
likelihood. If we consider all possible priors, then it recovers the
nonparametric maximum likelihood estimation (NPMLE) approach.


With $\hat{g}_{\rm perm}$ in hand, we approximate the target posterior
from \Cref{eq:exch-z-post},
\begin{equation*}
      \rmp^\star(\bz_n \mid \bx_n) \approx \prod_{i=1}^n \rmp(z_i \mid x_i, \hat{\rmg}_{\rm perm}),
\end{equation*}
where each posterior is
\begin{equation*}
      \rmp(z_i \g x_i, \hat{\rmg}_{\rm perm}) \propto \rmp(x_i \mid z_i) \hat{\rmg}_{\rm perm}(z_i).
\end{equation*}
NPMLE and the normal means model have been thoroughly studied in numerous works \citep{Kiefer1956,James1960,Jiang2009,Brown2009,Jiang2010,Xie2012EB,Efron2012,Efron2015deconvolution,Tan2016,Brown2018,Jiang2020,chen2022empirical,Soloff2021}.


If the likelihood is non-Gaussian, the same reasoning yields other nonparametric EB $\rmg$-modeling procedures. For example, with exponential data $x_i\mid z_i \sim \mathrm{Exp}(z_i)$, this approach recovers $\rmg$-modeling for the exponential sequence model \citep{Robbins1980exponential}. As another example, when $x_i$ is a sample variance and $z_i$ the true variance, we obtain the $\chi^2$ sequence model
$x_i \mid z_i \sim (z_i/k)\,\chi^2_k$ (for known $k$). The exchangeability assumption on $z_1, \cdots, z_n$ recovers the $\rmg$-modeling setting studied in \cite{ignatiadis2025empirical}.

\paragraph{Denoising an MNIST digit.} We now introduce a running example to illustrate the performance of Bayesian EB methods. The dataset is an image of "5" from the MNIST dataset of handwritten digits \citep{lecun2002gradient}. The $28\times28$ image is corrupted by i.i.d.\ Gaussian noise with variance equal to the empirical pixel variance of the clean image
\begin{equation*}
x_\omega\sim\cN(z_\omega^\star,\sigma_0^2),\qquad \omega\in[28]^2.
\end{equation*}
The goal is to recover the clean image $\bz^\star_{n,n}$ from the observed noisy image $\bx_{n,n}$. The maximum likelihood estimator is the simply the noisy image which has a MSE of $10.291$. The exchangeable sequence approach treats the problem as a simple normal means problem with known variance $\sigma_0^2$ and applies the NPMLE procedure. The resulting MSE is $3.61$--a substantial reduction from the error of the observed image. The first three panels in \Cref{fig:image-spatial} display the results. 

\subsection{Bayesian empirical Bayes for matrix recovery}
\label{sec-EBMR}

Let $\Omega=\N^2$ and $\cX=\R$.  We observe a noisy array $\bx_{n,p}=(x_{ij})_{i\in[n],\,j\in[p]}$ with latent signal $\bz_{n,p}=(z_{ij})_{i\in[n],\,j\in[p]}$ generated by
\begin{equation*}
      x_{ij}\sim\cN(z_{ij},\,\tau_{ij}^{-1}), \qquad \tau_{ij}>0.
\end{equation*}
Our goal is to estimate $\bz_{n,p}$ from $\bx_{n,p}$. We will use BEB to fit a structured prior that captures dependencies among entries.

We assume $\bz$ is \emph{separately exchangeable}: its distribution is invariant to independent permutations of rows and columns. Recall by the Aldous--Hoover theorem (\Cref{thm:aldous-hoover}), we can write the conditional distribution $\rmp(\bz_n \g \rmg_{\sep})$ as
\begin{equation}\label{eq:AH-sep}
u_i,v_j,u_{ij}\iid\unif[0,1],\qquad 
z_{ij}=\rmg_{\sep}(u_i,v_j,u_{ij}),
\end{equation}
where the invariant variable is a function $\rmg_{\sep}:[0,1]^3\to\R$. This conditional representation fully characterizes the finite-dimensional marginals of the ergodic distribution $\rmp(\bz\mid\rmg_{\sep})$, in the context of \Cref{ergodic-rep-thm}.

The corresponding Bayesian empirical Bayes model is
\begin{equation}   \label{eq:EBMR-sep}
\begin{aligned}
\rmg_{\sep} &\sim \mu^\star, \quad \\
u_i,v_j,u_{ij} &\iid \unif[0,1], \quad \quad \quad i \in [n], j \in [n]\\
z_{ij} &=\rmg_{\sep}(u_i,v_j,u_{ij}), \quad \\
x_{ij} &\sim \cN(z_{ij},\,\tau_{ij}^{-1}).
\end{aligned}
\end{equation}
We parameterize $\rmg_{\sep}$ by a neural network
$\rmg_{\sep,\theta}\in\cG_{\mathrm{NN}}$, a two-layer ReLU
feed-forward network mapping $[0,1]^3$ to $\R$.  We estimate $\theta$
by marginal maximum likelihood, now writing the conditional
distribution of $x_{ij}$ directly in terms of $\bu_n, \bv_p, \bu_{n,p}$,
\begin{equation}
      \label{eq:ebmr-MMLE}
  \hat\theta
\in
  \argmax_{\theta\in\Theta}
  \,
  \log
  \int \int \int
  \prod_{i=1}^n\prod_{j=1}^p
  \rmp\!\left(x_{ij}\mid\rmg_{\sep,\theta}(u_i,v_j,u_{ij})\right)
  \dd \bu_n \dd \bv_p \dd \bu_{n,p}.
\end{equation}
Exact evaluation of \Cref{eq:ebmr-MMLE} is intractable because the integral is high-dimensional.  We approximate it by maximizing an evidence lower bound (ELBO) under a structured variational family.  The estimate $\hat\rmg_{\sep,\theta}$ is the variational MMLE\@. See \Cref{sec-algorithms-ebmr} for details.

Given $\hat\rmg_{\sep,\theta}$, we approximate the empirical
Bayes posterior for $\bz_{n,p}$ with samples, through the posterior of
$\bu_n, \bv_p, \bu_{n,p}$. For each sample $s$,
\begin{equation*}
\begin{aligned}
   \bu_n^{(s)},\bv_p^{(s)},\bu_{n,p}^{(s)}
  &\sim
    \rmp(\bu_n, \bv_p, \bu_{n,p} \g
    \bx_{n,p},\hat\rmg_{\sep}), \\
  z_{ij}^{(s)}
  &=
    \hat\rmg_{\sep}(u_i^{(s)},v_j^{(s)},u_{ij}^{(s)}).   
\end{aligned}
\end{equation*}
The conditional posterior of the auxiliary latent variables is
\begin{equation}
    \rmp(\bu_n, \bv_p, \bu_{n,p} \g
    \bx_{n,p},\hat\rmg_{\sep})
  \propto
  \prod_{i=1}^n\prod_{j=1}^p
  \cN\!\left(x_{ij};
  \hat\rmg_{\sep}(u_i,v_j,u_{ij}),
  \tau_{ij}^{-1}\right),
  \label{eq:EBMR-posterior}
\end{equation}
which we sample from using the No-U-Turn Sampler (NUTS; \citealp{Hoffman2014}) or a variational approximation.


\paragraph{Related symmetries.} If the array is \emph{jointly
  exchangeable} then it is invariant when rows and columns are
permuted together. In this case, \Cref{eq:AH-sep} simplifies to
\begin{equation*}
    u_i,u_{ij}\iid\unif[0,1],\qquad 
z_{ij}=\rmg_{\joint}(u_i,u_j,u_{ij}),
\end{equation*}
with $\rmg_{\joint}:[0,1]^3\to\R$. We follow nearly the same procedure.

If the array is \emph{relatively exchangeable} conditioned on row covariates $\by_n=(y_i)_{i=1}^n$ with $y_i\in\cY$ and column covariates $\ba_p=(a_j)_{j=1}^p$ with $a_j\in\cA$, then \Cref{eq:AH-sep} expands to
\begin{equation*}
    u_i,v_j,u_{ij}\iid \unif[0,1],\qquad
z_{ij}=\rmg_{\rel}\bigl(y_i,a_j,u_i,v_j,u_{ij}\bigr), 
\end{equation*}
with $\rmg_{\rel}: \cY \times \cA \times [0,1]^3 \to \R$. This yields the covariate-informed Bayesian empirical Bayes procedure developed in \Cref{sec-caeb} of the Appendix.

If the array is \emph{fully exchangeable} then it is invariant to shuffling all of its cells. This case reduces to the algorithms of \Cref{sec-NM}, where $z_{ij}\iid\rmg_{\perm}$ for a random measure $\rmg_{\perm}$.


\paragraph{Denoising an MNIST digit (continued).} We revisit the example of denoising an MNIST "5"  from \Cref{fig:image-spatial}. The image is treated as a noisy matrix and we apply the proposed empirical Bayes matrix recovery (EBMR) methods under separate and joint exchangeability. We report $100\times\MSE$ between the posterior means $\hat{\bz}_{n,n}$ and the true pixel value $\bz^\star_{n,n}$. The fourth and fifth panels in \Cref{fig:image-spatial} show the results.  EBMR with a jointly exchangeable prior (EBMR-Joint) achieves the lowest error, followed by EBMR with a separately exchangeable prior (EBMR-Sep), then EBMR with a fully exchangeable prior (NPMLE). This ordering aligns with the symmetry hierarchy full $\Rightarrow$ separate $\Rightarrow$ joint: weaker (less restrictive) symmetry assumptions enlarge the class of admissible priors and increase flexibility. On the other hand, weaker symmetry assumptions also increase the sample and computational complexity for estimating $\rmg$. This creates a trade-off. In our setting, joint exchangeability offers a sweet spot between expressiveness and efficiency.

\subsection{Bayesian empirical Bayes for spatial regression}
\label{sec-spatial}
Let $\Omega = \R^d$ and $\cX=\R$.  We observe $\bx_n=\{x_\omega:\omega\in S_n\}$ at sites $S_n\subset\Omega$.  Each site has covariates $a_\omega=(a_{\omega,1},\ldots,a_{\omega,p})\in\R^p$.  The \textit{spatial regression model} is
\begin{equation}  \label{model:spatial-reg}
      x_\omega=\beta^\top a_\omega+z_\omega+\epsilon_\omega,\qquad
  \epsilon_\omega\sim\cN(0,\tau_\omega^{-1}),\qquad \tau_\omega>0,
\end{equation}
where $\beta\in\R^p$ are regression coefficients and $\bz=(z_\omega)_{\omega\in\Omega}$ is a latent process that captures spatial dependence.  We place a flat prior $\pi(\beta)\propto1$, so the estimator of $\beta$ coincides with weighted least squares. The resulting model along with \Cref{model:spatial-reg} coincide with the standard model in Bayesian spatial regression
\citep{handcock1993bayesian,Stein1999,DeOliveira1997spatial,Berger2001spatial,Banerjee2008,Banerjee2014spatial,Li2024spatial}.

We assume $\bz$ is \emph{stationary}: its distribution is invariant under translations $\phi_h:\omega\mapsto\omega+h$.  Following the ergodic decomposition in \Cref{eq:spatial_decomp}, any stationary process can be written as a mixture of \emph{ergodic stationary measures},
\begin{equation}
\rmg \sim\mu^\star,\qquad \bz\mid \rmg \sim \rmg,
\label{eq:stationary-mixture}
\end{equation}
where $\rmg$ is a stationary probability measure on $\R^\Omega$.  Following the BEB method, our goal is to estimate the invariant variable $\rmg$ from data.

\paragraph{Defining stationary measures with Gaussian processes.} The space of stationary measures is complex and so, in practice, we restrict the ergodic family $\cG$ to stationary ergodic Gaussian processes.  A Gaussian process $\GP(0,k)$ is stationary if its covariance function depends only on the displacement $k(\omega, \omega') = f(\omega - \omega') := k(\Delta)$. Each stationary covariance function $k$ with an atomless spectral density (as described below) defines an ergodic stationary Gaussian measure on $\R^\Omega$ \citep{maruyama1949harmonic}. So we can write
\begin{equation}
    k\sim \mu(k),\qquad \bz\mid k\sim \GP(0,k),
\label{eq:stationary-GP}
\end{equation}
where $\mu(k)$ is a distribution over stationary kernels.

How can we characterize the stationary covariance functions?  \textit{Bochner’s theorem} \citep{bochner2005harmonic} states that $k$ is stationary and positive definite if and only if it can be written as
\begin{equation}
      k(\Delta)=\int_{\R^d} e^{\,2\pi i\, s^\top \Delta}\, \psi(s)\,ds,
\label{eq:bochner}
\end{equation}
for some nonnegative integrable function $\psi$.  The function $\psi:\R^d\to[0,\infty)$ is called the \emph{spectral density}.  It describes how the variance of the process is distributed across spatial frequencies. The associated spectral measure is \emph{atomless} if it assigns zero mass to every singleton, i.e., $\Psi({s})=0$ for all $s\in\R^d$ (in particular, if $\Psi$ has density $\psi$, it is atomless). Bochner’s theorem gives a one-to-one correspondence between stationary Gaussian measures and their spectral densities.  We therefore identify each ergodic stationary measure $\rmg$ with its spectral density $\psi$:
\begin{equation}
  \psi\sim\mu^\star,\qquad \bz\mid\psi\sim\GP(0,k_\psi),
\label{eq:EB-spatial}
\end{equation}
where $k_\psi$ is the kernel defined by \eqref{eq:bochner}.  The empirical Bayes task is to estimate $\psi$ from the marginal distribution of the observed data.
\paragraph{Bayesian empirical Bayes for spatial regression.} We have defined the invariant variable $\rmg$ as the spectral density of a stationary kernel for a Gaussian process. With this definition, the BEB model is
\begin{equation}
\bz \sim \GP(0,k_\psi), \qquad
\beta \sim \pi(\beta)\propto1, \qquad
x_\omega \sim \cN(\beta^\top a_\omega+z_\omega,\tau_\omega^{-1}).
\label{eq:EB-spatial-full}    
\end{equation}
We estimate the invariant variable $\psi$ by maximizing the marginal likelihood,
\begin{equation}
\hat\psi
\in
\argmax_{\psi\in\cG}
\log
\int
\Bigl[\prod_{\omega\in S_n}\rmp(x_\omega\mid z_\omega,\beta)\Bigr]
\rmp(\bz_n\mid\psi)\pi(\beta)\,d\beta\,d\bz_n.
\label{eq:EB-spatial-MMLE}
\end{equation}
This is the empirical Bayes estimate of the stationary measure.  Given $\hat\psi$, we compute the posterior $\rmp(\beta,\bz_n\mid\bx_n,\hat\psi)$, which is Gaussian with known mean and covariance.  The BEB method thus fits the spectral density $\psi$, or equivalently the kernel $k_\psi$, that best explains the marginal distribution of the observed data.  

In practice, we model $\psi$ as a mixture of Gaussians, and fit it by maximum marginal likelihood. The details of this procedure, as well as how we perform simultaneous posterior inference, are given in \Cref{sec-alg-spatial} of the Appendix. 

In the spatial regression example, we model the spectral density of the Gaussian-process kernel as a mixture of Gaussian distributions. Under this model, the empirical Bayes procedure reduces to selecting the GP kernel parameters by maximum likelihood \citep{Rasmussen2006}. This approach is a common practice in the GP literature. It comes with strong theoretical guarantees and retains the conjugacy of the GP posterior \citep{Loh2000,Bachoc2014,Szab2015,Xu2017}. What we see here is that this approach can be interpreted as an instance of Bayesian empirical Bayes.

\paragraph{Denoising an MNIST digit (continued).} The image of a handwritten digit does not change its meaning if we shift the image, so stationarity is a reasonable assumption. We fit EB spatial regression with a constant covariate ($a_\omega\equiv1$) and a Gaussian-mixture spectral density with $K=10$, using the inference and sampling steps outlined in \Cref{sec-alg-spatial} of the Appendix. The rightmost panel of \Cref{fig:image-spatial} shows that EB spatial regression denoises better than NPMLE but not as well as EBMR-sep or EBMR-joint. Nonetheless, all EB methods improve over the MLE. Moreover, all BEB methods that use symmetries beyond exchangeability improve over the NPMLE.

\begin{figure}[t!]
\centering
\includegraphics[width=1\textwidth]{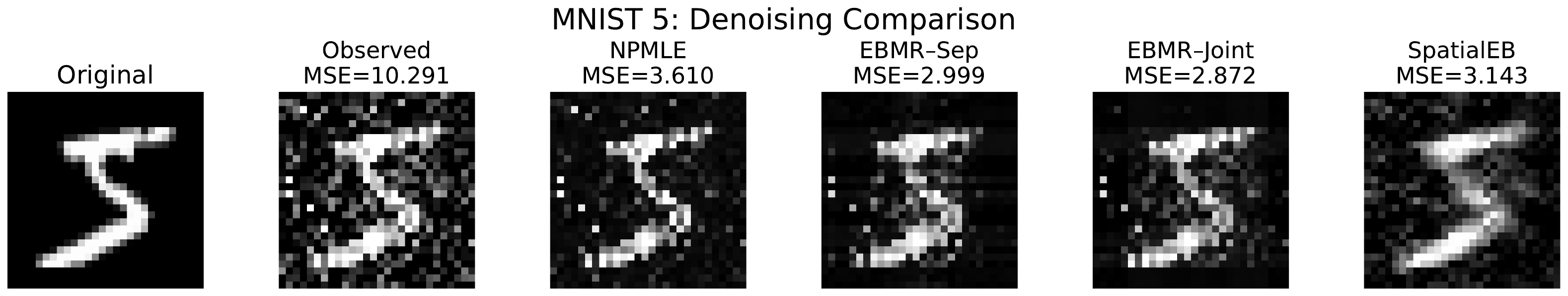}
\caption{\emph{Denoising a noisy MNIST “5” with Bayesian empirical Bayes methods. }
Top: original (left), noisy (middle), and NPMLE (right).
Bottom: EBMR-sep (left), EBMR-joint (middle), and EB spatial regression (right).
Errors are $100\times\MSE$ of posterior means; EBMR with a jointly exchangeable prior performs best, followed by EBMR with a separately exchangeable prior and EB spatial regression, then NPMLE.}
\label{fig:image-spatial}
\end{figure}



%% file: Tables/mse_table2.tex
\begin{table}[!t]
\centering
\caption{\emph{EBMR-Sep attains the lowest R-MSE in 50 of 60 settings.} Matrix-recovery performance for NPMLE, EBMR with a separately exchangeable prior (Sep), and EBMF across simulated settings. Reported values are the median relative MSE (R-MSE), defined as each method's MSE as a percentage of the MSE of the MLE. The lowest R-MSE among the three is shown in \textbf{bold}. An oracle that knows the true $\rmg$ and performs exact posterior inference is added as a reference.}
\label{tab:simul_EBMR}
\begingroup
\setlength{\tabcolsep}{3.5pt}
\renewcommand{\arraystretch}{1}
\scriptsize
\begin{adjustbox}{max width=\textwidth}
\begin{tabular}{ccccccccccccccccccccccc}
\toprule
\textbf{n} & \textbf{p} & $\boldsymbol{\tau}$ &
\multicolumn{4}{c}{\textbf{Linear}} &
\multicolumn{4}{c}{\textbf{Sine--log}} &
\multicolumn{4}{c}{\textbf{Sine--cos}} &
\multicolumn{4}{c}{\textbf{Tanh}} &
\multicolumn{4}{c}{\textbf{Reciprocal}} \\
\cmidrule(lr){4-7}\cmidrule(lr){8-11}\cmidrule(lr){12-15}\cmidrule(lr){16-19}\cmidrule(lr){20-23}
& & & \textbf{NPMLE} & \textbf{Sep} & \textbf{EBMF} & \textbf{Oracle} & \textbf{NPMLE} & \textbf{Sep} & \textbf{EBMF} & \textbf{Oracle} & \textbf{NPMLE} & \textbf{Sep} & \textbf{EBMF} & \textbf{Oracle} & \textbf{NPMLE} & \textbf{Sep} & \textbf{EBMF} & \textbf{Oracle} & \textbf{NPMLE} & \textbf{Sep} & \textbf{EBMF} & \textbf{Oracle} \\
\midrule
20 & 20 & 0.1 & 3.23 & \textbf{2.73} & 9.16 & 2.17 & \textbf{2.09} & 3.09 & 9.24 & 1.35 & 2.85 & \textbf{2.01} & 2.14 & 1.30 & 1.08 & \textbf{0.68} & 7.57 & 0.15 & 0.88 & \textbf{0.76} & 1.84 & 0.08 \\
20 & 20 & 0.25 & 7.28 & \textbf{6.42} & 10.15 & 4.74 & \textbf{3.83} & 4.81 & 9.95 & 3.04 & 5.64 & \textbf{4.90} & 5.36 & 2.84 & \textbf{0.96} & 1.41 & 8.63 & 0.46 & \textbf{0.75} & 1.04 & 4.61 & 0.23 \\
20 & 20 & 1 & 23.35 & \textbf{15.54} & 18.56 & 13.94 & 13.47 & \textbf{10.87} & 14.83 & 8.69 & 18.74 & \textbf{7.87} & 9.24 & 5.07 & 2.87 & \textbf{2.13} & 9.92 & 1.62 & 1.83 & \textbf{1.41} & 9.91 & 0.92 \\
20 & 20 & 4 & 53.24 & \textbf{33.18} & 43.95 & 29.51 & 36.01 & \textbf{25.97} & 31.50 & 22.58 & 48.74 & \textbf{12.78} & 13.19 & 9.06 & 8.93 & \textbf{6.20} & 13.07 & 5.33 & \textbf{4.65} & 4.74 & 10.72 & 3.18 \\
\midrule
20 & 50 & 0.1 & 2.88 & \textbf{2.50} & 8.26 & 2.20 & \textbf{2.00} & 2.03 & 10.38 & 1.37 & 2.49 & \textbf{1.98} & 2.06 & 1.38 & 0.65 & \textbf{0.48} & 7.75 & 0.18 & \textbf{0.61} & 0.63 & 1.75 & 0.09 \\
20 & 50 & 0.25 & 6.38 & \textbf{5.38} & 9.79 & 4.47 & 4.17 & \textbf{3.71} & 8.27 & 2.75 & 5.31 & \textbf{3.27} & 5.14 & 2.64 & 0.87 & \textbf{0.59} & 5.68 & 0.43 & 0.70 & \textbf{0.39} & 4.38 & 0.21 \\
20 & 50 & 1 & 20.13 & \textbf{12.43} & 22.78 & 11.73 & 14.11 & \textbf{9.53} & 18.03 & 8.07 & 17.94 & \textbf{6.45} & 20.55 & 4.71 & 2.29 & \textbf{1.87} & 8.05 & 1.58 & 1.73 & \textbf{1.17} & 7.35 & 0.82 \\
20 & 50 & 4 & 49.97 & \textbf{30.51} & 69.45 & 29.52 & 38.82 & \textbf{22.26} & 49.92 & 18.96 & 44.95 & \textbf{10.13} & 37.59 & 7.46 & \textbf{5.55} & 5.62 & 10.64 & 4.61 & \textbf{3.15} & 3.32 & 9.00 & 2.70 \\
\midrule
50 & 50 & 0.1 & 2.69 & \textbf{2.16} & 4.73 & 1.95 & 1.69 & \textbf{1.56} & 4.00 & 1.34 & 1.88 & \textbf{1.27} & 1.84 & 1.03 & 0.37 & \textbf{0.25} & 3.65 & 0.16 & 0.25 & \textbf{0.15} & 1.95 & 0.08 \\
50 & 50 & 0.25 & 6.21 & \textbf{4.13} & 6.00 & 3.81 & 3.86 & \textbf{3.17} & 5.13 & 2.73 & 4.49 & \textbf{2.32} & 4.61 & 1.94 & 0.65 & \textbf{0.53} & 4.05 & 0.40 & 0.38 & \textbf{0.30} & 4.83 & 0.21 \\
50 & 50 & 1 & 20.65 & \textbf{10.83} & 12.23 & 10.58 & 13.54 & \textbf{7.50} & 8.63 & 7.01 & 15.73 & \textbf{3.94} & 4.48 & 2.98 & 2.27 & \textbf{1.65} & 4.84 & 1.45 & 1.24 & \textbf{0.99} & 4.32 & 0.79 \\
50 & 50 & 4 & 50.99 & \textbf{27.49} & 37.36 & 27.34 & 38.32 & \textbf{19.57} & 23.11 & 17.21 & 42.34 & 6.60 & \textbf{6.28} & 5.48 & 8.07 & \textbf{4.98} & 7.77 & 4.41 & 4.03 & \textbf{2.98} & 5.09 & 2.51 \\
\bottomrule
\end{tabular}
\end{adjustbox}
\endgroup
\end{table}

%% file: Tables/mse_table4.tex
\begin{table}[t!]
\centering
\scriptsize
\begin{adjustbox}{max width=\textwidth}
\begin{tabular}{cc cccccc cccccc cccccc}
\toprule
$n$ & $p$ & \multicolumn{6}{c}{Linear} & \multicolumn{6}{c}{Logistic} & \multicolumn{6}{c}{Nonlinear} \\
\cmidrule(lr){3-8} \cmidrule(lr){9-14} \cmidrule(lr){15-20}
 &  & CIEB & SEP & Ridge & NPMLE & EBMF & Oracle & CIEB & SEP & Ridge & NPMLE & EBMF & Oracle & CIEB & SEP & Ridge & NPMLE & EBMF & Oracle \\
\midrule
20 & 20 & 35.21 & 16.95 & \textbf{16.06} & 77.19 & 24.66 & 10.68 & 34.38 & 16.75 & \textbf{13.17} & 82.69 & 20.83 & 8.00 & 33.05 & \textbf{17.85} & 25.68 & 81.10 & 20.88 & 8.95 \\
20 & 50 & 20.09 & \textbf{13.95} & 14.51 & 75.59 & 28.82 & 10.00 & 19.18 & 13.35 & \textbf{10.72} & 75.71 & 27.20 & 8.12 & 16.86 & \textbf{14.76} & 28.92 & 79.83 & 43.26 & 7.12 \\
20 & 100 & 14.73 & \textbf{13.32} & 13.58 & 76.71 & 23.27 & 10.00 & 12.51 & 12.37 & \textbf{9.75} & 76.27 & 22.40 & 8.24 & \textbf{11.58} & 12.86 & 26.91 & 78.65 & 31.35 & 6.21 \\
50 & 50 & 14.09 & \textbf{13.16} & 13.63 & 80.47 & 16.50 & 9.89 & 11.80 & 12.64 & \textbf{9.67} & 80.19 & 15.30 & 8.23 & \textbf{9.69} & 11.60 & 26.45 & 81.24 & 10.82 & 4.59 \\
50 & 100 & \textbf{11.19} & 11.97 & 13.34 & 82.64 & 17.87 & 9.51 & 9.63 & 11.33 & \textbf{9.20} & 77.83 & 15.69 & 8.08 & \textbf{7.00} & 10.68 & 25.20 & 79.03 & 11.71 & 4.44 \\
100 & 100 & \textbf{10.09} & 12.99 & 13.28 & 81.93 & 12.26 & 8.92 & \textbf{9.10} & 11.72 & 9.12 & 80.97 & 11.84 & 8.14 & \textbf{4.93} & 10.98 & 26.65 & 83.67 & 7.52 & 3.68 \\
\bottomrule
\end{tabular}
\end{adjustbox}
\caption{\emph{Covariate-informed EB attains the lowest R-MSE in large-sample settings.} CIEB (covariate-informed EB), SEP (EBMR with a separately exchangeable prior), a covariate ridge regression (Ridge), NPMLE, and EBMF across linear, logistic, and nonlinear generative models, with noise drawn from $\cN(0,1)$. Each cell reports the median R-MSE (lower is better) as a percentage of the MLE risk. The lowest among CIEB, SEP, Ridge, NPMLE, and EBMF is in \textbf{bold}. An oracle that is given the true $\rmg_{\rel}$ is shown as a lower-bound reference.}
\label{tab:simul_CIEB}
\end{table}

%% file: Tables/mse_table_full_2lines.tex
\begin{table}[!t]
\centering
\caption{\emph{Full matrix-recovery results across all settings.} Median relative MSE (R-MSE), each method's MSE as a percentage of the MLE's, for the separately exchangeable simulation of \Cref{sec-sim-EBMR}. The lowest R-MSE among NPMLE, EBMR-Sep, and EBMF is in \textbf{bold}. The oracle uses the true $\rmg$; Soft-Thr.\ and Rank-SVD are non-EB matrix-denoising baselines, reported where available.}
\label{tab:simul_EBMR_full}
\begingroup
\setlength{\tabcolsep}{5pt}
\renewcommand{\arraystretch}{0.95}
\scriptsize
\begin{tabular}{ccccccccc}
\toprule
\textbf{n} & \textbf{p} & $\boldsymbol{\tau}$ & \textbf{NPMLE} & \textbf{EBMR-Sep} & \textbf{EBMF} & \textbf{Oracle} & \textbf{Soft-Thr.} & \textbf{Rank-SVD} \\
\midrule
\multicolumn{9}{l}{\textit{Linear}} \\
20 & 20 & 0.1 & 3.23 & \textbf{2.73} & 9.16 & 2.17 & 16.49 & 100.00 \\
20 & 20 & 0.25 & 7.28 & \textbf{6.42} & 10.15 & 4.74 & 21.80 & 100.00 \\
20 & 20 & 1 & 23.35 & \textbf{15.54} & 18.56 & 13.94 & 31.45 & 100.00 \\
20 & 20 & 4 & 53.24 & \textbf{33.18} & 43.95 & 29.51 & 57.56 & 100.00 \\
20 & 50 & 0.1 & 2.88 & \textbf{2.50} & 8.26 & 2.20 & 12.67 & 100.00 \\
20 & 50 & 0.25 & 6.38 & \textbf{5.38} & 9.79 & 4.47 & 16.97 & 100.00 \\
20 & 50 & 1 & 20.13 & \textbf{12.43} & 22.78 & 11.73 & 24.23 & 100.00 \\
20 & 50 & 4 & 49.97 & \textbf{30.51} & 69.45 & 29.52 & 51.29 & 100.00 \\
50 & 50 & 0.1 & 2.69 & \textbf{2.16} & 4.73 & 1.95 & 8.73 & 81.42 \\
50 & 50 & 0.25 & 6.21 & \textbf{4.13} & 6.00 & 3.81 & 11.35 & 81.75 \\
50 & 50 & 1 & 20.65 & \textbf{10.83} & 12.23 & 10.58 & 18.25 & 82.81 \\
50 & 50 & 4 & 50.99 & \textbf{27.49} & 37.36 & 27.34 & 43.69 & 87.62 \\
\midrule
\multicolumn{9}{l}{\textit{Sine-log}} \\
20 & 20 & 0.1 & \textbf{2.09} & 3.09 & 9.24 & 1.35 & 8.73 & 100.00 \\
20 & 20 & 0.25 & \textbf{3.83} & 4.81 & 9.95 & 3.04 & 15.63 & 100.00 \\
20 & 20 & 1 & 13.47 & \textbf{10.87} & 14.83 & 8.69 & 25.97 & 100.00 \\
20 & 20 & 4 & 36.01 & \textbf{25.97} & 31.50 & 22.58 & 41.28 & 100.00 \\
20 & 50 & 0.1 & \textbf{2.00} & 2.03 & 10.38 & 1.37 & 8.16 & 100.00 \\
20 & 50 & 0.25 & 4.17 & \textbf{3.71} & 8.27 & 2.75 & 13.08 & 100.00 \\
20 & 50 & 1 & 14.11 & \textbf{9.53} & 18.03 & 8.07 & 19.85 & 100.00 \\
20 & 50 & 4 & 38.82 & \textbf{22.26} & 49.92 & 18.96 & 36.02 & 100.00 \\
50 & 50 & 0.1 & 1.69 & \textbf{1.56} & 4.00 & 1.34 & 6.43 & 81.56 \\
50 & 50 & 0.25 & 3.86 & \textbf{3.17} & 5.13 & 2.73 & 9.21 & 81.52 \\
50 & 50 & 1 & 13.54 & \textbf{7.50} & 8.63 & 7.01 & 14.72 & 82.00 \\
50 & 50 & 4 & 38.32 & \textbf{19.57} & 23.11 & 17.21 & 29.69 & 84.01 \\
\midrule
\multicolumn{9}{l}{\textit{Sine-cos}} \\
20 & 20 & 0.1 & 2.85 & \textbf{2.01} & 2.14 & 1.30 & 1.52 & 100.00 \\
20 & 20 & 0.25 & 5.64 & \textbf{4.90} & 5.36 & 2.84 & 3.81 & 100.00 \\
20 & 20 & 1 & 18.74 & \textbf{7.87} & 9.24 & 5.07 & 11.92 & 100.00 \\
20 & 20 & 4 & 48.74 & \textbf{12.78} & 13.19 & 9.06 & 23.18 & 100.00 \\
20 & 50 & 0.1 & 2.49 & \textbf{1.98} & 2.06 & 1.38 & 1.67 & 100.00 \\
20 & 50 & 0.25 & 5.31 & \textbf{3.27} & 5.14 & 2.64 & 3.95 & 100.00 \\
20 & 50 & 1 & 17.94 & \textbf{6.45} & 20.55 & 4.71 & 10.80 & 100.00 \\
20 & 50 & 4 & 44.95 & \textbf{10.13} & 37.59 & 7.46 & 16.89 & 100.00 \\
50 & 50 & 0.1 & 1.88 & \textbf{1.27} & 1.84 & 1.03 & 1.61 & 81.83 \\
50 & 50 & 0.25 & 4.49 & \textbf{2.32} & 4.61 & 1.94 & 3.63 & 81.61 \\
50 & 50 & 1 & 15.73 & \textbf{3.94} & 4.48 & 2.98 & 7.30 & 81.34 \\
50 & 50 & 4 & 42.34 & 6.60 & \textbf{6.28} & 5.48 & 11.40 & 81.71 \\
\midrule
\multicolumn{9}{l}{\textit{Tanh}} \\
20 & 20 & 0.1 & 1.08 & \textbf{0.68} & 7.57 & 0.15 & 7.53 & 100.00 \\
20 & 20 & 0.25 & \textbf{0.96} & 1.41 & 8.63 & 0.46 & 14.10 & 100.00 \\
20 & 20 & 1 & 2.87 & \textbf{2.13} & 9.92 & 1.62 & 23.07 & 100.00 \\
20 & 20 & 4 & 8.93 & \textbf{6.20} & 13.07 & 5.33 & 26.22 & 100.00 \\
20 & 50 & 0.1 & 0.65 & \textbf{0.48} & 7.75 & 0.18 & 6.78 & 100.00 \\
20 & 50 & 0.25 & 0.87 & \textbf{0.59} & 5.68 & 0.43 & 10.87 & 100.00 \\
20 & 50 & 1 & 2.29 & \textbf{1.87} & 8.05 & 1.58 & 14.93 & 100.00 \\
20 & 50 & 4 & \textbf{5.55} & 5.62 & 10.64 & 4.61 & 20.54 & 100.00 \\
50 & 50 & 0.1 & 0.37 & \textbf{0.25} & 3.65 & 0.16 & 5.60 & 81.30 \\
50 & 50 & 0.25 & 0.65 & \textbf{0.53} & 4.05 & 0.40 & 7.76 & 81.24 \\
50 & 50 & 1 & 2.27 & \textbf{1.65} & 4.84 & 1.45 & 10.23 & 81.67 \\
50 & 50 & 4 & 8.07 & \textbf{4.98} & 7.77 & 4.41 & 14.18 & 81.52 \\
\midrule
\multicolumn{9}{l}{\textit{Reciprocal}} \\
20 & 20 & 0.1 & 0.88 & \textbf{0.76} & 1.84 & 0.08 & 1.83 & 100.00 \\
20 & 20 & 0.25 & \textbf{0.75} & 1.04 & 4.61 & 0.23 & 4.58 & 100.00 \\
20 & 20 & 1 & 1.83 & \textbf{1.41} & 9.91 & 0.92 & 13.32 & 100.00 \\
20 & 20 & 4 & \textbf{4.65} & 4.74 & 10.72 & 3.18 & 22.39 & 100.00 \\
20 & 50 & 0.1 & \textbf{0.61} & 0.63 & 1.75 & 0.09 & 1.91 & 100.00 \\
20 & 50 & 0.25 & 0.70 & \textbf{0.39} & 4.38 & 0.21 & 4.50 & 100.00 \\
20 & 50 & 1 & 1.73 & \textbf{1.17} & 7.35 & 0.82 & 10.70 & 100.00 \\
20 & 50 & 4 & \textbf{3.15} & 3.32 & 9.00 & 2.70 & 14.87 & 100.00 \\
50 & 50 & 0.1 & 0.25 & \textbf{0.15} & 1.95 & 0.08 & 1.81 & 81.53 \\
50 & 50 & 0.25 & 0.38 & \textbf{0.30} & 4.83 & 0.21 & 4.09 & 81.46 \\
50 & 50 & 1 & 1.24 & \textbf{0.99} & 4.32 & 0.79 & 8.03 & 81.33 \\
50 & 50 & 4 & 4.03 & \textbf{2.98} & 5.09 & 2.51 & 10.32 & 81.25 \\
\bottomrule
\end{tabular}
\endgroup
\end{table}